\documentclass[aps,prb,floatfix,twocolumn,groupedaddress,amsmath]{revtex4-2}

\usepackage{txfonts}
\usepackage{graphicx}
\usepackage{hyperref}
\hypersetup{
    colorlinks=true,
    linkcolor=blue,
    filecolor=magenta,      
    urlcolor=blue,
    citecolor=blue
    }
\begin{document}

\title{Shot Noise in Mesoscopic Systems: from Single Particles to Quantum Liquids}

\author{Kensuke Kobayashi$^{1,2,3}$ and Masayuki Hashisaka$^{4,5}$}
\affiliation{$^1$Institute for Physics of Intelligence (IPI) \& Department of Physics, Graduate School of Science, The University of Tokyo, Hongo 7-3-1, Bunkyo-ku, Tokyo 113-0033, Japan}
\affiliation{$^2$Trans-scale Quantum Science Institute, The University of Tokyo, Bunkyo-ku, Tokyo 113-0033, Japan}
\affiliation{$^3$Graduate School of Science, Osaka University, 1-1 Machikaneyama, Toyonaka, Osaka 560-0043, Japan}
\affiliation{$^4$NTT Basic Research Laboratories, NTT Corporation, 3-1 Morinosato-Wakamiya, Atsugi, Kanagawa 243-0198, Japan}
\affiliation{$^5$JST, PRESTO, 4-1-8 Honcho, Kawaguchi, Saitama 332-0012, Japan}


\begin{abstract}
Shot noise, originating from the discrete nature of electric charge, is generated by scattering processes. Shot-noise measurements have revealed microscopic charge dynamics in various quantum transport phenomena. In particular, beyond the single-particle picture, such measurements have proved to be powerful ways to investigate electron correlation in quantum liquids. Here, we review the recent progress of shot-noise measurements in mesoscopic physics. This review summarizes the basics of shot-noise theory based on the Landauer-B\"{u}ttiker formalism, measurement techniques used in previous studies, and several recent experiments demonstrating electron scattering processes. We then discuss three different kinds of quantum liquids, namely those formed by, respectively, the Kondo effect, the fractional quantum Hall effect, and superconductivity. Finally, we discuss current noise within the framework of nonequilibrium statistical physics and review related experiments. We hope that this review will convey the significance of shot-noise measurements to a broad range of researchers in condensed matter physics.
\end{abstract}

\maketitle
\tableofcontents

\section{Introduction}
Semiconductor mesoscopic systems have been extensively studied since the establishment of microfabrication techniques~\cite{DattaETMS,ImryIMP,IhnEQTMSS} in the 1980s. These systems allow us to artificially realize and control various quantum phenomena of electron charge and spin. In this review, we summarize the basics and recent progress of shot-noise measurements in mesoscopic physics. While conductance, the most fundamental transport property, provides information on the time-averaged electron transport, shot noise offers more in-depth insights into non-equilibrium electron dynamics. 

Shot noise, one of the most important topics in mesoscopic physics, has been theoretically studied since the early 1990s~\cite{deJong1997,BlanterPR2000,MartinBook}. However, initially, shot-noise measurements were not commonly performed in experiments, because of technical difficulties. What impressed researchers with the importance of shot-noise measurements was the detection of fractionally charged quasiparticles in fractional quantum Hall (QH) systems, which led to the Nobel Prize in Physics in 1998 for the discovery of the fractional QH effect~\cite{SaminadayarPRL1997,de-PicciottoNature1997}. Then, several excellent reviews written by theorists around 2000 have extensively promoted  shot-noise research~\cite{deJong1997,BlanterPR2000,MartinBook}. This review will introduce various experiments, including those performed by us~\cite{FerrierNatPhys2016,HashisakaPRL2015}, reported since these early reviews. Although there already exists an instructive review recently written by experimentalists~\cite{PiatrushaJETP2018}, it is worth reviewing the shot-noise measurements over a broad range of mesoscopic systems, from experiments understood within the single-particle picture to those targeting quantum many-body physics.

The central idea of this review is as follows. Current and noise, corresponding to the average and variance of the number of electrons passing through a conductor per unit time, respectively, provide different information on a transport phenomenon. For example, because the shot-noise intensity is given as the product of the current and the effective charge of a charge carrier, the effective charge can be evaluated by measuring both the current and shot noise. Actually, combining these measurements has revealed a two-particle scattering process in the Kondo effect, fractional charges in fractional QH systems, and Cooper pairs in superconducting junctions. This review focuses on such a combination of conductance and shot-noise measurements.

This review is organized as follows. In Section~\ref{sec:current_noise}, we discuss the basics of the current-noise theory within the Landauer-B\"{u}ttiker formalism. Section~\ref{sec:techniques} presents the experimental techniques for current-noise measurements. Section~\ref{sec:shotexample} introduces several experiments in which the transport phenomena can be understood within a single-particle picture, including those on a quantum point contact (QPC), two-channel or multichannel systems, and fermion quantum optics. Section~\ref{sec:quantumliquid} describes several shot-noise measurements performed on quantum liquids, or equivalently, quantum many-body states, namely the Kondo states, the fractional QH states, and superconductors. Section~\ref{sec:fluctuationtheorem} introduces a noise study based on the fluctuation theorem, a different approach from that using the Landauer-B\"{u}ttiker picture. Section~\ref{sec:closing} summarizes this review with reference to future experimental issues.

\section{Basics of current-noise theory}
\label{sec:current_noise}
\subsection{Classical current-noise theory}
\label{sec:classicalnoise}

\begin{figure}[!b]
\center 
\includegraphics[width=8.5cm]{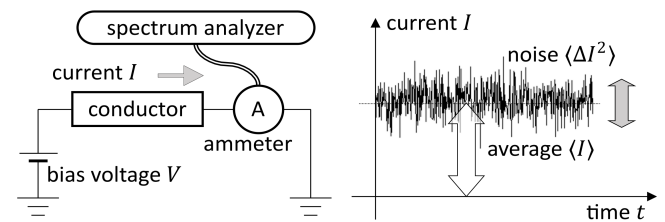}
\caption{Current- and noise-measurement setup. A constant bias voltage $V$ is applied to a conductor and current $I$ is measured with a time-resolved high-precision ammeter. The current noise $\langle \Delta I^2 \rangle$ is evaluated by using a spectrum analyzer.} \label{noisemeasurement}
\end{figure}

Suppose a bias voltage $V$ is applied to a conductor, for example, a resistor, as shown in Fig.~\ref{noisemeasurement}. We monitor the time $t$ dependence of current $I(t)$ with a high-precision ammeter. Besides a time-averaged value $\langle I(t) \rangle$ of current, there always exists a fluctuation (noise) $\Delta I(t) \equiv I(t) -\langle I(t) \rangle$ around it.

Let us consider the Fourier transform of $\Delta I(t)$ over the measurement time interval $-\tau/2\leq t \leq \tau/2$: $\Delta I(\omega) \equiv \int_{-\tau/2}^{\tau/2}dt \Delta I(t)e^{i \omega t} $, where $\omega \equiv 2\pi f$ is the angular frequency for frequency $f$. The time-averaged variance of the current noise $\Delta I(t)$, given by $\langle \Delta I(t)^2\rangle \equiv \lim_{\tau\rightarrow \infty} \frac{1}{\tau} \int_{-\tau/2}^{\tau/2} \Delta I(t)^2 dt $, satisfies the following relation known as the Parseval theorem. 
\begin{equation}
\langle \Delta I(t)^2 \rangle
= \frac{1}{2\pi}\lim_{\tau\rightarrow \infty}\frac{1}{\tau}\int_{-\infty}^{\infty}  \vert \Delta I(\omega) \vert^2 d\omega.
\end{equation}
The power spectral density (PSD) $S(\omega)$ of the current is defined as
\begin{equation}
\begin{split}
&S(\omega) \equiv \lim_{\tau\rightarrow \infty} 
\frac{2}{\tau}\vert \Delta I(\omega) \vert^2  \\
&=\lim_{\tau\rightarrow \infty} 
\frac{2}{\tau}\int_{-\tau/2}^{\tau/2}dt \int_{-\tau/2}^{\tau/2}dt'
 \Delta I(t)\Delta I(t') e^{i \omega (t-t')}.
\label{Definition_PSD_classic}
\end{split}
\end{equation}
Since $I(t)$ and $S(\omega)$ are real, $\Delta I(-\omega) =[\Delta I(\omega)]^*$ and $S(\omega)=S(-\omega)$. Therefore, we can reduce the angular frequency $\omega$ to the non-negative range by redefining $S(\omega)$ as twice of itself, which accounts for the factor 2 in Eq.~(\ref{Definition_PSD_classic}). Henceforth, we use $\omega \in [0, \infty]$ in principle~\cite{NyquistPR1928,JohnsonPR1928,BeenakkerPT2003}. Equation (\ref{Definition_PSD_classic}) satisfies the following relation:
\begin{equation}
\langle \Delta I(t)^2 \rangle =\frac{1}{2\pi}\int_{0}^{\infty} S(\omega) d\omega.
\label{eq_PSD2}
\end{equation}
This equation quantifies the current-noise intensity with the PSD measured with a spectrum analyzer (see Fig.~\ref{noisemeasurement}).

In some textbooks and technical literature, PSD is often defined, based on Eq.~(\ref{eq_PSD2}), as~\cite{ZielNoise1986}
\begin{equation}
S(f) =\frac{2\langle \Delta I(t)^2\rangle_f}{\Delta f},
\label{noisedef2}
\end{equation}
where $\langle \Delta I(t)^2\rangle_f$ is the current noise measured around the frequency $f$ with the bandwidth of $\Delta f$.

The current is nothing but the average of charge $Q$, or the number of electrons $N$, passing through a device per unit time:
\begin{equation}
\langle I \rangle=\frac{\langle Q \rangle}{\tau}=\frac{e\langle N \rangle}{\tau}, \quad [\textrm{A}]=\left[\frac{\textrm{C}}{\textrm{s}}\right],
\label{Naverage}
\end{equation}
where $e$ is the elementary charge and $\tau$ is the measurement time. Meanwhile, the current noise corresponds to the time-averaged variance $\Delta Q^2$ or $\Delta N^2$ as
\begin{equation}
S(f)=\frac{2\langle \Delta Q^2 \rangle}{\tau}
=\frac{2e^2 \langle \Delta N^2 \rangle}{\tau},\quad
\left[\frac{\textrm{A}^2}{\textrm{Hz}}\right]
=\left[\frac{\textrm{C}^2}{\textrm{s}}\right].
\label{Nvariance}
\end{equation}
Comparing the experimental results of these two quantities, we can extract information that is not accessible by standard dc-current measurements alone, such as the charge of carriers.

\subsubsection{Thermal and shot noise}
When the system shown in Fig.~\ref{noisemeasurement} is in equilibrium at $V=0$, the average current is zero ($\langle I \rangle=0$). However, even in this case, there is a finite current noise referred to as thermal noise or Johnson-Nyquist noise at finite temperature~\cite{JohnsonPR1928,NyquistPR1928}. The thermal noise $S_\textrm{th}$ is described as 
\begin{equation}
S_\textrm{th} = 4 k_{\rm{B}} T_{\rm{e}} G,
\label{thermalnoise}
\end{equation}
where $G$, $T_{\rm{e}}$, and $k_{\rm{B}}$ are the conductance, electron temperature, and Boltzmann constant, respectively. Nyquist derived Eq.~(\ref{thermalnoise}) from the second law of thermodynamics to explain the results of Johnson's current-noise measurements~\cite{JohnsonPR1928} and indicated its link with black-body radiation~\cite{NyquistPR1928}. 
In Eq.~(\ref{thermalnoise}), the conductance $G$ as well as $T_{\rm{e}}$ appears. Since the conductance characterizes the linear response of the system to the external bias ($I=GV$) and Joule heat ($GV^2$), we see that Eq.~(\ref{thermalnoise}) reflects the fluctuation-dissipation relation.
Later, in Sect.~\ref{subsubsec:examples}, we derive the same result [Eq.~(\ref{Eq:JNLandauer})] based on the scattering theory, a different approach from the Nyquist's one.

It is noteworthy that Johnson discussed the evaluation of $k_{\rm{B}}$ from the measured thermal noise~\cite{JohnsonPR1928}. This discussion was a pioneering attempt for the precise evaluation of the Boltzmann constant in metrology~\cite{QuMetrologia2017}.

\begin{figure}[!t]
\begin{center}
\includegraphics[width=7cm]{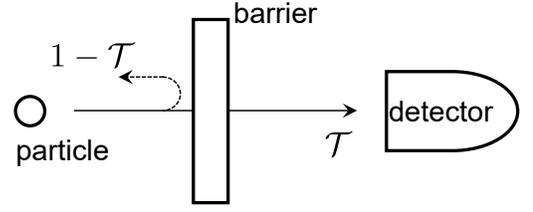}
\end{center}
\caption{Scattering of a particle at a potential barrier. The transmission and reflection probabilities are ${\cal T}$ and $1-{\cal T}$, respectively.} \label{PartitionProcess}
\end{figure}

Next, let us consider shot-noise generation under a non-equilibrium condition. Here, we assume that $I$ is carried by electron tunneling through a potential barrier (scatterer), as sketched in Fig.~\ref{PartitionProcess}. When the transmission probability $\cal{T}$ is small, the shot-noise intensity is described as
\begin{equation}
S_\textrm{shot}=2e\vert \langle I \rangle\vert
\label{Schottky}.
\end{equation}
The numerical factor 2 comes from the definition of PSD at positive frequencies, as explained earlier [see Eq.~(\ref{Definition_PSD_classic})]. Schottky derived this expression to investigate the flow of electrons in a vacuum tube~\cite{SchottkyAP1918}.

To understand the meaning of Eq.~(\ref{Schottky}), let us consider that $N$ particles emitted from a source impinge on the barrier, and each particle is either independently transmitted or reflected with a probability of ${\cal T}$ or $1-{\cal T}$, respectively. The detector measures the transmitted particles. The probability $P_{N}(N_1)$ of detecting $N_1$ particles is given by the binomial distribution as
\begin{equation}
P_{N}(N_1) = \frac{N!}{N_1!(N-N_1)!}{\cal T}^{N_1}(1-{\cal T})^{N-N_1}.
\end{equation}
The average $\langle N_1 \rangle$ and variance $\langle
\Delta N_1^2 \rangle$ are given by
\begin{equation}
\begin{split}
\langle N_1 \rangle &= N{\cal T},\\ 
\langle
\Delta N_1^2 \rangle &\equiv  \langle (N_1- \langle N_1 \rangle)^2
\rangle \\
&= N{\cal T}(1-{\cal T}) = \langle N_1 \rangle (1-{\cal T}).
\end{split}
\end{equation}

When the transmission probability is very small (${\cal T} \ll 1$), both $\langle N_1 \rangle$ and $\langle \Delta N_1^2 \rangle$ are equal to $N{\cal T}$, and this is nothing but the signature of the Poisson distribution. Using Eqs.~(\ref{Naverage}) and (\ref{Nvariance}), we obtain $S_\textrm{shot}/\vert \langle I \rangle\vert = 2e\langle\Delta N_1^2 \rangle/\langle N_1 \rangle=2e$. Thus, the shot noise reflects the discrete nature of charge carriers. Shot noise is sometimes referred to as partition noise since it is generated when a current is partitioned into transmitted and reflected parts. 

It is useful to introduce the Fano factor $F$~\cite{FanoPR1947}, a dimensionless parameter that quantifies the current noise:
\begin{equation}
F \equiv \frac{\langle \Delta N_1^2 \rangle}{\langle N_1 \rangle} = \frac{S_\textrm{shot}}{2e\vert \langle I \rangle\vert}.
\label{Fano_classic}
\end{equation}
By definition, $F = 1$ for the Poisson distribution. In this case, scattering events are independent of each other, namely there is no correlation between them.

By comparing Eqs.~(\ref{thermalnoise}) and (\ref{Schottky}), one can notice that noise properties in equilibrium and non-equilibrium situations are qualitatively different. Particularly, elementary charge $e$ appears only in the shot-noise formula [Eq.~(\ref{Schottky})], indicating that the shot noise serves as a unique probe for charge transport. 

As an interesting historical note, the non-equilibrium shot noise~\cite{SchottkyAP1918} was found ten years earlier than the equilibrium thermal noise~\cite{JohnsonPR1928,NyquistPR1928}, which might reflect the inherence of the non-equilibrium in nature.

\subsection{Noise in quantum transport}
\label{subsec:noise_in_quantum_transport}
Conductance through a mesoscopic system can be understood using our discussion, referred to as the Landauer-B\"uttiker formalism~\cite{DattaETMS,ImryIMP,IhnEQTMSS}. In this subsection, we introduce the current-noise theory using the same framework~\cite{BlanterPR2000}.

\subsubsection{Scattering approach}
Here, from the pedagogical perspective, we consider a simple two-terminal device coupled to a single conduction channel on both the left and right sides of the device. The theoretical descriptions until Sect.~\ref{subsubsec:examples} are taken from Ref.~[\onlinecite{KatoBussei2014}]. Note that the scattering approach can be straightforwardly generalized to multiterminal and multichannel cases~\cite{BlanterPR2000}. We present the general result in Sect.~\ref{subsub:generalshot}.

\begin{figure}[!b]
\begin{center}
\includegraphics[width=8.5cm]{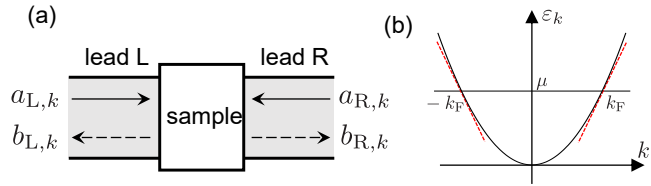}
\end{center}
\caption{(Color online) (a) Two-terminal scattering model. The conduction channels of injected ($a_{\textrm{L}, k}$, $a_{\textrm{R}, k}$) and scattered ($b_{\textrm{L}, k}$, $b_{\textrm{R}, k}$) electrons are represented by solid and dashed arrows, respectively. (b) Parabolic dispersion relation in the leads. Dotted lines indicate the linear approximation near the Fermi surface $k \simeq \pm k_\textrm{F}$ [see Eq.~(\ref{Eq_linearapprox})].} \label{fig_setup}
\end{figure}

Figure~\ref{fig_setup}(a) shows a schematic of the setup. The spinless Hamiltonian of the left (L) and right (R) leads, which are regarded as one-dimensional free electron systems, is expressed using the creation ($c^\dagger_k$) and annihilation ($c_k$) operators as
\begin{equation}
{\cal H} = \sum_k (\varepsilon_k-\mu) c^\dagger_k c_k,
\label{eq:hamiltonian_freeelectron}
\end{equation}
where $\varepsilon_k=\hbar^2k^2/2m$ ($m$, electron mass; $k$, wavenumber of electron; $\hbar$, reduced Planck constant) is the kinetic energy of an electron and $\mu$ is the chemical potential. Here, we perform a linear approximation to the parabolic band dispersion such that
\begin{equation}
\varepsilon_k -\mu= \pm \hbar v_\textrm{F}(k \mp k_\textrm{F}),
\label{Eq_linearapprox}
\end{equation}
as shown in Fig.~\ref{fig_setup}(b), considering only low-energy excitations near the Fermi surface. 
Here, $v_\textrm{F}=\hbar k_\textrm{F}/m$ is the Fermi velocity, and $k_\textrm{F}$ is the Fermi wavenumber.

There exist right- and left-moving electrons in lead L. The annihilation (creation) operators of the former and the latter are described as $a_{\textrm{L}, k}$ ($a^\dagger_{\textrm{L}, k}$) and $b_{\textrm{L}, k}$ ($b^\dagger_{\textrm{L}, k}$), respectively. Within the linear approximation, the Hamiltonian can be written by modifying Eq.~(\ref{eq:hamiltonian_freeelectron}) as
\begin{equation}
\begin{split}
{\cal H} &= \sum_k \hbar v_\textrm{F}(k-k_\textrm{F}) a_{\textrm{L}, k}^\dagger a_{\textrm{L}, k} \\
&+\sum_k (-\hbar v_\textrm{F})(k+k_\textrm{F}) b_{\textrm{L}, k}^\dagger b_{\textrm{L}, k}.
\end{split}
\end{equation}

We define the current operator at position $x$ in lead L as
\begin{equation}
\hat{I}(x)=\frac{\hbar e}{2im}\left( \hat{\psi}^\dagger(x) \frac{d\hat{\psi}(x)}{dx} -  \frac{d\hat{\psi}^\dagger(x)}{dx}\hat{\psi}(x) \right),
\end{equation}
where $\hat{\psi}(x)= \sum_k (1/\sqrt{L}) \exp(ikx) c_k$ is the field operator, and $L$ is the length of the lead. $\hat{I}(x)$ can be expressed using $c^\dagger_k$ and $c_k$ as
\begin{equation}
\hat{I}(x)=\frac{\hbar e}{L}\sum_{k, k'} \frac{k+k'}{2m}c^\dagger_k c_{k'} e^{i(k'-k)x}.
\end{equation}
By considering the contribution only around $k = \pm k_\textrm{F}$ and using $a_{\textrm{L}, k}$ and $b_{\textrm{L}, k}$ instead of $c_k$, we obtain the following formula:
\begin{equation}
\hat{I}(x)=\frac{ev_\textrm{F}}{L}\sum_{k, k'} (a_{\textrm{L}, k}^\dagger a_{\textrm{L}, k'} - b_{\textrm{L}, k}^\dagger  b_{\textrm{L}, k'}) e^{i(k'-k)x}.
\end{equation}
With the assumption that the sample is connected to the lead at $x=0$, the current flowing from lead L into the sample becomes
\begin{equation}
\hat{I}_\textrm{L}=\hat{I}(x=0)=\frac{ev_\textrm{F}}{L}\sum_{k, k'} (a_{\textrm{L}, k}^\dagger a_{\textrm{L}, k'} - b_{\textrm{L}, k}^\dagger  b_{\textrm{L}, k'}). 
\label{Eq_current_operator}
\end{equation}

The scattering process between the incoming ($a_{\alpha, k}$) and outgoing ($b_{\alpha, k}$) electrons in lead $\alpha$ ($\alpha=\textrm{L}$ or $\textrm{R}$) is described as
\begin{equation}
\begin{pmatrix}
b_{\textrm{L}, k} \\
b_{\textrm{R}, k} \\
\end{pmatrix}
=S
\begin{pmatrix}
a_{\textrm{L}, k} \\
a_{\textrm{R}, k} \\
\end{pmatrix}.
\label{Eq_Scattering_Matrix}
\end{equation}
The components of the $S$ matrix are given by
\begin{equation}
S=
\begin{pmatrix}
s^{\textrm{L}\textrm{L}}(k) & s^{\textrm{L}\textrm{R}}(k)  \\
s^{\textrm{R}\textrm{L}}(k) & s^{\textrm{R}\textrm{R}}(k)  \\
\end{pmatrix}
=
\begin{pmatrix}
r & t'  \\
t & r'  \\
\end{pmatrix}.
\end{equation}
Note that, to satisfy the commutation relation $[a_{\alpha, k}, a_{\alpha', k'}^\dagger]=[b_{\alpha, k}, b_{\alpha', k'}^\dagger]=\delta_{\alpha, \alpha'}\delta_{k, k'}$, the $S$ matrix must be unitary, namely $|t|^2=|t'|^2=1-|r|^2=1-|r'|^2$.

Using the $S$ matrix, we express the current operator [see Eq.~(\ref{Eq_current_operator})] as
\begin{equation}
\hat{I}_\textrm{L}=\frac{ev_\textrm{F}}{L}\sum_{\alpha=\textrm{L}, \textrm{R}}
\sum_{\beta=\textrm{L}, \textrm{R}}
\sum_{k, k'} a_{\alpha, k}^\dagger  A_{\textrm{L}}^{\alpha \beta}(k, k')  a_{\beta, k'},
\label{Eq_current}
\end{equation}
where
\begin{equation}
A_{\textrm{L}}^{\alpha \beta}(k, k') =\delta_{\textrm{L},\alpha}\delta_{\textrm{L},\beta}- \left[ s^{\textrm{L}\alpha}(k)  \right]^* 
s^{\textrm{L}\beta}(k').
\end{equation}

\subsubsection{Landauer formula}
\label{sec:LandauerFormula}
Here, we derive the conductance formula. Assuming that incident electrons are in thermal equilibrium and taking their statistical average $\langle \cdots \rangle$, we obtain 
\begin{equation}
\langle a_{\alpha, k}^\dagger  a_{\beta, k'}  \rangle  = \delta_{\alpha, \beta}\delta_{k, k'}f_\alpha(k).
\label{Eq_StAverage}
\end{equation}
$f_\alpha(k)$ is the Fermi-Dirac distribution function in lead $\alpha$:
\begin{equation}
f_\alpha(k)=\frac{1}{\exp \left[(\varepsilon_k-\mu_\alpha)/k_\textrm{B} T_\alpha \right]+1},
\end{equation}
where $T_\alpha$ and $\mu_\alpha$ are temperature and the chemical potential, respectively, in lead $\alpha$. The statistical average of Eq.~(\ref{Eq_current}) are calculated as
\begin{equation}
\begin{split}
\langle \hat{I}_\textrm{L} \rangle &=\frac{ev_\textrm{F}}{L}\sum_k \sum_\alpha A_\textrm{L}^{\alpha\alpha}(k, k)f_\alpha(k) \\
&=\frac{e}{2\pi\hbar}\int_{-\infty}^{\infty} d\varepsilon  \sum_\alpha A_\textrm{L}^{\alpha\alpha}(\varepsilon , \varepsilon )f_\alpha(\varepsilon).
\end{split}
\end{equation}
Note that we replace the summation $(1/L)\sum_k \cdots$ with the integral $\int dk/(2\pi)\cdots$, assuming sufficiently large $L$, and using Eq.~(\ref{Eq_linearapprox}). The relations $A_\textrm{L}^{\textrm{L}\textrm{L}}=1-(s^{\textrm{L}\textrm{L}})^* s^{\textrm{L}\textrm{L}}=1-|r|^2=|t|^2\equiv {\cal T}$ and $A_\textrm{L}^{\textrm{R}\textrm{R}}=(s^{\textrm{L}\textrm{R}})^* s^{\textrm{L}\textrm{R}}=-|t'|^2=-|t|^2=- {\cal T}$, where ${\cal T}$ is the transmission probability, lead to
\begin{equation}
\langle \hat{I}_\textrm{L} \rangle =\frac{e}{2\pi\hbar}\int_{-\infty}^{\infty} d\varepsilon {\cal T(\varepsilon)}[f_\textrm{L}(\varepsilon)-f_\textrm{R}(\varepsilon)].
\label{Eq_current_Landauer}
\end{equation}

For simplicity, let us assume that ${\cal T}$ is energy independent and the system is at absolute zero temperature.
When lead L is biased with $V$, the Fermi-Dirac distribution function in each lead can be written as $f_\textrm{R}(\varepsilon)=\Theta(-\varepsilon)$ and $f_\textrm{L}(\varepsilon)=\Theta(-\varepsilon+eV)$, where $\Theta(x)$ is the Heaviside function. In this case, $f_\textrm{L}(\varepsilon)-f_\textrm{R}(\varepsilon)$ is one only when $0<\varepsilon<eV$ and is zero otherwise, resulting in $\langle \hat{I}_\textrm{L} \rangle =\frac{e^2}{2\pi\hbar}{\cal T}V$. Thus, we obtain the well-known Landauer's conductance formula as
\begin{equation}
G=\frac{\langle \hat{I}_\textrm{L} \rangle}{V} =\frac{e^2}{h} {\cal T},
\label{LandauerConductance}
\end{equation}
where $h=2\pi\hbar$ is Planck constant.
If the channel is spin degenerate, the equation is modified to $G=\frac{2e^2}{h} {\cal T}$.

When the system is at finite temperature, Eq.~(\ref{Eq_current_Landauer}) is described as  
\begin{equation}
G=\frac{e^2}{h}\int d\varepsilon {\cal T}(\varepsilon)\left(-\frac{df}{d\varepsilon} \right),
\label{LandauerConductanceFiniteT}
\end{equation}
where
\begin{equation}
f(\varepsilon)\equiv\frac{1}{{\exp \left[(\varepsilon-\mu)/k_\textrm{B} T_\textrm{e} \right]+1}}.
\end{equation}
To obtain this formula, we use $f_\textrm{L}(\varepsilon)-f_\textrm{R}(\varepsilon)=f(\varepsilon-eV)-f(\varepsilon)\simeq \left(-\frac{df}{d\varepsilon}\right)eV$. 
When ${\cal T}$ is energy-independent, we again obtain Eq.~(\ref{LandauerConductance}) at finite temperature.

\subsubsection{Current noise}
\label{sec:noisetheory}
We define the time evolution of the current operator in the Heisenberg representation
\begin{equation}
\hat{I}_\textrm{L}(t)=\exp\left(\frac{iHt}{\hbar}\right)\hat{I}_\textrm{L}\exp\left(-\frac{iHt}{\hbar}\right),
\end{equation}
and introduce the current-noise operator
\begin{equation}
\Delta\hat{I}_\textrm{L}(t) =\hat{I}_\textrm{L}(t)-\langle \hat{I}_\textrm{L}(t)\rangle.  
\end{equation}
Using this operator, we describe the second-order current-current correlation function as
\begin{equation}
\begin{split}
C(t, t')&\equiv\langle \Delta\hat{I}_\textrm{L}(t) \Delta\hat{I}_\textrm{L}(t') \rangle \\
&= \langle \hat{I}_\textrm{L}(t) \hat{I}_\textrm{L}(t') \rangle - \langle \hat{I}_\textrm{L}(t) \rangle\langle\hat{I}_\textrm{L}(t') \rangle.
\end{split}
\label{correlation_func}
\end{equation}
Note that the current operators $\hat{I}_\textrm{L}(t)$ and $\hat{I}_\textrm{L}(t')$ are not commutative. Following Eq.~(\ref{Definition_PSD_classic}), the current-noise PSD $S(\omega)$ is given by 
\begin{equation}
S(\omega) \equiv \lim_{\tau\rightarrow \infty} 
\frac{2}{\tau}\int_{-\tau/2}^{\tau/2}dt \int_{-\tau/2}^{\tau/2}dt'
C(t, t')  e^{i \omega (t-t')}. 
\label{Definition_PSD_quantum}
\end{equation}

Unlike the classical case discussed in Sect.~\ref{sec:classicalnoise}, $S(\omega)$ in Eq.~(\ref{Definition_PSD_quantum}) is not necessarily a real number, and $S(\omega)= S^*(-\omega)$ holds instead of $S(\omega) = S(-\omega)$. The real part of $S(\omega)$ can be expressed as $\textrm{Re}[S(\omega)]=\left[S(\omega)+S(-\omega)\right]/2\equiv S_{\textrm{sym}}(\omega)$, where $S_{\textrm{sym}}(\omega)$ is referred to as symmetrized noise~\cite{BlanterPR2000}. While the imaginary part of $S(\omega)$ is sometimes important at high frequencies, particularly in the quantum-noise regime~\cite{DeblockScience2003} [see Fig.~\ref{fig3_2}(a)], in this review we focus on the noise at low frequencies, where $S(0)=S_{\textrm{sym}}(0)$ generally holds.

If the Hamiltonian ${\cal H}$ is time-independent, $C(t, t')$ depends only on the time difference $\Delta t = t-t'$, and $C(t, t')=C(\Delta t)$ equals zero at large $\Delta t$ [also see the discussion in Sect.~\ref{sec:basics_of}]. In this case, Eq.~(\ref{Definition_PSD_quantum}) is modified to
\begin{equation}
\begin{split}
S(\omega) &= \lim_{\tau\rightarrow \infty} 
\frac{2}{\tau}\int_{-\tau/2}^{\tau/2}dt' \int_{-\infty}^{\infty}d(\Delta t)C(\Delta t)
 e^{i \omega\Delta t} \\
&= 2 \int_{-\infty}^{\infty}d(\Delta t)C(\Delta t) e^{i \omega\Delta t}.
\label{Definition_PSD_quantum2}
\end{split}
\end{equation}
Thus, the noise PSD is formulated as the Fourier transform of $C(\Delta t)$. The zero-frequency noise is expressed as
\begin{equation}
S\equiv S(0) =  2 \int_{-\infty}^{\infty}dt \left[\langle \hat{I}_\textrm{L}(t) \hat{I}_\textrm{L}(0) \rangle - \langle \hat{I}_\textrm{L}(t) \rangle\langle\hat{I}_\textrm{L}(0) \rangle \right].
\label{s_zero}
\end{equation}
By substituting $a_{\alpha, k}(t) = a_{\alpha, k}\exp(-i \varepsilon_k t/\hbar)$, Eq.~(\ref{Eq_current}) becomes
\begin{equation}
\begin{split}
\hat{I}_\textrm{L}&=\frac{ev_\textrm{F}}{L}\sum_{\alpha=\textrm{L}, \textrm{R}}
\sum_{\beta=\textrm{L}, \textrm{R}}
\sum_{k, k'} a_{\alpha, k}^\dagger  A_{\textrm{L}}^{\alpha \beta}(k, k')  a_{\beta, k'}  \\
&\times \exp \left[\frac{i(\varepsilon_k-\varepsilon_{k'})t}{\hbar}\right].
\label{Eq_current_time}
\end{split}
\end{equation}
Therefore, Eq.~(\ref{s_zero}) can be described as 
\begin{equation}
\begin{split}
&S = 2 \int_{-\infty}^{\infty}dt \left(\frac{ev_\textrm{F}}{L} \right)^2
\sum_{k, k', k'', k'''} \sum_{\alpha, \beta, \alpha', \beta'} \\
&A_{\textrm{L}}^{\alpha \beta}(k, k') A_{\textrm{L}}^{\alpha' \beta'}(k'', k''')  
[\langle a_{\alpha, k}^\dagger a_{\beta, k'} a_{\alpha', k''}^\dagger a_{\beta', k'''} \rangle \\
&-\langle a_{\alpha, k}^\dagger a_{\beta, k'} \rangle  \langle a_{\alpha', k''}^\dagger a_{\beta', k'''} \rangle
]  \exp \left[\frac{i(\varepsilon_k-\varepsilon_{k'})t}{\hbar}\right].
\label{Eq36}
\end{split}
\end{equation}
Wick's theorem and Eq.~(\ref{Eq_StAverage}) lead to
\begin{equation}
\begin{split}
&\langle a_{\alpha, k}^\dagger a_{\beta, k'} a_{\alpha', k''}^\dagger a_{\beta', k'''} \rangle
-\langle a_{\alpha, k}^\dagger a_{\beta, k'} \rangle  \langle a_{\alpha', k''}^\dagger a_{\beta', k'''} \rangle \\
&= \langle a_{\alpha, k}^\dagger a_{\beta', k'''} \rangle  \langle a_{\beta, k'} a_{\alpha', k''}^\dagger \rangle \\
&=\delta_{\alpha, \beta'}\delta_{k, k'''}\delta_{\beta, \alpha'}\delta_{k', k''}f_\alpha(k)[1-f_\beta(k')].
\end{split}
\end{equation}
The second term on the leftmost side of this equation is the exchange term that takes the statistical nature of particles into account. The resultantly obtained factor $f_\alpha(k)[1-f_\beta(k')]$ represents the fermionic nature of electrons.

By replacing the summation of the wavenumbers $(1/L)\sum_k \cdots$ with the integral $\int dk/(2\pi)\cdots$ and changing the integration variable from $k$ to $\varepsilon$ using Eq.~(\ref{Eq_linearapprox}) in Eq.~(\ref{Eq36}), we obtain
\begin{equation}
\begin{split}
S&=\frac{2e^2}{(2\pi\hbar)^2}\int d\varepsilon \int d\varepsilon' \int^{\infty}_{-\infty}dt \\
&\sum_{\alpha,\beta}A_\textrm{L}^{\alpha\beta}(\varepsilon, \varepsilon')
A_\textrm{L}^{\beta\alpha}(\varepsilon', \varepsilon)f_\alpha(\varepsilon)[1-f_\beta(\varepsilon')]
\exp \left[ \frac{i(\varepsilon-\varepsilon')t}{\hbar}\right].
\end{split}
\end{equation}
In the end, using the relation $\int^{\infty}_{-\infty}dt e^{i(\varepsilon-\varepsilon')t/\hbar}=2\pi \hbar\delta(\varepsilon-\varepsilon')$, the general formula for the current noise is written as 
\begin{equation}
S=\frac{e^2}{\pi\hbar} \int d\varepsilon
\sum_{\alpha, \beta}A_\textrm{L}^{\alpha\beta}(\varepsilon, \varepsilon)
A_\textrm{L}^{\beta\alpha}(\varepsilon, \varepsilon) \\
 f_\alpha(\varepsilon) \left[1-f_\beta(\varepsilon)\right].
\label{NoiseSingleChannel}
\end{equation}

\subsubsection{Derivations of thermal and shot noise}
\label{subsubsec:examples}
Both the thermal- and shot-noise formulae are derived from Eq.~(\ref{NoiseSingleChannel}). When the system is in equilibrium, or $|eV|\ll k_\textrm{B}T_\textrm{e}$, the thermal noise dominates over the shot noise. Using the three relations, $eV=0$, 
\begin{equation}
f(\varepsilon)\left[1-f(\varepsilon)\right]=k_\textrm{B}T_\textrm{e} \left(-\frac{\partial f}{\partial \varepsilon}\right),
\end{equation}
and 
\begin{equation}
\sum_{\alpha,\beta}A_\textrm{L}^{\alpha\beta}(\varepsilon, \varepsilon)A_\textrm{L}^{\beta\alpha}(\varepsilon, \varepsilon)=2{\cal T}(\varepsilon),
\end{equation}
Eq.~(\ref{NoiseSingleChannel}) is modified to
\begin{eqnarray}
S=\frac{2e^2k_\textrm{B}T_\textrm{e}}{\pi\hbar}\int d\varepsilon {\cal T}(\varepsilon)\left(-\frac{df}{d\varepsilon} \right).
\end{eqnarray}
By comparing it with Eq.~(\ref{LandauerConductanceFiniteT}), we obtain
\begin{equation}
S=4k_\textrm{B}T_\textrm{e}G.
\label{Eq:JNLandauer}
\end{equation}
This is the thermal-noise formula introduced above [see Eq.~(\ref{thermalnoise})].

Let us consider the case of $eV \neq 0$ at zero temperature, where we have $f_\textrm{L}(\varepsilon)=\Theta(-\varepsilon+eV)$ and $f_\textrm{R}(\varepsilon)=\Theta(-\varepsilon)$. When $eV>0$, $f_\alpha(\varepsilon)[1-f_\beta(\varepsilon)]\neq 0$ holds only when $\alpha = \textrm{L}$, $\beta = \textrm{R}$, and $0<\varepsilon<eV$. Hence, zero-frequency noise $S$ is given by
\begin{equation}
S=\frac{e^2}{\pi\hbar} A_\textrm{L}^{\textrm{L}\textrm{R}}(\varepsilon, \varepsilon)
A_\textrm{L}^{\textrm{R}\textrm{L}}(\varepsilon, \varepsilon)\times |eV|.
\label{Eq44}
\end{equation}
Suppose that the energy dependence of ${\cal T}$ is negligibly small.
Since
\begin{equation}
A_\textrm{L}^{\textrm{L}\textrm{R}}A_\textrm{L}^{\textrm{R}\textrm{L}}=|t|^2(1-|t|^2)={\cal T}(1-{\cal T}),   
\end{equation}
Eq.~(\ref{Eq44}) can be described as
\begin{equation}
S=\frac{e^2}{\pi\hbar}{\cal T}(1-{\cal T}) |eV| = 2e\vert\langle I \rangle\vert(1-{\cal T}).
\label{Eq:zeroTshot}
\end{equation}
At finite temperature, Eq.~(\ref{Eq:zeroTshot}) is modified to
\begin{equation}
\begin{split}
S &= 4 k_\textrm{B}T_\textrm{e}G \\
&+ 2e\vert\langle I \rangle\vert (1-{\cal T})
 \left[  \coth\left(\frac{eV}{2k_\textrm{B}T_\textrm{e}}\right) - \frac{2k_\textrm{B}T_\textrm{e}}{eV} \right].
\label{ShotTheory1}
\end{split}
\end{equation}

In the weak transmission limit (${\cal T}\ll 1$) at zero temperature, Eq.~(\ref{Eq:zeroTshot}) becomes
\begin{equation}
S=\frac{2e^2}{h} {\cal T} \vert eV\vert=2e\vert\langle I \rangle\vert.
\end{equation}
This equation corresponds to the classical Schottky-type shot-noise formula [see Eq.~(\ref{Schottky})]. Similarly, the finite-temperature shot-noise formula in the weak transmission limit is written as
\begin{equation}
S = 2e \vert \langle I \rangle\vert  \coth\left(\frac{eV}{2k_\textrm{B}T_\textrm{e}}\right).
\label{ShotTunnel}
\end{equation}

Equation (\ref{Eq:zeroTshot}) tells that a conductor with ${\cal T}=1$ has no current noise at zero temperature. The noiseless feature explicitly indicates that charge current fed from a reservoir does not fluctuate at zero temperature due to the fermionic nature of electrons.

\subsubsection{General formula}
\label{subsub:generalshot}
While so far we have assumed a conductor with a single conduction channel for simplicity [see Fig.~\ref{fig_setup}], there can exist many channels in actual mesoscopic systems.
Here, we present a generalized formula in multichannel cases. Assuming that the transmission probability ${\cal T}_n$ $(n=1, 2, 3\cdots)$ is energy independent, we obtain the current $\langle I \rangle$ and the conductance $G$ at zero temperature as 
\begin{equation}
\langle I \rangle =GV,  \quad
G = \frac{2e^2}{h}\sum_n {\cal T}_n.
\label{Landauer}
\end{equation}
This is the well-known Landauer formula, in which $G$ is given as the sum of contributions from the parallel channels. The factor 2 represents the spin degeneracy at zero magnetic field. The low-frequency noise $S$ is described as
\begin{equation}
S=\frac{2e^2}{\pi\hbar} \sum_n{\cal T}_n(1-{\cal T}_n) \vert eV\vert =2e\vert \langle I \rangle\vert F,
\label{shot_47}
\end{equation}
where $F$ is the Fano factor defined in Eq.~(\ref{Fano_classic}). In the present case, $F$ is given by
\begin{equation}
F = \frac{\sum_n {\cal T}_n(1-{\cal T}_n)}{\sum_n {\cal T}_n}.
\label{FanoTheory}
\end{equation}
Equations (\ref{shot_47}) and (\ref{FanoTheory}) explain that the current noise is given as the sum of noise contributions from the parallel channels, similarly to the case of $\langle I \rangle$.

Current noise at finite temperature is
\begin{equation}
S = 4 k_\textrm{B}T_\textrm{e}G + 2e\vert\langle I \rangle\vert F
 \left[  \coth\left(\frac{eV}{2k_\textrm{B}T_\textrm{e}}\right) - \frac{2k_\textrm{B}T_\textrm{e}}{eV} \right].
\label{ShotTheory}
\end{equation}
This equation is the most commonly used shot-noise formula in experiments.

As clearly shown in Eq.~(\ref{ShotTheory}), current noise is related to both temperature ($k_\textrm{B}T_\textrm{e}$) and bias ($eV$) in a mixed way. While Landauer argued that thermal noise and shot noise could be non-dividable~\cite{LandauerPRB1993},  it is convenient to regard them as additive independent noise in many cases. In this review, the term ``shot noise'' means the quantity obtained by subtracting the first term ($4 k_\textrm{B}T_\textrm{e}G$) from Eq.~(\ref{ShotTheory}), which represents the excess noise generated by a finite bias.

\begin{figure}[!t]
\center 
\includegraphics[width=8cm]{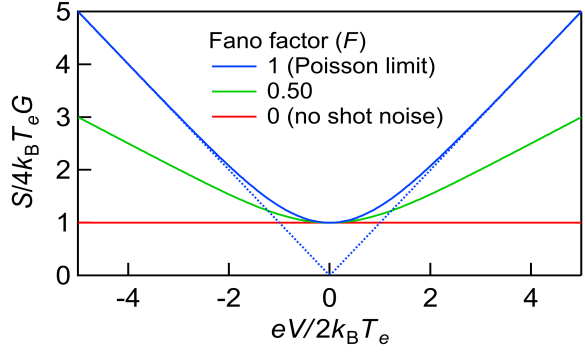}
\caption{(Color online) $S/(4 k_\textrm{B}T_\textrm{e}G)$ plotted as a function of $X \equiv eV/2k_\textrm{B}T_\textrm{e}$ for $F=0, 0.5$, and 1. The dotted line is the asymptotic line in the Poisson limit.}
\label{ShotTheoryFig}
\end{figure}

Eq.~(\ref{ShotTheory}) tells that the dimensionless quantity $S/(4 k_\textrm{B}T_\textrm{e}G)$ is a function of $X \equiv eV/2k_\textrm{B}T_\textrm{e}$ as
\begin{equation}
\frac{S}{4k_\textrm{B}T_\textrm{e}G} = 1 + F\left[X \coth(X) -1\right].
\label{eq:shotnoisenormalized}
\end{equation}
Figure~\ref{ShotTheoryFig} displays $S/(4 k_\textrm{B}T_\textrm{e}G)$ for the cases of $F=0, 0.5$, and 1. The $F=1$ case is sometimes referred to as the Poisson limit, where we observe that $S/4k_\textrm{B}T_\textrm{e}G \rightarrow |X|$ for $|X|\rightarrow \infty$, namely that current noise at high bias or low temperature corresponds to the classical Schottky-type shot noise. 

\section{Noise measurement}
\label{sec:techniques}
\rm
Compared to standard conductance measurements, current-noise measurements have not been widely performed because of technical difficulties. The main problem is that the current-noise intensity in mesoscopic devices is often too small to measure with a commercially available ammeter. A variety of experimental techniques have been used to solve this problem. In this section, we first explain the basics of current-noise measurements and then introduce several techniques that have provided accurate measurements.

\subsection{Current-noise PSD}
\label{sec:spectrum density}

\begin{figure*}[t]
\begin{center}
\includegraphics[width=16cm]{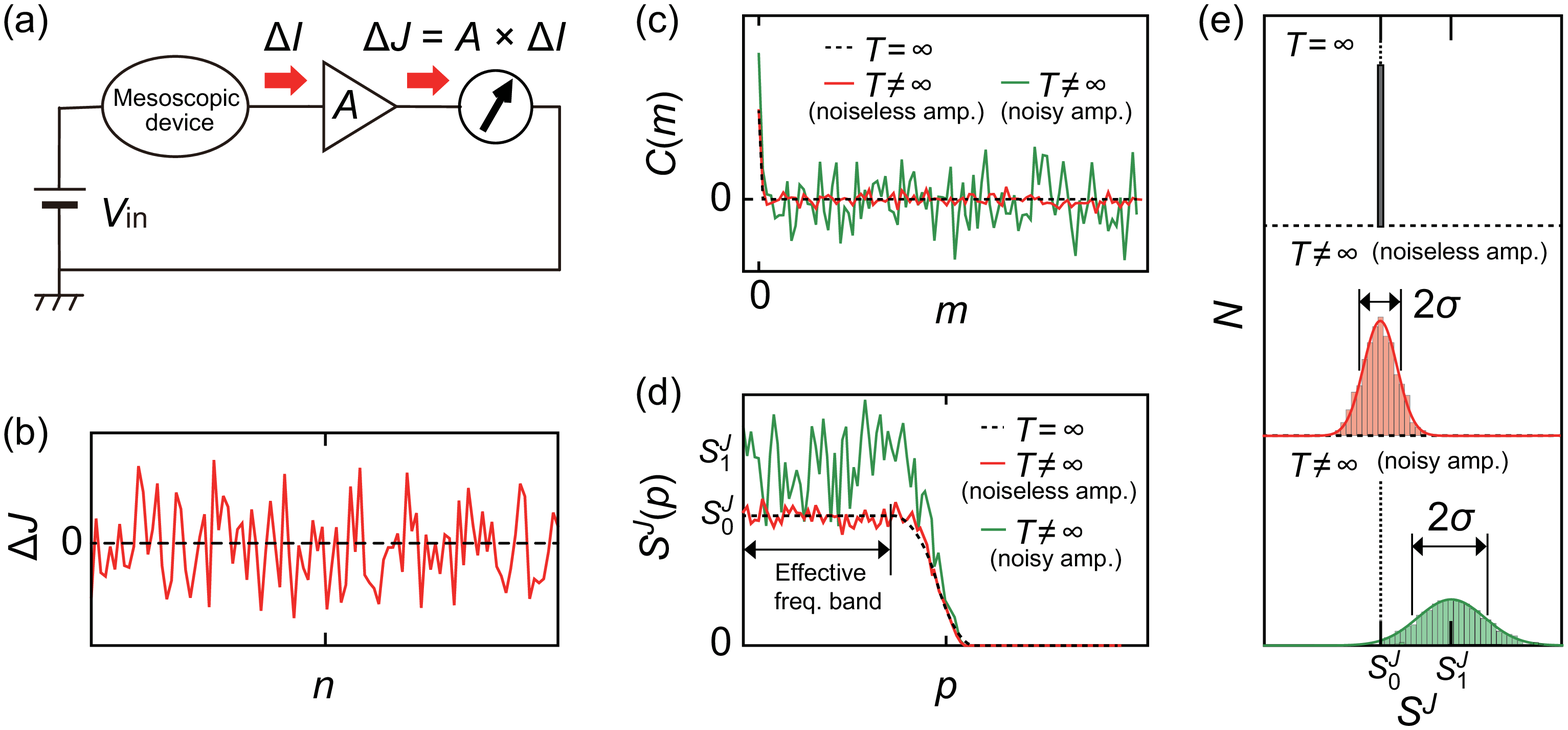}
\end{center}
\caption{(Color online) (a) Schematic of measurement setup using a current amplifier with gain $A$ and an ammeter. (b) Schematic of time-domain current-noise data $\Delta J_n$. Noise data are plotted as a function of the measurement number $n$. (c)(d) Schematic of $C(m)$ (c) and $S^J(p)$ (d). Dashed black lines are for the $N_T = \infty$ case. Solid red and green lines are for the $N_T \neq \infty$ case with a noiseless measurement setup and a noisy setup, respectively. (e) Schematic of the histogram analysis for data in the effective frequency band. $S^J(p)$ takes a fixed value over the frequency band when $N_T = \infty$ (black), while $S^J(p)$ scatters around $S^J_0=\langle S^J(p)\rangle$ with a variance $\sigma$ when $N_T \neq \infty$ (red and green). Extrinsic noise in the measurement setup enhances $\sigma$ as well as $\langle S^J(p)\rangle$ (from $S^J_0$ to $S^J_1$) and thus decreases measurement accuracy.} \label{fig3_1}
\end{figure*}

Generally, in electronic transport experiments, one applies an input voltage $V_{\rm{in}}$ or current $I_{\rm{in}}$ to a mesoscopic device (``sample,'' hereafter) and measures the response to evaluate the sample's transport properties. In a conductance $G = \langle I(t)\rangle /V$ measurement, the time average $\langle I(t)\rangle$ of the output current $I(t)$ is often measured with a $V_{\rm{in}}$ applied. In contrast, in current-noise measurements, the measured quantity is not $\langle I(t)\rangle$ but the variance $\langle [I(t)-\langle I(t)\rangle]^2\rangle \equiv \langle \Delta I(t)^2\rangle$.

Let us consider a current $I_{\alpha}(t)$ outputs from a terminal $\Omega_{\alpha}$ of a sample. The magnitude of the current noise $\Delta I_{\alpha}(t) \equiv I_{\alpha}(t)-\langle I_{\alpha}(t)\rangle$ is often evaluated by its power spectral density (PSD) $S_{\alpha\alpha}(f) = 2\langle \Delta I_{\alpha}(t)^{2}\rangle_{f} / \Delta f$ [Eq.~(\ref{noisedef2})]. As explained in Sect.~\ref{sec:current_noise},  $S_{\alpha\alpha}(f)$ is given by the Fourier transform of the noise auto-correlation function $C_{\alpha\alpha}(\tau)$ as [see Eqs.~(\ref{noisedef2}), (\ref{correlation_func}), and (\ref{Definition_PSD_quantum2})]
\begin{equation}
\begin{split}
S_{\alpha\alpha}(f)&=2\int_{-\infty}^{\infty}C_{\alpha\alpha}(\tau) e^{2\pi if\tau} d\tau, 
\label{auto_Sbb}
\end{split}
\end{equation}
\begin{equation}
\begin{split}
C_{\alpha\alpha}(\tau)&=\lim_{T\rightarrow \infty}
\frac{1}{T}\int_{-T/2}^{T/2} \Delta I_{\alpha}(t)\Delta I_{\alpha}(t+\tau)dt.
\label{autocorr_beta}
\end{split}
\end{equation}

We can also evaluate the correlation between current noise in different terminals $\Omega_{\alpha}$ and $\Omega_{\beta}$ by the cross-PSD $S_{\alpha\beta}(f) = 2\langle \Delta I_{\alpha}(t)\Delta I_{\beta}(t)\rangle _{f}/ \Delta f$, which is the Fourier transform of the cross-correlation function $C_{\alpha\beta}(\tau)$ of $I_{\alpha}(t)$ and $I_{\beta}(t)$:
\begin{equation}
\begin{split}
S_{\alpha\beta}(f)&=2\int_{-\infty}^{\infty}C_{\alpha\beta}(\tau) e^{2\pi if\tau} d\tau, 
\label{crosscorr_Sbc}
\end{split}
\end{equation}
\begin{equation}
\begin{split}
C_{\alpha\beta}(\tau)&=\lim_{T\rightarrow \infty} 
\frac{1}{T}\int_{-T/2}^{T/2} \Delta I_{\alpha}(t)\Delta I_{\beta}(t+\tau)dt.
\label{crosscorr_beta}
\end{split}
\end{equation}

\subsection{Basics of current-noise measurements}
\label{sec:basics_of}
\rm

When the noise auto-correlation function [Eq.~(\ref{autocorr_beta})] of a current $I(t)$ is a delta-type function, the noise PSD, $S^{I}(f)$, is independent of frequency; in this case, $S^{I}(f)$ is referred to as ``white noise'' (see discussion in Sect.~\ref{sec:noisetheory}). In this subsection, we discuss a virtual measurement of white current noise. 

Because $S^{I}(f)$ is usually too small (typically of the order of $10^{-28} \rm{A}^{2}/\rm{Hz}$) to measure with a standard ammeter, we consider amplifying noise $\Delta I$ to $\Delta J = A \times \Delta I$ using an amplifier with gain $A$. Figure~\ref{fig3_1}(a) shows a schematic of a measurement setup using an amplifier. We evaluate $S^{J}(f) = 2\langle \Delta J(t)^{2}\rangle_{f} / \Delta f$, where $\Delta J(t) \equiv J(t)-\langle J(t)\rangle$ is the measured current noise, to estimate the intrinsic noise $S^{I}(f)$.

Here, we briefly discuss an ideal measurement using a noiseless amplifier and ammeter. By a current-noise measurement for $T$ seconds with a sampling rate $r$, one obtains time-domain data $\Delta J_n~(n = 0, 1,,, N_{T} - 1)$, where $N_{T} = r \times T$ is the total number of data points. We calculate the noise auto-correlation function $C(m)~(m = 0, 1,,, N_{T} - 1)$ by replacing the integral in Eq.~(\ref{autocorr_beta}) with the sum of the products among the data, as follows. 
\begin{equation}
\begin{split}
C(m)=\frac{1}{N_{T}}\sum_{n=0}^{N_{T}-1}\Delta J_n\Delta J_{\rm{mod(\it{n+m, N_T})}}.
\label{p_sum}
\end{split}
\end{equation}
One obtains the current-noise PSD $S^{J}(p)~(p = 0, 1,,,N_{T}-1)$ by the discrete Fourier transform  
\begin{equation}
\begin{split}
S^{J}(p)=\frac{2}{N_{T}}\sum_{m=0}^{N_{T}-1}C(m)e^{2\pi ipm/N_{T}}.
\label{sj_sum}
\end{split}
\end{equation}
The frequency resolution is $1/T$ (Hz), and the upper limit of the frequency band is $1/T \times N_{T} = r$ (Hz).

First, let us consider the case of an infinitely fast $r = \infty$ and long $T = \infty$ measurement. In this case, because $C(m)$ is a delta-type function, $S^{J}$ becomes a white-noise spectrum over the whole frequency band.

In actual experiments, $r$ is finite, and the measurement has to be completed in a finite time $T$. In this case, we need to analyze a finite number of discrete time-domain data [see Fig.~\ref{fig3_1}(b)]. Here, we first consider the case of finite $r$ while assuming $T = \infty$. Because $C(m)$ is a discrete delta-type function [black dashed line in Fig.~\ref{fig3_1}(c)], $S^{J}(p)$ takes a fixed value $S^{J}_0$ [black dashed line in Fig.~\ref{fig3_1}(d)] in the ``effective frequency band'', which is determined by the upper-frequency limit of the measurement (typically about $0.4 \times r$ Hz): the factor 0.4 reflects a low-pass filtering generally applied to prevent aliasing errors. The histogram analysis of the data in the effective band is shown in the upper panel in Fig.~\ref{fig3_1}(e), where all the data points are on $S^{J}_0$. One can accurately estimate $S^{I}$ from the measured $S^{J}_0$ as $S^{I} = S^{J}_0/A^{2}$, where $A$ is the gain of the amplifier. 

Let us consider a current-noise measurement in a finite time ($T \neq \infty$, and hence $N_{T} \neq \infty$). In this case, to avoid the influence of the data truncation, one needs to multiply the time-domain data by a window function, e.g., Hanning window. In contrast to the $T = \infty$ case, $C(m \neq 0)$ fluctuates around $C = 0$ [red line in Fig.~\ref{fig3_1}(c)], causing the fluctuation of $S^{J}(p)$ in the effective frequency band [red line in Fig.~\ref{fig3_1}(d)]. The middle panel in Fig.~\ref{fig3_1}(e) shows the histogram analysis of the $S^{J}(p)$ data. The current-noise intensity is given by the peak $S^{J}(p)$ value $S^{J}_0 = \langle S^{J}(p)\rangle$, and the accuracy of the analysis can be evaluated as the standard deviation $\sigma \equiv \langle [S^{J}(p) -\langle S^{J}(p)\rangle]^{2}\rangle ^{1/2}$. Because $\sigma$ decreases in inverse proportion to $\sqrt{N_{T}}=\sqrt{r\times T}$, the accuracy is improved by increasing $r$ or $T$. 

So far, we have discussed current-noise measurements using a noiseless amplifier and ammeter. Conversely, below, we consider the influence of extrinsic noise generated in these measurement devices. When the input-referred noise of the amplifier and the ammeter is given by $S^{I}_{\rm{amp}}$ and $S^{I}_{\rm{meas}}$, respectively, the relation between $S^I$ in a sample and the measured noise $S^J$ is described as 
\begin{equation}
\begin{split}
S^{J}=A^2(S^I+S^{I}_{\rm{amp}})+S^{I}_{\rm{meas}}.
\label{inputnoise}
\end{split}
\end{equation}
When the gain $A$ is large enough to hold $A^2\times S^{I}_{\rm{amp}} \gg S^{I}_{\rm{meas}}$, $S^{I}_{\rm{amp}}$ dominates the extrinsic noise in the measurement setup.

The extrinsic noise enhances both $C(m)$ peak at $m = 0$ and fluctuation at $m \neq 0$ [green line in Fig.~\ref{fig3_1}(c)], resulting in the increase in $\langle S^{J}(p)\rangle$ (from $S^{J}_0$ to $S^{J}_1$) and the fluctuation of $S^{J}(p)$ in the effective frequency band [green line in Fig.~\ref{fig3_1}(d)]. The lowest panel in Fig.~\ref{fig3_1}(e) displays the histogram representation of the $S^{J}(p)$ data. The measurement accuracy for $S^{I}=S^{J}_0/A^{2}-S^{I}_{\rm{amp}}$ drops due to the increase in $\sigma$. Although the accuracy can be improved by increasing $T$ and/or $r$ as in the noiseless-measurement case, $S^{I}_{\rm{amp}}$ is often larger than $S^{I}$ so that it takes a long time to obtain high accuracy. Thus, the extrinsic noise in the measurement devices degrades the efficiency of current-noise measurements.

When one uses two current amplifiers in series, the relation between $S^I$ and $S^J$ is given by
\begin{equation}
\begin{split}
S^{J}=A_2^2[A_1^2(S^I+S^{I}_{\rm{amp1}})+S^{I}_{\rm{amp2}}]+S^{I}_{\rm{meas}}.
\label{doubleamp}
\end{split}
\end{equation}
Here, $A_1$ and $S^{I}_{\rm{amp1}}$, respectively, are the gain and the input-referred current noise of the first amplifier, and $A_2$ and $S^{I}_{\rm{amp2}}$ are those of the second one. When $A_1$ is large enough to hold $A_1^2\times S^{I}_{\rm{amp1}} \gg S^I_{\rm{amp2}}$, the influence of $S^{I}_{\rm{amp2}}$, as well as $S^{I}_{\rm{meas}}$, can be neglected, and $S^{I}_{\rm{amp1}}$ dominates the system performance.

\subsection{Noise sources in a mesoscopic device}
\begin{figure}[tb]
\begin{center}
\includegraphics[width=7cm]{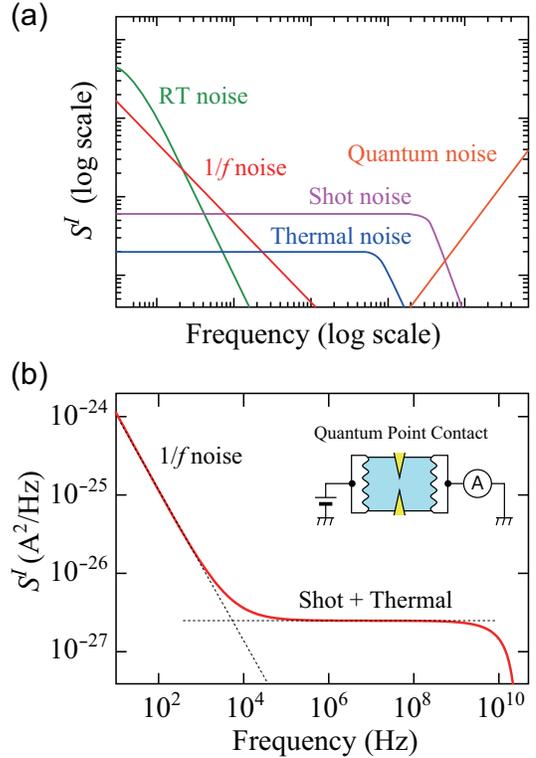}
\end{center}
\caption{(Color online) (a) Schematic of noise PSDs of several noise sources. (b) Simulated current-noise PSD in a spin-degenerated QPC of the conductance $0.5 \times 2e^2/h$. We assumed that temperature is 100 mK, source-drain bias is 100 $\rm{\mu}$V, and the $1/f$ corner frequency is 10 kHz~\cite{AguadoPRL2000,GustavssonPRL2007}.}
\label{fig3_2}
\end{figure}

Generally, a mesoscopic device has a variety of current-noise origins, each of which has its characteristic PSD [Fig.~\ref{fig3_2}(a)]. One important example is $1/f$ noise (red line), which originates from the trapping of electrons in unintentionally formed discrete levels in a sample~\cite{McWhorter1957,HoogePR1969,ZielNoise1986}. While we have considered a measurement for white noise above, we take the frequency dependence into account below.

Figure~\ref{fig3_2}(b) shows a representative current-noise PSD in a QPC fabricated in a two-dimensional electron system (2DES)~\cite{AguadoPRL2000,GustavssonPRL2007}. Whereas shot noise and thermal noise, respectively, usually have broadband spectra up to gigahertz frequencies depending on the applied bias ($eV$) and temperature ($k_{\rm{B}}T_{\rm{e}}$), they can be regarded as white noise at low frequencies (typically below a few hundred megahertz). Indeed, in Fig.~\ref{fig3_2}(b), one observes that the current noise is almost independent of frequency from 100 kHz to 100 MHz, where shot noise and thermal noise are dominant. Most shot-noise measurements evaluate the PSD in this white-noise regime.

At very low frequencies, we observe that the PSD monotonically increases with decreasing frequency due to $1/f$ noise and random telegraph (RT) noise [Fig.~\ref{fig3_2}(a)]. This prevents us from evaluating the shot noise and thermal noise. The frequency at which the $1/f$-noise intensity is comparable to that of white noise is often referred to as the corner frequency ($f_{\rm{c}}$). For example, $f_{\rm{c}}$ is typically about 100 kHz for a sample fabricated in a GaAs-based heterostructure. Besides, the quantum noise [Fig.~\ref{fig3_2}(a)], which is out of the scope of this review, becomes dominant at very high frequencies. 

\subsection{Calibration of noise-measurement systems}
\label{sec:calibration_of_noise-measurement}
In estimating $S^I$ from the measured noise $S^J$, it is essential to know precisely the gain $A$ and input-referred noise $S^I_{\rm{amp}}$ of the amplifier [see Eq.~(\ref{inputnoise})]. We also need to know the unintentional attenuation of the current noise in the wiring between the sample and measurement instruments. Meeting these requirements requires the calibration of an experimental setup. Below, we introduce two techniques often used for calibrating noise-measurement setups.  

\subsubsection{Calibration by thermal-noise measurement}
\label{sec:johnson_noise_thermometry}
 When a sample is in thermal equilibrium (no bias applied), thermal noise $S^I_{\rm{th}}$ dominates current noise $S^I$. The magnitude of the low-frequency thermal noise is described as
\begin{equation}
\begin{split}
S^I_{\rm{th}}=4k_{\rm{B}}T_{\rm{e}}\rm{Re}(\it{Y}),
\label{johnson_y}
\end{split}
\end{equation}
where $T_{\rm{e}}$ is electron temperature, $Y$ is admittance of the sample, and ${\rm{Re}}(Y)$ is the real part of $Y$, namely dc conductance $G$.

Let us consider the change in $S^J$ induced by varying $T_{\rm{e}}$ or $G$~\cite{DiCarloRSI2006,HashisakaRSI2009,ArakawaAPL2013,LeeRSI2021}. Suppose that $S^J$ is given by $A^2(S^I+S^{I}_{\rm{amp}})$ [see Eq.~(\ref{inputnoise})]. An increase in temperature from $T_{\rm{e}}$ to $T_{\rm{e}}+\Delta T_{\rm{e}}$ results in an increase in $S^I_{\rm{th}}$ and hence $S^J$ to $S^J +\Delta S^J$, where $\Delta S^J \equiv S^J(T_{\rm{e}}+\Delta T_{\rm{e}})-S^J(T_{\rm{e}})$. The gain $A$ can be evaluated as $A=[\Delta S^J/4k_{\rm{B}}\Delta T_{\rm{e}}{\rm{Re}}(Y)]^{1/2}$. Note that $A$ is the total gain of the whole measurement system, including the amplifier's gain and the attenuation in the wiring, and we describe $``A"$ for both gains with and without the attenuation for simplicity. $S^I_{\rm{amp}}$ can be evaluated by extrapolating the $T_{\rm{e}}$ dependence of $S^J$ to $T_{\rm{e}} \rightarrow0$ [$S^I_{\rm{amp}}= S^J(T_{\rm{e}} \rightarrow 0)/A^2$]. 

\subsubsection{Calibration by shot-noise measurement}
\label{sec:shot_noise_thermometry}

Shot noise is caused by stochastic charge-scattering processes in a conductor [see Eqs.~(\ref{ShotTunnel}) and (\ref{ShotTheory})]. When the scattering process is well known, shot noise is also useful for calibrating the measurement system. 

Let us consider simple scattering processes of transmission probability ${\cal{T}}\rm{\ll1}$. When a current flowing through the scatterer is given by $\langle I\rangle$, shot noise $S^I_{\rm{shot}}$ at zero temperature is given by 
\begin{equation}
\begin{split}
S^I_{\rm{shot}}=2e\langle I\rangle.
\label{shot_calib}
\end{split}
\end{equation}
At finite temperature, Eq.~(\ref{shot_calib}) is modified to
\begin{equation}
\begin{split}
S^I_{\rm{shot}}=2e\langle I\rangle\left[\coth\left(\frac{eV}{2kT_{\rm{e}}}\right)-\frac{2kT_{\rm{e}}}{eV}\right].
\label{shot_calib_finite}
\end{split}
\end{equation}
Note that Eq.~(\ref{shot_calib_finite}) is obtained by subtracting 4$k_{\rm{B}}T_{\rm{e}}G$ from $S^I$ on the right-hand side of Eq.~(\ref{ShotTunnel}) (see discussion in Sect.~\ref{subsub:generalshot}). From the measured $\langle I\rangle$ dependence of $S^J$, one can estimate the amplifier's gain as $A = (\Delta S^J/2e\Delta \langle I\rangle)^{1/2}$, where $\Delta S^J$ and $\Delta \langle I\rangle$ are the changes in $S^J$ and $\langle I\rangle$, respectively. Electron temperature is evaluated from a fit to the experimental data with Eq.~(\ref{shot_calib_finite})~\cite{SpietzScience2003,OtaJPCM2017,TikhonovPRB2020}. The amplifier's noise $S^I_{\rm{amp}}$ is obtained by substituting $S^I_{\rm{th}} = 4k_{\rm{B}}T_{\rm{e}}G$ into $S^I_{\rm{amp}}= S^J(\langle I\rangle=0)/A^2-S^I_{\rm{th}}$. Thus, one can evaluate both $A$ and $S^I_{\rm{amp}}$ from the shot-noise measurements. 

\subsection{Examples of current-noise measurements}
\label{sec:measurement_systems}
Noise-measurement techniques can be categorized into several groups. This section describes the concept, advantages, and drawbacks of each category by introducing examples from past experiments. 

Before starting the discussion, we summarize some assumptions. First, we assume that a sample is placed in a cryostat to focus on quantum transport at low temperatures.  The current noise generated in the sample is taken from the cryostat through coaxial cables whose length is typically about a few meters. If the cables directly connect the sample to measurement instruments, the sample is ac grounded through their stray capacitance of about a few hundred picofarads, causing the attenuation of current noise. For an accurate current-noise measurement, it is crucial to suppress the attenuation. Second, while above we have assumed current-noise measurements using an ammeter and a current amplifier, as shown in Fig.~\ref{fig3_1}(a), below we consider converting $\Delta I$ to voltage noise $\Delta V$ to measure it with an oscilloscope or spectrum analyzer with a broad frequency band and a wide dynamic range. Third, it is essential to meet standard requirements for low-temperature experiments, e.g., minimizing heat inflow into the low-temperature environment and screening external electromagnetic disturbances.

\subsubsection{Measurement setup using a voltage amplifier}
\label{sec:voltage_amp}

\begin{figure}[tb]
\begin{center}
\includegraphics[width=8cm]{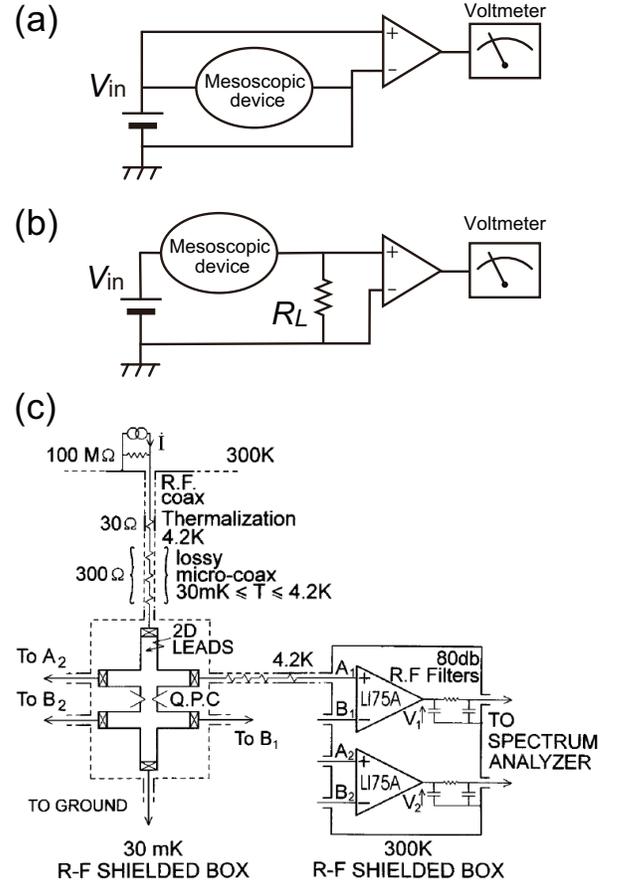}
\end{center}
\caption{(a) Schematic of current-noise measurement using a voltage amplifier and a high-speed voltmeter. (b) Impedance conversion with a load resistance $R_{\rm{L}}$. (c) Example of current-noise measurements using voltage amplifiers. Current noise generated in a QPC causes voltage noise across the QPC. The voltage noise is measured through different pairs of voltage probes and analyzed to evaluate their cross-correlation. Reprinted figure with permission from Ref.~[\onlinecite{KumarPRL1996}]. {\copyright} (1996) American Physical Society.}
\label{fig3_3}
\end{figure}

Current noise $\Delta I$ causes voltage noise $\Delta V = \Delta I~\times ~R$ between the input and output terminals of a sample (resistance $R$). One of the simplest experimental techniques for evaluating $\Delta I$ is to measure $\Delta V$ with a voltage amplifier and a high-speed voltmeter, as shown in Fig.~\ref{fig3_3}(a). Figure~\ref{fig3_3}(c) shows a schematic of the measurement setup using this technique reported in Ref.~[\onlinecite{KumarPRL1996}]. The sample is a QPC fabricated in a 2DES in an AlGaAs/GaAs heterostructure. Voltage noise $\Delta V=\Delta I/G_{\rm{QPC}}$ between the input and output terminals flows through coaxial cables and amplified to $\Delta W = A \times \Delta V$ at room temperature. Here, $G_{\rm{QPC}}$ is the two-terminal conductance of the QPC, and $A$ is the amplifier's gain. The spectrum analyzer converts the time-domain data $\Delta W(t)$ to the noise auto-correlation PSD $S^W(f) \equiv 2\langle \Delta W(t)^2\rangle_{f}/\Delta f$. In this setup, the relation between $S^I$ and $S^W$ is given by 
\begin{equation}
\begin{split}
S^W=A^2\times (S^I/G_{\rm{QPC}}^2+S^V_{\rm{input}})+S^V_{\rm{output}},
\label{sw}
\end{split}
\end{equation}
where $S^V_{\rm{input}}$ is the PSD of the input-referred voltage noise of the amplifier, and $S^V_{\rm{output}}$ is the extrinsic noise raised after the amplification.

In the experimental setup shown in Fig.~\ref{fig3_3}(c), both time-domain data $\Delta W_1(t)$ and $\Delta W_2(t)$, on the right- and left-hand sides of the Hall-bar, respectively, are analyzed to evaluate their cross-correlation $S^W_{12}(f) \equiv 2\langle \Delta W_1(t)\Delta W_2(t)\rangle_{f}/\Delta f$ [see Eqs.~(\ref{crosscorr_Sbc}) and (\ref{crosscorr_beta})]~\cite{KumarPRL1996,SampietroRSI1999}. When both the gain and input noise of the two amplifiers are the same ($A_1 = A_2 = A$ and  $S^V_{\rm{input1}}=S^V_{\rm{input2}}=S^V_{\rm{input}}$), $S^W_{12}$ is described as
\begin{equation}
\begin{split}
S^W_{\rm{12}}\simeq A^2\times (S^I/G_{\rm{QPC}}^2+S^V_{\rm{input}}).
\label{sw_cross}
\end{split}
\end{equation}
Note that the output noise $S^V_{\rm{output1}}$ and $S^V_{\rm{output2}}$ are washed out for large $N_T$ because they do not correlate, hence the cross-correlation measurement suppresses the influence of the extrinsic noise.

The above measurement setup can be made using commercially available amplifiers and a spectrum analyzer. Because of its simplicity, this method is useful for some current-noise measurements; it was applied for measuring shot noise generated by spin accumulation in a tunnel-magneto-resistance device~\cite{ArakawaPRL2015}, for example. On the other hand, it only works at very low frequencies because sample resistance $R$ and the stray capacitance $C_{\rm{Coax}}$ of coaxial cables form an RC low-pass filter to set an upper-frequency limit [cut-off frequency $f_{RC} = 1/(2\pi RC)$]. The RC filtering is problematic, particularly when $f_{RC}$ is lower than the corner frequency of the $1/f$ noise. In this case, the $1/f$ noise buries other noises over the entire range of the measurable frequency band, preventing us from evaluating the shot noise. 

The frequency bandwidth can be expanded by shunting the output terminal of a sample to ground with a load resistance $R_{\rm{L}}$ smaller than $R$ [see Fig.~\ref{fig3_3}(b)]~\cite{DelattreNatPhys2009,OkazakiAPL2013,ArakawaPRL2015,OkazakiNatCom2016}. The drawback is that this method suppresses the magnitude of the voltage noise by a factor of $[R_{\rm{L}}/(R + R_{\rm{L}})]^2$, which degrades the resolution of the current-noise measurement.

\subsubsection{Measurement setup using an inductor-capacitor resonant circuit}
\label{sec:LC_resonance}
\begin{figure}[tb]
\begin{center}
\includegraphics[width=8cm]{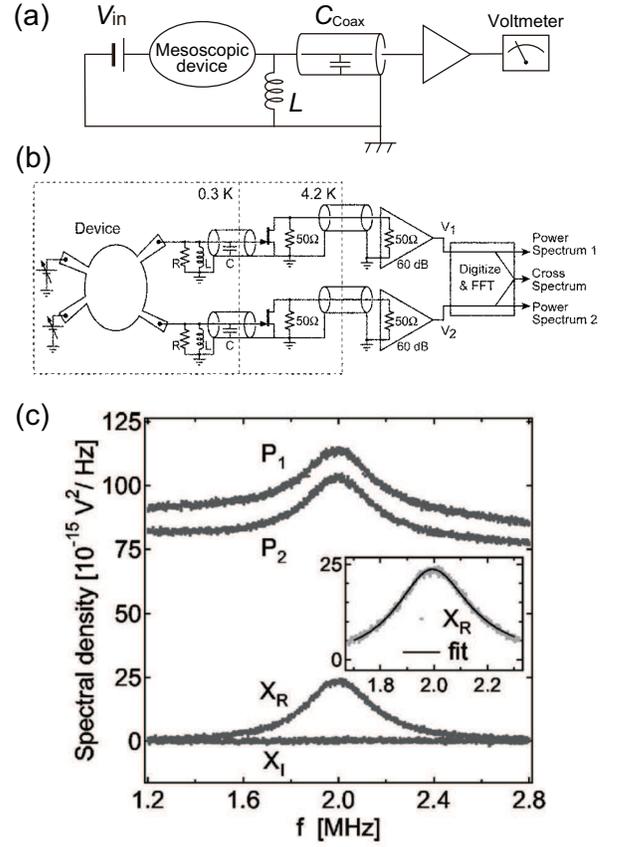}
\end{center}
\caption{(a) Schematic of measurement setup using an LC resonant circuit and a common-source voltage amplifier. (b)(c) Experimental setup with two noise-measurement lines (b) and obtained representative current-noise PSDs (c). Reprinted from Ref.~[\onlinecite{DiCarloRSI2006}], with the permission of AIP Publishing.}
\label{fig3_4}
\end{figure}

In the experimental setup explained above [see Fig.~\ref{fig3_3}(a)], the resistance of a sample ($R\sim h/e^2 \approx 26~\rm{k}\Omega$) and the stray capacitance ($C_{\rm{Coax}} \sim$ 100 pF) typically gives $f_{RC} \simeq 100$ kHz. This $f_{RC}$ value is sometimes not high enough for some current-noise measurements. A method using an inductor-capacitor (LC) resonant circuit has been widely used to solve this problem~\cite{DiCarloRSI2006,HashisakaRSI2009,ArakawaAPL2013,LeeRSI2021}. Figure~\ref{fig3_4}(a) shows the typical measurement setup. The dc output current flows to ground through the inductor $L$ at low temperature. The inductor forms an LC resonant circuit with $C_{\rm{Coax}}$ to have a high impedance $Z_0$ at the resonance frequency $f_{\rm{LC}}=1/(2\pi\sqrt{LC_{\rm{Coax}}})$. Current noise $\Delta I$ generated in a sample causes voltage noise $\Delta V = \Delta I~\times~Z_0$ near $f_{\rm{LC}}$. By choosing an appropriate value of parameter $L$, one can set $f_{\rm{LC}}$ much higher than the $1/f$ corner frequency $f_c$ and thus enable the evaluation of the shot noise. This method is suitable for measuring current noise in various mesoscopic devices, such as QPCs~\cite{DiCarloPRL2008,HashisakaPRB2008, MuroPRB2016}, QH devices~\cite{de-PicciottoNature1997,BartolomeiScience2020}, and quantum dots (QDs)~\cite{McClurePRL2007,FerrierNatPhys2016}. 

Whereas this technique successfully excludes the influence of $1/f$ noise, it has a narrow frequency bandwidth. The narrow bandwidth decreases the number of data points available for the analysis, which increases the standard deviation $\sigma$ of $S^I$ data (see discussion in Sect.~\ref{sec:basics_of}). A cryogenic low-noise amplifier is often used to compensate for the degradation of the resolution.

Note that one needs to carefully calibrate the measurement setup at low temperature because $Z_0$ of the LC circuit and the performance of a cryogenic amplifier differ from those at room temperature (see discussion in Sect.~\ref{sec:calibration_of_noise-measurement}).

Figure~\ref{fig3_4}(b) is an example of experimental setups using LC resonant circuits~\cite{DiCarloRSI2006}. Current noise generated in a sample causes voltage noise near the resonance frequency $f_{\rm{RLC}} \approx 2$ MHz of the resistor-inductor-capacitor (RLC) circuit. The voltage noise is amplified by a homemade cryogenic common-source amplifier at 4.2 K and then taken out of the cryostat using a 50-$\Omega$ impedance-matched coaxial cable. The output current noise is again amplified at room temperature and recorded by an analog-to-digital converter (digitizer) for the FFT analysis. The low-noise performance of the cryogenic amplifier contributes to increasing the resolution. The resolution is further improved by evaluating the cross-correlation between the two measurement lines. 

Figure~\ref{fig3_4}(c) displays representative experimental results of the auto-correlation PSD $P_i$ ($i$=1 or 2) and the real ($X_R$) and imaginary ($X_I$) parts of the cross-correlation PSD. Both $P_i$ and $X_R$ show RLC resonance line shapes. The shot noise generated in the sample is estimated from the peak height, which can be evaluated by a Lorentzian fit, as shown for $X_R$ in the inset.

\subsubsection{Measurement setup using a transimpedance amplifier}
\label{sec:transimpedance}

\begin{figure}[tb]
\begin{center}
\includegraphics[width=8cm]{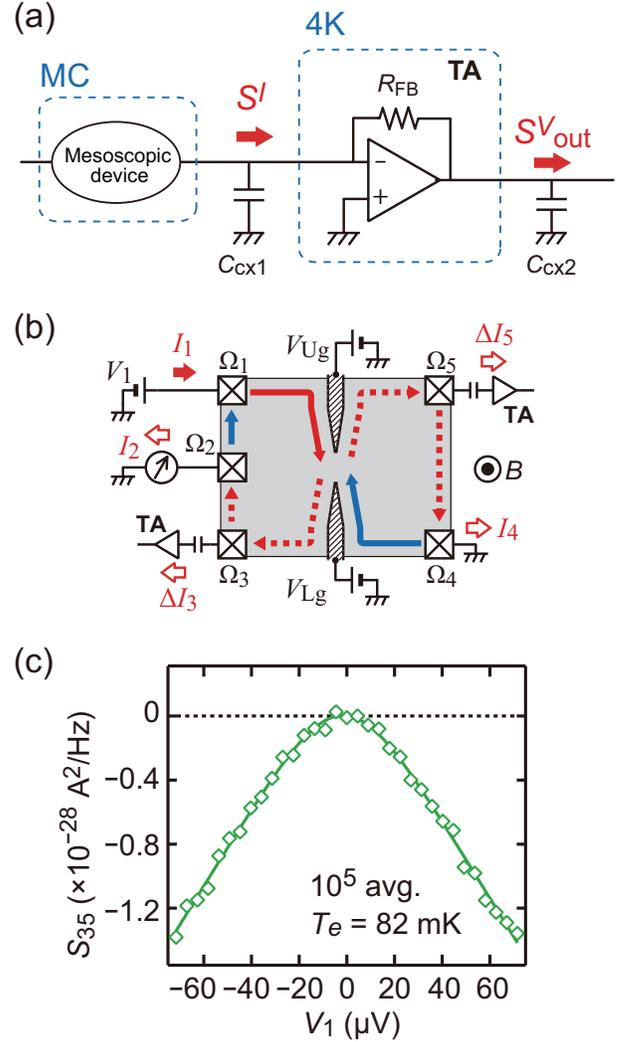}
\end{center}
\caption{(Color online) (a) Schematic of a measurement setup using a TA at 4 K. The TA converts $S^I$, generated in a sample placed on a mixing-chamber (MC) plate in a dilution refrigerator, to $S^V_{\rm{out}}(f) = \vert Z_{\rm{trans}}(f)\vert^2 S^I(f)$. (b) Current-noise measurement for a QPC in a QH system. Current $I_1$ flows into the sample through the ohmic contact $\rm{\Omega}_1$ to impinge on the QPC. The shot noise generated at the QPC is evaluated by measuring the reflected ($I_3$) and transmitted ($I_5$) currents and evaluating their cross-correlation. (c) Representative cross-correlation data $S_{35}=2\langle {\rm{Re}}[\Delta I_3(t)\Delta I_5(t)]\rangle_f/\Delta f$. Reproduced from Ref.~[\onlinecite{HashisakaRSI2014}], with the permission of AIP Publishing.}
\label{fig3_5}
\end{figure}

In the two methods introduced above, current noise is converted to voltage noise and then amplified by a voltage amplifier. On the other hand, a transimpedance amplifier (TA), which converts a current noise $\Delta I$ to voltage noise $\Delta V$ with high transimpedance $Z_{\rm{trans}} = \Delta V/\Delta I \approx R_{\rm{FB}}$, can also be used for current-noise measurements~\cite{HashisakaRSI2014}. Here, $R_{\rm{FB}}$ is the feedback resistance. Figure~\ref{fig3_5}(a) shows a schematic of a measurement setup using a TA. The TA converts $S^I(f)$ generated in a sample to $S^V_{\rm{out}}(f) = \vert Z_{\rm{trans}}(f)\vert^2 S^I(f)$.

Compared with the LC resonant circuit (see Sect.~\ref{sec:LC_resonance}), the wider frequency bandwidth of a TA enables us to use many data points for the histogram analysis, enhancing the resolution of current-noise measurements [see Fig.~\ref{fig3_1}(e)]. For example, in Ref.~[\onlinecite{HashisakaRSI2014}], although the input-referred noise of a TA is relatively high (higher than that of the common-source HEMT amplifier in Ref.~[\onlinecite{DiCarloPRL2008}]), the resolution is as good as that of a measurement setup using an LC resonant circuit~\cite{DiCarloPRL2008}.

It is important to note that this method is advantageous for two-current cross-correlation measurements because the TA's low-input impedance suppresses crosstalk caused by capacitive couplings. For example, when the input impedance of the amplifiers is 10 k$\Omega$, the capacitive coupling of 1 pF induces crosstalk of about 6 \% at 1 MHz (voltage noise $\Delta V_A$ in one of the measurement-lines leads to $\Delta V_B = 0.06\Delta V_A$ in the other line), while it is only 0.06 \% in the case of 100 $\Omega$ input impedance. 

Figure~\ref{fig3_5}(b) shows a schematic of a shot-noise measurement on a QPC fabricated in a QH system using TAs (see Sect.~\ref{sec:current_noise_chiral_edge}), and Fig.~\ref{fig3_5}(c) shows a representative result. Current noise $\Delta I_3$ and $\Delta I_5$ at ohmic contacts $\Omega_3$ and $\Omega_5$ were measured, and their cross-correlation $S_{35}$ was evaluated. The experimental data (diamonds) agrees very well with the theoretical curve (solid line), demonstrating the high reliability of this measurement technique.

While the above experiment used TAs based on HEMTs~\cite{HashisakaRSI2014}, a TA using a superconducting-quantum-interference device (SQUID) can also be employed for current-noise measurements~\cite{JehlRSI1999,JehlNature2000,TranJJAP2017}. The advantage of the SQUID is that it can be placed on a mixing-chamber plate, namely close to a sample, because of its small energy consumption. However, it cannot be used at finite magnetic fields due to the breakdown of the superconductivity.

\subsubsection{High-frequency shot-noise measurements}
\label{sec:high_frequency}

While above we have introduced shot-noise measurements in the white-noise limit, other experiments have demonstrated shot-noise measurements at higher frequencies. The gigahertz shot-noise intensity corresponds to the number of gigahertz photons generated by charge scattering; hence, measuring it is important for understanding the correlation between electrons and photons in mesoscopic systems. Such high-frequency measurements can provide unique information on correlated electron systems, for example, the Josephson frequency in fractional QH systems~\cite{KapferScience2019,BisogninNatCom2019}.

\begin{figure}[htb]
\begin{center}
\includegraphics[width=8cm]{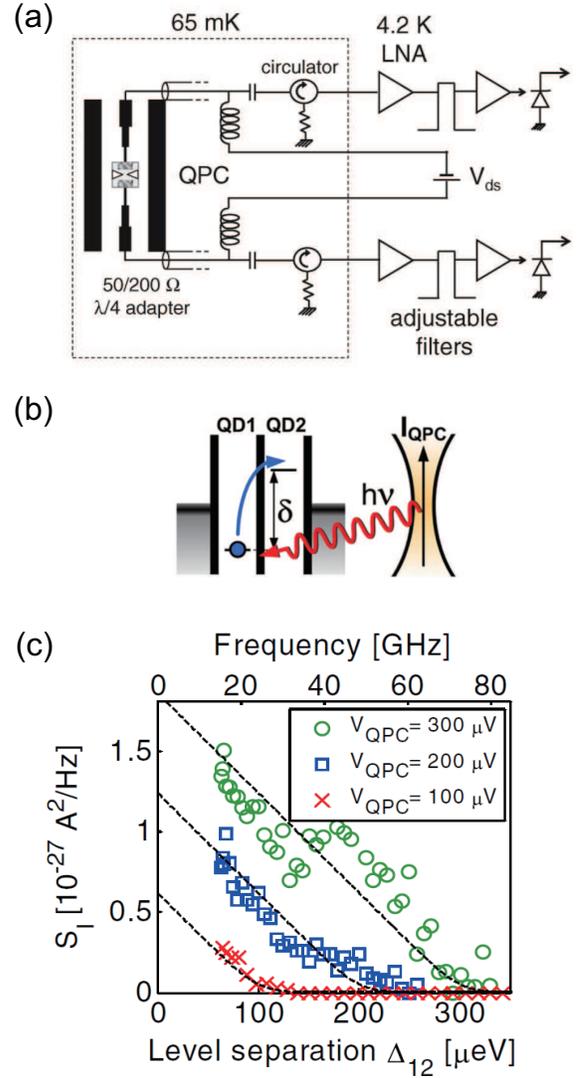}
\end{center}
\caption{(Color online) (a) Schematic of a current-noise measurement at gigahertz frequencies. Reprinted figure with permission from Ref.~[\onlinecite{ZakkaPRL2007}] Copyright (2007) by the American Physical Society. Shot noise generated in a QPC placed on a coplanar waveguide is amplified by a cryogenic LNA and then measured at room temperature. Circulators prevent the backflow of microwave photons from room temperature. (b) Schematic of a frequency-selective shot-noise measurement using a DQD photon detector. The DQD detects microwave photons emitted from the QPC, the energy of which corresponds to the level separation $\delta$. (c) Representative shot-noise PSD measured using the setup shown in (b). The measurements were performed at several source-drain biases ($V_{\rm{QPC}}$) applied to the QPC. Panels (b) and (c) are reprinted with permission from Ref.~[\onlinecite{GustavssonPRL2007}]. {\copyright} (2007) American Physical Society.} 
\label{fig3_6}
\end{figure}

Figure~\ref{fig3_6}(a) presents a schematic of a high-frequency shot-noise measurement (from 4 to 8 GHz)~\cite{ZakkaPRL2007}. Shot noise is generated at a QPC placed on a coplanar waveguide due to a dc source-drain bias applied through bias-tee circuits. The high-frequency shot noise enhanced by the reflections at both ends of the coplanar waveguide is amplified by a cryogenic low-noise-amplifier (LNA) and a room-temperature amplifier before it is detected as rf photons by photodiodes. Although LNAs usually generate large extrinsic noise, circulators installed at low temperatures prevent its backflow to the sample. Thus, cryogenic high-speed measurement techniques has allowed rf shot-noise detection.

High-frequency shot noise can also be measured using an on-chip photon detector~\cite{AguadoPRL2000,DeblockScience2003,OnacPRL2006,GustavssonPRL2007}. Figure~\ref{fig3_6}(b) shows a schematic of such a measurement using a double quantum dot (DQD), which detects the shot noise generated in a nearby QPC~\cite{GustavssonPRL2007}. When the DQD absorbs a photon emitted from the QPC, the energy of which corresponds to the energy-level separation $\delta$ in the DQD, an electron located in the left QD (QD1) is transferred to the right QD (QD2) to be measured as a current flowing through the DQD. This process allows us to perform frequency-selective shot-noise detection by tuning $\delta$ with gate voltages. Figure~\ref{fig3_6}(c) shows representative shot-noise PSDs measured as a function of the level separation ($\delta = \Delta_{\rm{12}}$). For the three different source-drain biases applied to the QPC, the PSDs agree well with the theoretical shot-noise curves (dashed lines) over a broad frequency range (from 15 to 80 GHz). Similar on-chip rf shot-noise detection has also been demonstrated using carbon nanotubes~\cite{OnacPRL_2_2006} and semiconductor nanowires~\cite{GustavssonPRB2008}. High-frequency shot noise has also been measured by bolometric-detection techniques using a 2DES as a detector~\cite{HashisakaPRB2008,JompolNatCom2015}.

\subsubsection{Counting of single electrons}
\label{sec:FCS}
If one monitors all the electrons flowing through a sample, the obtained time-domain data provide perfect information on the probability distribution of the electron scattering process. Although such a measurement is difficult for general mesoscopic devices, it has been achieved for QD devices, thanks to their small number of transmitted electrons per unit time and charge detectors sensitive enough to the charging effect in QDs~\cite{GustavssonPRL2006,FujisawaScience2006,BelzigPRB2005,KaasbjergPRB2015}.

\begin{figure}[htb]
\begin{center}
\includegraphics[width=8cm]{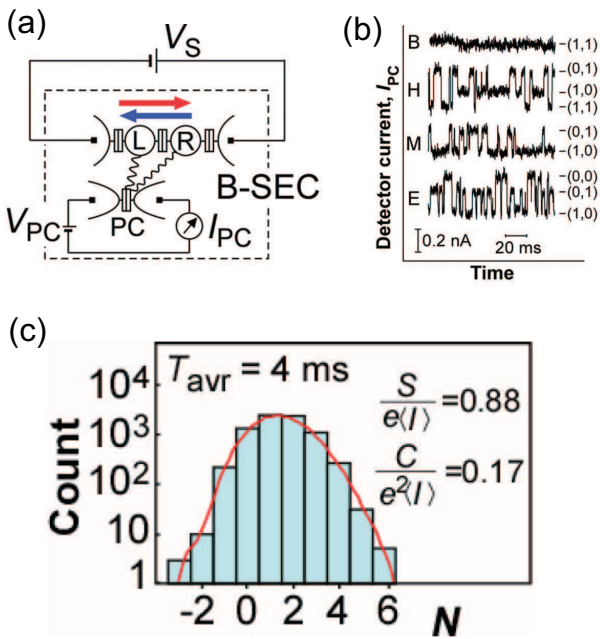}
\end{center}
\caption{(Color online) (a) Schematic of an electron counting device. Current $I_{\rm{PC}}$ flowing through a point contact (PC) varies depending on the electron occupation in the DQD placed near the PC, forming a charge detector. (b) Representative time-domain signals of $I_{\rm{PC}}$ taken in different regimes of the DQD (B, blockade regime; H, hole-like transport regime; M, middle regime; and E, electron-like transport regime). Stepwise fluctuation indicates the change in the charge state in the DQD. (c) Representative result of a histogram analysis for the number of electron transmissions during a time interval. Second-order noise $S=e^2\langle\delta N^2\rangle /T_{\rm{avg}}$ and third-order noise $C=e^3\langle\delta N^3\rangle /T_{\rm{avg}}$ are estimated from the fit, where $\delta N = N - \langle N\rangle$ and $T_{\rm{avg}}$ is the averaging time. Panels are reprinted with permission from Ref.~[\onlinecite{FujisawaScience2006}]. {\copyright} (2006)  American Association for the Advancement of Science.}
\label{fig3_7}
\end{figure}

Figure~\ref{fig3_7}(a) shows a schematic of a counting-statistics experiment for a series DQD~\cite{FujisawaScience2006}. The QPC asymmetrically coupled to the DQD operates as a charge detector because the transmitted current $I_{\rm{PC}}$ flowing through it depends on the electron numbers ($n$, $m$) in the left and right QDs, respectively. Figure~\ref{fig3_7}(b) shows representative time-domain data of $I_{\rm{PC}}$ measured under different DQD conditions. The stepwise fluctuation of $I_{\rm{PC}}$ reflecting changes in ($n$, $m$) enables us to monitor the one-by-one electron transport through the DQD. Figure~\ref{fig3_7}(c) shows the result of a histogram analysis for the transmitted current. One observes that about two electrons are transmitted through the DQD per unit time and that the electron number has a finite variance due to the shot-noise generation. Moreover, the asymmetric distribution indicates finite skewness, namely third-order noise generated in the transmission process. Thus, even higher-order cumulants can be evaluated in counting-statistics experiments.

\section{Examples of shot noise studies}
\label{sec:shotexample}
The purpose of this review is to show what can be learned from a combination of current and current-noise measurements. We present several shot-noise experiments that are helpful for this purpose. Before discussing the quantum many-body phenomena in Sect.~\ref{sec:quantumliquid}, in this section, we introduce experiments, most of which can be understood within the Landauer formalism based on the single-particle picture. Although excellent studies were performed in the 1990s as well, we mainly focus on experiments that have been conducted since the previous reviews published around 2000~\cite{deJong1997,BlanterPR2000,MartinBook}. 

\subsection{Single-channel transport through a QPC}
\label{subsec:QPC}
The shot-noise formula [Eq.~(\ref{ShotTheory})] has been quantitatively verified in experiments on QPCs, where the number of conduction channels and their transmission probabilities can be precisely controlled. Therefore, here we first address the shot noise in a QPC.

\begin{figure}[tbhp]
\center 
\includegraphics[width=8cm]{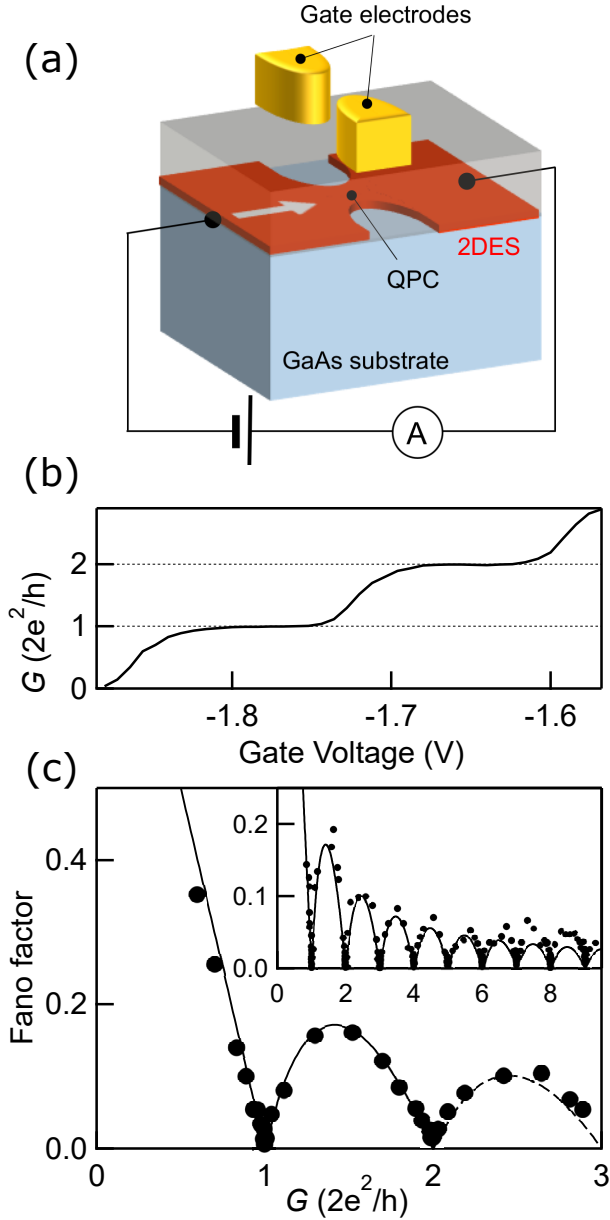} 
\caption{(Color online) (a) Schematic of a QPC fabricated in a 2DES. Two large 2DES regions are connected through a point-like constriction formed using split-gate electrodes. (b) Conductance quantization through a QPC measured at 100 mK and 0 T. Conductance $G$ varies as a function of gate voltage, showing stepwise behavior in a unit of $2e^{2}/h$~\cite{MuroPRB2016}. (c) Measured Fano factor (circles) plotted as a function of the conductance. The solid line is the theoretical curve [see Eq.~(\ref{FanoTheory})]. (inset) Fano factor over a wide range of conductance up to $10\times 2e^2/h$. Panels (b) and (c) are reproduced from Ref.~[\onlinecite{MuroPRB2016}]. } \label{QPC_Fano}
\end{figure}

A QPC is often fabricated in a 2DES formed in a GaAs/AlGaAs heterostructure using a pair of gate electrodes (``split-gate'' electrodes). A negative split-gate voltage depletes the 2DES underneath the electrodes, as shown in Fig.~\ref{QPC_Fano}(a), and then decreases the width of the 2DES constriction. The constriction works as a point contact between the two large 2DES regions. In a high-mobility 2DES, the electron mean free path can be much longer than the length of the constriction (typically $\simeq 1~\mu$m) so that electrons are ballistically transmitted through it. When the constriction width is comparable to the Fermi wavelength ($\simeq$ 40 nm for electrons in a typical 2DES), only a small number of conduction channels exist in the constriction, and the transmission probability ${\cal{T}}$ of each channel can be varied continuously as a function of the gate voltage. In such a case, the conductance shows a stepwise or ``quantized'' behavior due to the electron's wave nature; therefore, the constriction is called a QPC. 

Figure~\ref{QPC_Fano}(b) shows typical conductance-quantization data obtained from a QPC formed in a high-mobility ($1,000-2,000$~m$^2/$Vs) 2DES~\cite{MuroPRB2016}. When the gate voltage is increased from the pinch-off voltage ($\simeq -1.9~$V), conductance $G$ increases in a stepwise manner with a unit of $2e^2/h$, where factor 2 reflects the spin degeneracy. The conductance plateaus around $-1.8$~ and $-1.65$~V indicate that each conduction channel is fully transmitted or reflected. Such conductance quantization, which is fully explained by Eq.~(\ref{Landauer}), was first reported in 1988~\cite{vanWeesPRL1988}. Today, one can observe more than 20 conductance steps ($\sim20\times 2e^2/h$) for a high-quality QPC~\cite{vanWeesPRL1988,RosslerNJP2011}.

Immediately after the first QPC experiment~\cite{vanWeesPRL1988}, shot noise in a QPC was intensively studied both theoretically~\cite{LesovikJETP1989,ButtikerPRL1990,ButtikerPRB1992,MartinPRB1992} and experimentally~\cite{ReznikovPRL1995,KumarPRL1996,LiuNature1998,NakamuraPRB2009,MuroPRB2016}. Figure~\ref{QPC_Fano}(c) displays the shot-noise data, namely the Fano factor (circles) estimated from the bias $V$ dependence of the current noise [see Eq.~(\ref{ShotTheory})], observed in the QPC, the conductance $G$ of which is shown in Fig.~\ref{QPC_Fano}(b)~\cite{MuroPRB2016}. The experimental data agree very well with the theoretical curve (solid curve) calculated using Eq.~(\ref{FanoTheory}); for example, both experimental and theoretical values are zero at $2e^2/h$ and $4e^2/h$. The inset of Fig.~\ref{QPC_Fano}(c) shows the measured Fano factor over a wider range of $G$ (up to $\simeq 10\times 2e^2/h$), again demonstrating the agreement between the experiment and theory. These data indicate that the shot noise is generated only in the channel of intermediate transmission probability ($0 < {\cal T}_n < 1$).

It is important to note that at low temperatures, the electron occupation probability in the leads is either 0 or 1 at each energy (see Sect.~\ref{sec:LandauerFormula}). In this case, the impinging current does not fluctuate, and hence, the excess noise, or the shot noise, directly reflects the scattering process at the sample (here, a QPC)~\cite{OliverScience1999,BeenakkerPT2003}.

Because the shot noise generated in a QPC is well explained by considering electron partitioning in each conduction channel, the stepwise conductance trace in Fig.~\ref{QPC_Fano}(b) and the shot-noise result in Fig.~\ref{QPC_Fano}(c) provide the same information. In contrast, below we introduce several experiments on two-channel transport, where the shot noise provides information different from the conductance.

\subsection{Two-channel transport}
Let us consider spin-dependent transport through the lowest one-dimensional subband in a QPC. Here, the transmission probabilities of spin-up and spin-down electrons are given by ${\cal{T}}_\uparrow$ and ${\cal{T}}_\downarrow$, respectively. The conductance $G$ through the QPC is described as
\begin{equation}
G = \frac{e^2}{h}({\cal{T}}_\uparrow+{\cal{T}}_\downarrow).
\label{eq:spinpol_G}
\end{equation}
From Eq.~(\ref{FanoTheory}), the Fano factor $F_\textrm{sp}$ is written as
\begin{equation}
F_\textrm{sp} =\frac{(1-{\cal{T}}_\uparrow){\cal{T}}_\uparrow+(1-{\cal{T}}_\downarrow){\cal{T}}_\downarrow}{{\cal{T}}_\uparrow+{\cal{T}}_\downarrow}.
\label{Fano_Spin}
\end{equation}
Because we can evaluate both ${\cal{T}}_\uparrow$ and ${\cal{T}}_\downarrow$ by solving these two equations, the combination of conductance ($G$) and shot-noise ($F_\textrm{sp}$) measurements enables a fully quantitative estimation of the transmission probabilities.

\begin{figure}[tbp]
\center 
\includegraphics[width=7.5cm]{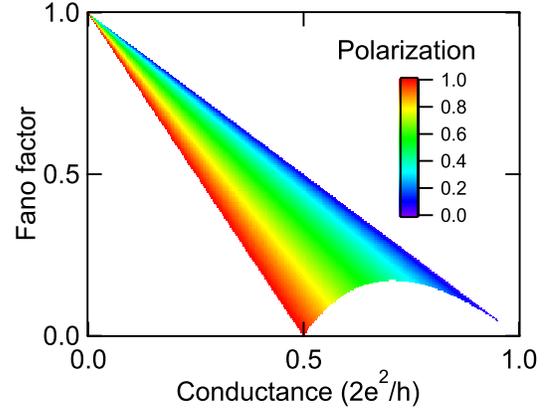} 
\caption{(Color online) Spin polarization $P$ and the Fano factor $F$ as a function of the conductance $G$. Compare it with Fig.~\ref{QPC_KohdaNatComm}(a).} \label{QPC_FanoSpin}
\end{figure}

The spin polarization $P$ of the transmitted current can be defined as $P\equiv \vert{\cal{T}}_\uparrow-{\cal{T}}_\downarrow\vert/({\cal{T}}_\uparrow+{\cal{T}}_\downarrow)$. When $P=0$, namely ${\cal{T}}_\uparrow = {\cal{T}}_\downarrow = {\cal{T}}_0$, we obtain $F = 1-{\cal{T}}_0$ from Eq.~(\ref{Fano_Spin}). On the other hand, when $P>0$, we find 
\begin{equation}
F_\textrm{sp}=\frac{(1-{\cal{T}}_\uparrow){\cal{T}}_\uparrow+(1-{\cal{T}}_\downarrow){\cal{T}}_\downarrow}{{\cal{T}}_\uparrow+{\cal{T}}_\downarrow}< 1-\frac{{\cal{T}}_\uparrow+{\cal{T}}_\downarrow}{2}.
\end{equation}
This relation tells that the spin polarization always decreases the shot-noise intensity, as visually presented in Fig.~\ref{QPC_FanoSpin}.

\subsubsection{Spin polarization}
\label{secInGaAs}
One of the authors of this review performed a shot-noise measurement to evaluate the spin-polarized transport in a solid-state Stern-Gerlach type experiment~\cite{KohdaNatComm2012}. The device was a QPC in a 2DES in an InGaAsP/InGaAs heterostructure, where the spin-orbit interaction is significant. In this device, the Rashba spin-orbit interaction emerges due to the potential modulation at the edges fabricated by chemical etching. Electrons propagating through the constriction undergo the spatial modulation of the effective magnetic field induced by the spin-orbit interaction, resulting in the separation of propagation trajectories between spin-up and spin-down electrons. 

\begin{figure}[tbp]
\center 
\includegraphics[width=8.5cm]{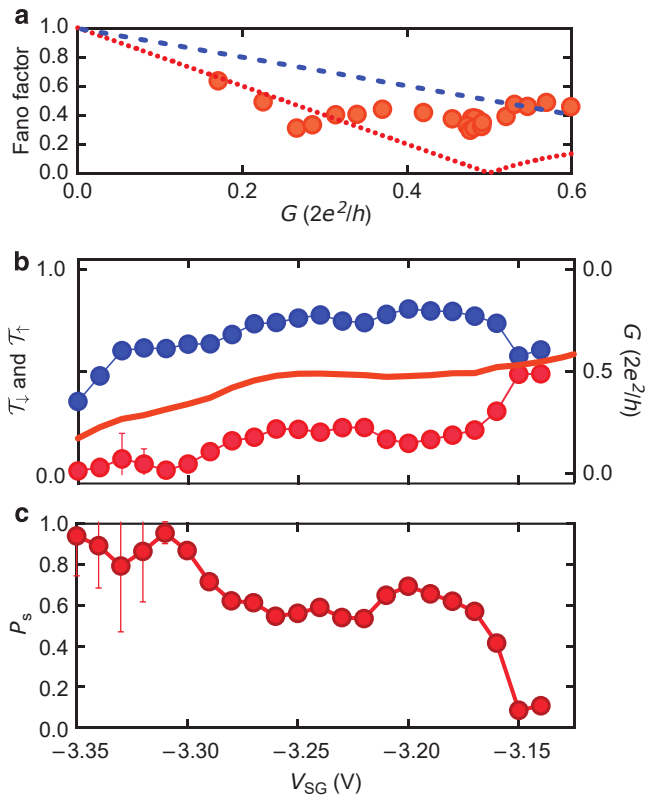} 
\caption{(Color online) (a) Fano factor plotted as a function of conductance $G$ through a InGaAs-based QPC. Experimental results are marked with filled circles. The dashed blue curve presents the theoretical calculation assuming $P=0$, while the dotted red one is that assuming $P=1$. See also Fig.~\ref{QPC_FanoSpin}. (b) $G$ as a function of side gate voltage $V_{\textrm{SG}}$ (solid curve; right axis). Filled circles show the estimated ${\cal{T}}_\downarrow$ and ${\cal{T}}_\uparrow$ (left axis). (c) $V_{\textrm{SG}}$ dependence of $P$ ($=P_{\textrm{S}}$ in the figure). Figures are reprinted from Ref.~[\onlinecite{KohdaNatComm2012}]. {\copyright} (2012) The Author(s).} \label{QPC_KohdaNatComm}
\end{figure}

Figure~\ref{QPC_KohdaNatComm} shows the results of measurements performed at a zero magnetic field at 4.2 K. The solid curve in Fig.~\ref{QPC_KohdaNatComm}(b) shows the QPC conductance $G$ as a function of the side-gate voltage ($V_{\textrm{SG}}$). The conductance shows a plateau at $~0.5(2e^2/h)$ between $V_{\textrm{SG}}=-3.15$~V and $-3.25$~V, suggesting the lifting of the spin degeneracy at a zero magnetic field. Figure~\ref{QPC_KohdaNatComm}(a) plots the Fano factor $F$ evaluated by shot-noise measurements as a function of $G$. The measured $F$ is smaller than the theoretical curve calculated assuming spin degeneracy, namely $P=0$ (dashed line) and is close to the one assuming $P=1$ (dotted line). The observed Fano-factor reduction is the signature of spin polarization. When we solve Eqs.~(\ref{eq:spinpol_G}) and (\ref{Fano_Spin}) together using the measured $G$ and $F$ values, ${\cal{T}}_\downarrow$ and ${\cal{T}}_\uparrow$ are evaluated as shown in Fig.~\ref{QPC_KohdaNatComm}(b). They differ from each other below $G = 0.5(2e^2/h)$. Figure~\ref{QPC_KohdaNatComm}(c) summarizes the $V_{\textrm{SG}}$ dependence of $P$, where one observes $P \simeq 0.7$ at $0.5(2e^2/h)$ and its further increase at lower $G$.

The theoretical simulation well supports the observed spin-polarized transport~\cite{KohdaNatComm2012}; it demonstrates that the narrowing of the transport channel enhances the reflection rate of the spin-down electrons more than that of the spin-up ones, resulting in the spin polarization of the transmitted current. 

The above results point to the potential of the Stern-Gerlach-type device as a spin-polarized-current source in spintronics applications, and at the same time, demonstrate the usefulness of combining conductance and shot-noise measurements for analyzing two-channel transport.

\subsubsection{0.7 conductance anomaly}
Since the mid 1990s, many experiments reported a peculiar conductance behavior of QPCs fabricated in standard GaAs/AlGaAs heterostructures. An unexpected plateau-like structure appears below the first conductance plateau at $2e^2/h$, often near $0.7\times 2e^2/h$; therefore, the behavior is referred to as the ``0.7 (conductance) anomaly''. Various theoretical and experimental studies have attempted to find the origin of the 0.7 anomaly (e.g. Refs.~[\onlinecite{IqbalNature2013}] and~[\onlinecite{BauerNature2013}]). However, even now, its origin and the mechanism are not completely understood. Here, we introduce a few theoretical and experimental works related to the shot-noise studies (for details of the 0.7 anomaly, see the special section in \textit{Journal of Physics: Condensed Matter} published in 2008~\cite{PepperJPC2008}). 

In 1996, Thomas \textit{et al.} experimentally demonstrated that the 0.7 anomaly continuously changes to a spin-polarized conductance plateau at $e^2/h$ by applying an in-plane high magnetic field~\cite{ThomasPRL1996}. This observation suggests that the spontaneous spin polarization at a zero magnetic field is responsible for the 0.7 anomaly. Another important observation is the resemblance between the 0.7 anomaly and the Kondo effect observed in QDs~\cite{DiCarloPRL2006}. In contrast to QDs, a QPC does not have a well-defined localized state; however, theories have discussed the idea that spin-dependent localized states could appear even in a QPC~\cite{RejecNature2006} and that the Kondo effect via the localized state could be the origin of the 0.7 anomaly~\cite{CronenwettPRL2002}. 

Shot-noise studies have been performed to obtain more profound insight into this phenomenon~\cite{RochePRL2004,DiCarloPRL2006,NakamuraPRB2009}. The experiments have found that the 0.7 anomaly causes the Fano factor reduction, as seen in Fig.~\ref{QPC_FanoSpin}. One possible explanation for this observation is spin-polarized transport, as in the case of the Stern-Gerlach-type experiment (see the last subsection). However, in contrast to the Stern-Gerlach-type experiment that can be understood within the single-particle picture, the 0.7 anomaly manifests the presence of a spin-related many-body effect and requires more careful analysis to identify its origin. We expect that recent progress in Kondo physics [see Sect.~\ref{Subsec:KondoNoise}] may provide a better understanding of the 0.7 anomaly.

\subsubsection{Spin current}
While we have discussed shot-noise measurements on spin-polarized transport in semiconductor devices, they have also been employed to evaluate spin-polarized transport in metallic devices. Such experiments are of particular importance for spintronics, where spin current, a flow of spin angular momentum, is the central issue. 

\begin{figure}[!b]
\center
\includegraphics[width=7cm]{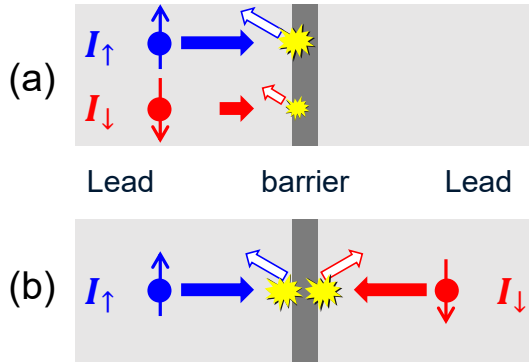} 
\caption{(Color online) (a) Shot-noise generation by scattering of spin-up and spin-down charge currents at a tunnel barrier. (b) Scattering of a pure spin current. In this case, although the net charge current flowing through the barrier is zero because $I_\uparrow=-I_\downarrow$, shot noise is generated proportionally to $|\langle I_\uparrow \rangle |+|\langle I_\downarrow \rangle |$.}
\label{SpinCurrentBarrier}
\end{figure}

Here, we take Schottky's discussion in 1918~\cite{SchottkyAP1918} one step further. Because an electron carries charge and spin, the discrete nature of spin, as well as that of charge, may cause current noise. Therefore, it is natural to ask whether shot noise is generated when a tunnel barrier scatters spin current, as shown in Fig.~\ref{SpinCurrentBarrier}(a). 

In the zero-temperature case where the Fano factor is one, shot noise generated at a tunnel barrier is described as $S=2e(|\langle I_\uparrow\rangle |+|\langle I_\downarrow\rangle |)$ within the single-particle picture, where $I_\uparrow$ ($I_\downarrow$) is the charge current of spin-up (spin-down) electrons [Fig.~\ref{SpinCurrentBarrier}(a)]. Charge and spin currents are defined as $I_\textrm{C}= I_\uparrow+ I_\downarrow$ and $I_\textrm{S}= I_\uparrow- I_\downarrow$, respectively. Suppose that a pure spin current impinges on the barrier ($I_\textrm{C}=0$ and $I_\textrm{S}>0$). In this case, although no net current flows through the barrier because $I_\uparrow=-I_\downarrow$, finite shot noise is generated proportionally to $|\langle I_\uparrow \rangle |+|\langle I_\downarrow \rangle |$ [Fig.~\ref{SpinCurrentBarrier}(b)]. This equation manifests the idea that the shot noise directly measures the amount of spin current.

Based on this idea, one of the authors of this review conducted an experiment to detect the shot noise associated with spin currents~\cite{ArakawaPRL2015}. In this experiment, a spin-polarized current ($I_\uparrow\neq I_\downarrow$) flowing in a non-magnetic metal was applied to a tunnel junction; the spin-polarized current was fed from a ferromagnetic metal through the other tunnel junction. Whereas the experiment was performed not for a pure spin current but a spin-polarized charge current due to a technical issue, the spin-current-induced shot noise was successfully detected in the experiment. While there have been numerous theoretical explanations of spin-current noise since the early 2000s~\cite{MishchenkoPRB2003,BelzigPRB2004,LamacraftPRB2004,MeairPRB2011}, this experiment was its first demonstration to the best of our knowledge. The spin-current detection by shot-noise measurement is promising for studying various spintronics issues, such as spin-transfer torque and thermal spin phenomena.

\subsubsection{Edge mixing}
Orbitals, like spins, sometimes lead to two-channel transport. Shot-noise measurements are helpful in investigating such two-orbital (or two-pseudo-spin) transport. Here, as an example, we introduce shot-noise measurements performed on graphene $pn$ junctions in QH regimes.

Graphene is a typical two-dimensional system that behaves as a zero-gap semiconductor with Dirac-like linear band dispersion, where the polarity of charge carriers can be controlled by applying a gate voltage. When we prepare $p$- and $n$-type regions, where charge carriers are holes and electrons, respectively, in a single graphene device, a $pn$ junction is formed at their boundary. A graphene $pn$ junction has been extensively studied as a promising candidate for observing various intriguing phenomena, such as Klein tunneling~\cite{KatsnelsonNatPhys2006} and the Veselago lens~\cite{CheianovScience2007}.

\begin{figure}[!b]
\center 
\includegraphics[width=6cm]{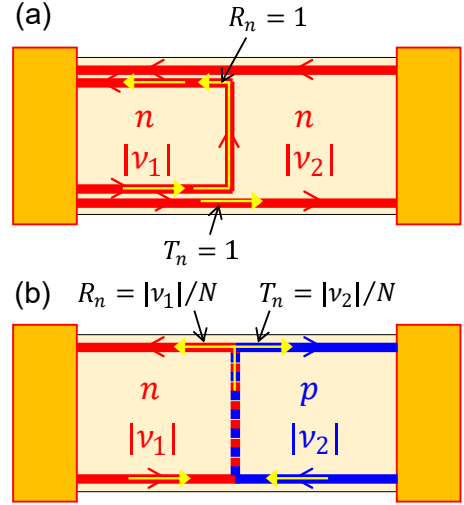} 
\caption{(Color online) (a) Schematic of QH edge channels in a unipolar graphene device. The edge channels are either fully transmitted or reflected at a boundary of different QH states. (b) Schematic of edge channels in a bipolar graphene device. Because of the opposite chiralities in the $p$- and $n$-type regions, the edge channels copropagate along the junction and equilibrate with each other~\cite{AbaninScience2007}.} 
\label{Fig_GraphenePN}
\end{figure}

When graphene is in the QH regime, unique two-channel transport appears at the $pn$ junction, as demonstrated in both experiments~\cite{WilliamsScience2007, OzyilmazPRL2007} and theories~\cite{AbaninScience2007}. In a QH system, charge carriers propagate along unidirectional one-dimensional channels referred to as edge channels, as discussed in Sect.~\ref{subsec:qhe}. Here, we consider a graphene device, where two different QH regions (Landau-level filling factor $\nu_1$ and $\nu_2$) form a boundary at the center of a sample. In Figs.~\ref{Fig_GraphenePN}(a) and (b), the arrows schematically show the $n$- and $p$-type edge channels in samples with (a) unipolar ($nn$) and (b) bipolar ($pn$) junctions. In the former case, each edge channel is either fully transmitted or reflected at the junction and hence generates no shot noise. In the latter case, on the other hand, edge channels fed from the left and right reservoirs encounter each other at the junction bottom, reflecting the opposite chirality in the $p$- and $n$-type regions. The edge channels copropagate along the junction and then separate again at the junction top. The $pn$ junction mixes the potentials between the $p$- and $n$-type edge channels during the copropagation. 

Let us discuss transport properties of graphene QH junctions in more detail. In the case of a unipolar junction between the $\nu_1$ and $\nu_2$ states, the two-terminal conductance $G_{\textrm{uni}}$ of the sample is given by
\begin{equation}
G_{\textrm{uni}}=\min(\vert \nu_1\vert, \vert\nu_2\vert)\frac{e^2}{h},
\label{Eq_GrapheneNNQH}
\end{equation}
where $\min(\vert \nu_1\vert, \vert\nu_2\vert)$ is the lower value between $\vert \nu_1\vert$ and $\vert\nu_2\vert$ that corresponds to the number of transmission channels through the sample. On the other hand, in the case of a bipolar junction, the $\nu_1$ and $\nu_2$ edge channels are mixed at the junction, as shown in Fig.~\ref{Fig_GraphenePN}(b). If the charge excitations are evenly redistributed across all the copropagating channels, we can describe the transmission and reflection probabilities of the channels fed from the left contact as ${\cal{T}}_n = \vert \nu_2 \vert/N$ and ${\cal{R}}_n=\vert \nu_1 \vert/N$, respectively (here, $N = \vert \nu_1 \vert +\vert \nu_2 \vert$). In this case, $G_{\textrm{bi}}$ is expressed as
\begin{equation}
G_{\textrm{bi}}=\frac{e^2}{h}\sum_n^{\vert \nu_1 \vert}{\cal{T}}_n =\frac{\vert \nu_1 \vert \vert \nu_2 \vert}{\vert \nu_1 \vert +\vert \nu_2 \vert}\frac{e^2}{h}.
\label{Eq_GraphenePNQH}
\end{equation}

Suppose that the edge mixing is caused by elastic charge-scattering processes between the copropagating channels. In this case, the $pn$ junction works as a beam splitter for charge carriers, like a standard QPC does in a GaAs/AlGaAs heterostructure. The Fano factor, which quantifies the shot-noise intensity generated at the junction, is described as~\cite{AbaninScience2007}
\begin{equation}
F=\frac{\vert \nu_1 \vert\vert \nu_2 \vert}{(\vert \nu_1 \vert+\vert \nu_2 \vert)^2}.
\label{eq:fano_pn}
\end{equation}
For example, Eq.~(\ref{eq:fano_pn}) predicts $F = 1/4$ for $(\nu_1, \nu_2)=(\pm 2, \mp 2)$ and $F = 3/16$ for $(\nu_1, \nu_2)=(\pm 2, \mp 6), (\pm 6, \mp 2)$. This equation contrasts with the case of a unipolar junction [Fig.~\ref{Fig_GraphenePN}(a)], where no carrier partitioning occurs to generate the shot noise.

\begin{figure*}[!t]
\center 
\includegraphics[width=12cm]{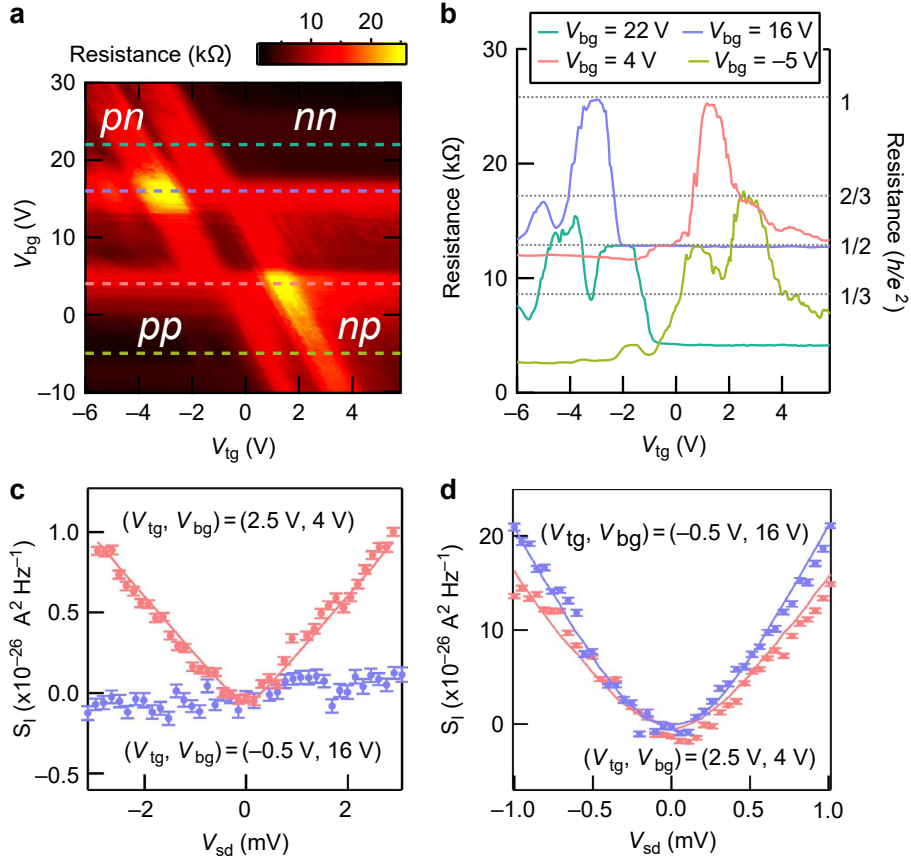} 
\caption{(Color online) (a) Color plot of the two-terminal resistance as a function of $V_{\textrm{tg}}$ and $V_{\textrm{bg}}$ at 8 T. (b) Cross sections of the color plot at $V_{\textrm{bg}}=22, 16, 4$, and $-5$ V. (c) Measured shot noise $S_I$ as a function of $V_{\textrm{sd}}$ at 8 T at $(V_{\textrm{tg}}, V_{\textrm{bg}})=(2.5~\textrm{V}, 4~\textrm{V})$ and $(-0.5~\textrm{V}, 16~\textrm{V})$, which correspond to $(\nu_1, \nu_2)=(6, -2)$ and $(2, 2)$, respectively. The solid curve is the numerical fit. (d) $S_I$ as a function of $V_{\textrm{sd}}$ at 0 T under the same gate-voltage conditions as in (c). The solid curves are the numerical fits. Figures are reproduced from Ref.~[\onlinecite{MatsuoNatComm2015}]. {\copyright} (2015) The Author(s).} 
\label{Fig_GraphenePNExp}
\end{figure*}

Several experiments demonstrated that the two-terminal conductance of graphene $pn$ junctions is well explained by Eq.~(\ref{Eq_GraphenePNQH})~\cite{WilliamsScience2007,OzyilmazPRL2007,MatsuoSciRep2015}. Moreover, one of the authors of this review measured the shot noise in a narrow ($< 10~\mu$m) $pn$-junction device to confirm the relation in Eq.~(\ref{eq:fano_pn})~\cite{MatsuoNatComm2015}. In the shot-noise experiment, the QH junction was formed by applying a back-gate voltage ($V_{\textrm{bg}}$) to tune the carrier density over the whole graphene device and a top-gate voltage ($V_{\textrm{tg}}$) to modify the density in the half area of the device. Figure~\ref{Fig_GraphenePNExp}(a) shows a color plot of the measured two-terminal resistance $R$ as a function of $V_{\textrm{tg}}$ and $V_{\textrm{bg}}$ near the Dirac point ($V_{\textrm{tg}}\simeq 0$ V and $V_{\textrm{bg}}\simeq 10$ V), where the formations of the $nn$, $pn$, $np$, and $pp$ junctions are observed. Figure~\ref{Fig_GraphenePNExp}(b) shows the resistance traces along the cross sections in Fig.~\ref{Fig_GraphenePNExp}(a) at $V_{\textrm{bg}}=22, 16, 4$, and $-5$ V. The observed quantized resistances at $R=h/2e^2$ and $h/6e^2$ in the $nn$ and $pp$ regimes exhibit the formation of unipolar QH junctions [Fig.~\ref{Fig_GraphenePN}(a)]. On the other hand, $ R=h/e^2$, $\frac{2}{3}h/e^2$, and $\frac{1}{3}h/e^2$ plateaus in the $pn$ and $np$ regimes demonstrate bipolar QH junctions [Fig.~\ref{Fig_GraphenePN}(b)]. These results are well explained by Eqs.~(\ref{Eq_GrapheneNNQH}) and (\ref{Eq_GraphenePNQH})~\cite{WilliamsScience2007, OzyilmazPRL2007,AbaninScience2007}.

Figure~\ref{Fig_GraphenePNExp}(c) shows the current-noise data measured for the unipolar $(\nu_1, \nu_2)=(2, 2)$ [$(V_\textrm{tg}, V_\textrm{bg}) = (-0.5~\textrm{V}, 16~\textrm{V})$, $R=\frac{1}{2}\frac{h}{e^2}$] junction and the bipolar $(\nu_1, \nu_2)=(6, -2)$ [$(V_\textrm{tg}, V_\textrm{bg}) = (2.5~\textrm{V}, 4~\textrm{V})$, $R=\frac{2}{3}\frac{h}{e^2}$] one. Shot noise is absent in the unipolar case, which agrees with the above explanations for Fig.~\ref{Fig_GraphenePN}(a). The absence of shot noise was observed at $(\nu_1, \nu_2)=(-2, -2)$ and $(2, 6)$, too. In contrast, finite shot noise is observed in the bipolar case. The Fano factor evaluated by the numerical fit is $0.18\pm0.01$, close to the theoretical value of $3/16=0.1875$ [see Eq.~(\ref{eq:fano_pn})]. We also obtained $F=0.18\pm 0.01$ at $(\nu_1, \nu_2)=(2, -6)$. These observations show that the edge mixing in the narrow $pn$ junction can be regarded as the elastic charge-scattering or ``beam-splitting'' process between the channels. While it is difficult to fabricate a QPC in graphene, a zero-gap semiconductor with linear dispersion, the above results indicate that a $pn$ junction works as a beam splitter, which is a fundamental building block for fermion quantum optics in condensed matter (for examples, see Refs.~[\onlinecite{WeiSciAdv2017}] and~[\onlinecite{MorikawaAPL2015}]).

Note that entirely different experimental results are obtained at a zero magnetic field, namely in non-quantum-Hall systems. Figure~\ref{Fig_GraphenePNExp}(d) shows the shot-noise data under the same gate-voltage conditions as in Fig.~\ref{Fig_GraphenePNExp}(c). Finite shot-noise generation is observed in both unipolar and bipolar junctions, indicating the difference from the quantum-Hall-junction case. While theories predict $F=1-1/\sqrt{2}\sim 0.29$~\cite{CheianovPRB2006} at a zero field, we observed $F\sim 0.5$ due to the influence of disorder in the sample~\cite{LewenkopfPRB2008}.

A closely related study of a graphene QH $pn$ junction was reported by Kumada \textit{et al.} at the same time~\cite{KumadaNatComm2015}. They measured the $pn$-junction-width dependence of the shot-noise intensity to observe a monotonic decrease with increasing junction width. This observation indicates that the copropagating channels relax to the thermal equilibrium state after a long propagating distance, and in the long-channel limit, the $pn$ junction behaves as a floating ohmic contact.

While we have discussed junction devices in graphene so far, here, we briefly mention the shot noise in a uniform graphene device at a zero magnetic field. Theory predicts shot-noise generation in the ballistic region even when the graphene is ideally homogeneous, and the Fano factor at the Dirac point is $1/3$ in a short and wide graphene strip~\cite{TworzydloPRL2006}. Interestingly, the Fano factor of $F=1/3$ equals that of disordered metals in the classical diffusive regime. The $F=1/3$ shot noise in graphene has been under scrutiny in several experiments~\cite{DanneauPRL2008,DiCarloPRL2008,FayPRB2011,TanPRB2013,SahuPRB2019}.

\subsection{Multiple-parameter case}
\label{subsec:multichannel}

In the last subsection, we discussed transport phenomena dominated by two parameters: the transmission probabilities of spin-up ($\cal{T}_\uparrow$) and spin-down ($\cal{T}_\downarrow$) electrons. This subsection shows how shot-noise measurements have been applied to cases of three or more parameters, for example, in the quantum-Hall-effect breakdown regime (Sect.~\ref{subsub:QHEBD}) or tunnel-junction devices with multiple tunneling paths (Sect.~\ref{subsub:MTJ}). Generally, transport properties in such nonequilibrium and/or large systems are difficult to evaluate quantitatively. However, shot-noise measurements sometimes provide critical information to solve such complicated problems.

\subsubsection{Breakdown of the QH effect}
\label{subsub:QHEBD}

Here, we introduce experiments in which three different measurements---conductance, shot-noise, and resistively-detected nuclear-magnetic-resonance (RD-NMR) measurements---were performed to investigate the breakdown of the QH effect, a typical non-equilibrium phenomenon in mesoscopic systems~\cite{ChidaPRB2012,HashisakaPRB2020}. 

\begin{figure*}[tb]
\center 
\includegraphics[width=15cm]{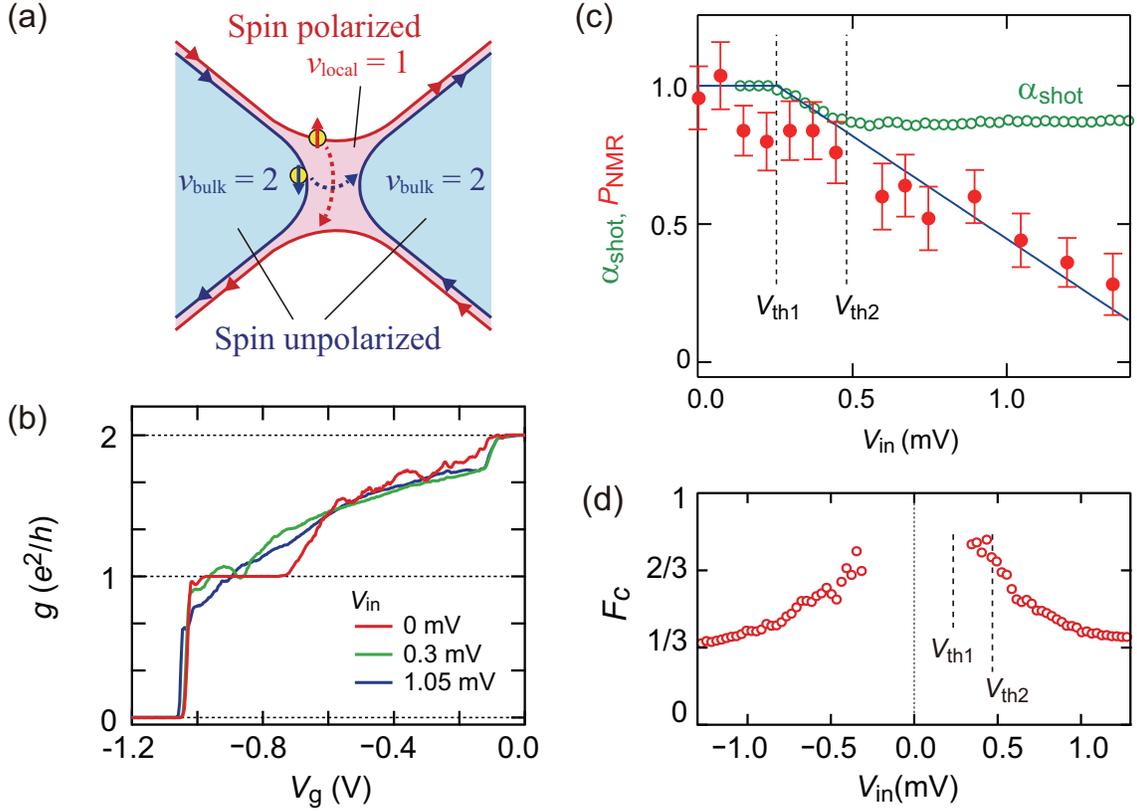}
\caption{(Color online) (a) Schematic of a spin-polarized $\nu=1$ QH state locally formed at a narrow constriction in a bulk $\nu=2$ QH system. The $\nu=1$ and $\nu=2$ edge channels are represented by solid red and blue arrows, respectively. Possible spin-conserved inter-channel tunnelings are shown by the dashed red and blue arrows. (b) Differential conductance $g$ measured as a function of $V_{\rm{g}}$. Conductance plateau at $g = e^2/h$ observed at $V_{\rm{in}} = 0$ breaks down at finite bias. (c) $V_{\rm{in}}$ dependence of $\alpha_{\rm{shot}}$ and $P_{\rm{NMR}}$ evaluated by shot-noise measurement and RD-NMR measurement, respectively, at $V_{\rm{g}}=-0.96$ V. $\alpha_{\rm{shot}}$ starts to decrease from $\alpha_{\rm{shot}} = 1$ at $V_{\rm{in}} = V_{\rm{th1}}$ and saturates at $\alpha_{\rm{shot}} \simeq 0.9$ above $V_{\rm{in}} = V_{\rm{th2}}$, while $P_{\rm{NMR}}$ monotonically decreases with increasing $V_{\rm{in}}$. (d) $V_{\rm{in}}$ dependence of Fano factor $F_C$ estimated by solving equations for the measured conductance, shot noise, and Knight shift of NMR. Reproduced figure with permission from Ref.~[\onlinecite{HashisakaPRB2020}]. {\copyright} (2020) American Physical Society.}
\label{fig4_1}
\end{figure*}

As discussed in Sect.~\ref{subsec:qhe}, the QH effect is a phenomenon in which the Hall conductance of a 2DES is quantized in units of $e^2/h$ under a high perpendicular magnetic field. In a QH regime, the bulk region of a 2DES becomes insulating (incompressible) due to the complete occupation of the Landau levels below the Fermi energy, and chiral one-dimensional channels are formed at the edge of the 2DES. Accordingly, the longitudinal resistance becomes zero, and the Hall conductance is quantized, reflecting the absence of electron backscattering along the edge channels. When one applies a high source-drain voltage to a QH system, the Hall conductance deviates from the quantized value, while the longitudinal resistance takes on a finite value. This nonlinear behavior is referred to as quantum-Hall-effect (QHE) breakdown~\cite{NachtweiPhysicaE1999}.

It is well known that the quantized Hall resistance of the QH state is used as a resistance standard, and an accurate Hall-resistance measurement should apply a current as high as possible within the linear-response regime. In this context, it is essential to understand the QHE-breakdown mechanism, which is the cause of the nonlinear behaviors at finite bias. Many experiments have been conducted on the breakdown mechanism in macroscopic Hall-bar or Corbino-type samples of several $\rm{\mu}$m to mm in size~\cite{NachtweiPhysicaE1999}. Current noise in such a macroscopic sample has also been measured to observe the precursory phenomenon of the QHE breakdown, generation of finite excess noise in the linear-response regime~\cite{ChidaPRB2013}. 

The QHE breakdown has also been studied as a representative nonequilibrium phenomenon in a mesoscopic system~\cite{NachtweiPhysicaE1999}. Notably, the breakdown of a spin-polarized QH state has come under scrutiny as a possible source of nuclear spin polarization in GaAs-based heterostructures. In this context, breakdown phenomena in locally-formed mesoscopic QH systems have often been examined in experiments~\cite{WaldPRL1994,DixonPRB1997,YusaNature2005,MasubuchiAPL2006,CorcolesPRB2009,ChidaPRB2012,HennelPRL2016,FauziPRB2017,HashisakaPRB2020}. Current-noise measurements are even more potent for investigating such small systems than they are for macroscopic systems. 

Below we discuss the QHE breakdown of a local $\nu=1$ system formed in a bulk $\nu=2$ system~\cite{ChidaPRB2012,HashisakaPRB2020}. Figure~\ref{fig4_1}(a) shows a schematic of such a local $\nu=1$ system. When one applies a negative split-gate voltage $V_{\rm{g}}$ to form a narrow constriction in the $\nu=2$ system, zero-bias conductance through the constriction varies as a function of $V_{\rm{g}}$, as shown by the red trace in Fig.~\ref{fig4_1}(b). The conductance plateau at $e^2/h$ ($-1.0~{\rm{V}}<{\it{V}}_{\rm{g}}<-0.7~{\rm{V}}$) indicates the formation of the local $\nu=1$ state due to the decrease in electron density in the constriction. When a high source-drain bias $V_{\rm{in}}$ is applied, the transmitted current varies nonlinearly to break down the conductance plateau [see green and blue traces in Fig.~\ref{fig4_1}(b)]. Intuitively, the most-likely mechanism for such a nonlinear behavior is the spin-conserving tunneling of spin-down electrons between the $\nu=2$ edge channels or that of spin-up electrons between the $\nu=1$ channels [schematically shown in Fig.~\ref{fig4_1}(a)].

Here, we formulate the transmitted current $I_t$ across the constriction as $I_t = I_{\uparrow}{\cal{T}}_{\uparrow}+I_{\downarrow}{\cal{T}}_{\downarrow}$, where $I_{\uparrow(\downarrow)}$ is the spin-up (spin-down) current impinging on the constriction and ${\cal{T}}_{\uparrow(\downarrow)}$ is the transmission probability of the spin-up (spin-down) electrons. If we assume that the inter-channel tunneling current is carried by stochastic electron tunneling, i.e., with no correlation [see Fig.~\ref{fig4_1}(a)], we can evaluate ${\cal{T}}_{\uparrow}$ and ${\cal{T}}_{\downarrow}$ by solving Eqs.~(\ref{eq:spinpol_G}) and (\ref{Fano_Spin}) together and estimate the spin polarization $\alpha_{\rm{shot}} \equiv ({\cal{T}}_{\uparrow}-{\cal{T}}_{\downarrow})/({\cal{T}}_{\uparrow}+{\cal{T}}_{\downarrow})$. Open green circles in Fig.~\ref{fig4_1}(c) shows the $V_{\rm{in}}$ dependence of $\alpha_{\rm{shot}}$. In the linear-response regime at low bias ($V_{\rm{in}}<V_{\rm{th1}}$), we observe $\alpha_{\rm{shot}}=1$ because spin-up electrons are fully transmitted through the constriction (${\cal{T}}_{\uparrow}=1$) while spin-down electrons are completely reflected (${\cal{T}}_{\downarrow}=0$). In the nonlinear regime ($V_{\rm{in}}>V_{\rm{th1}}$), $\alpha_{\rm{shot}}$ decreases with increasing $V_{\rm{in}}$, and when $V_{\rm{in}}$ is further increased ($V_{\rm{in}}>V_{\rm{th2}}$), $\alpha_{\rm{shot}}$ saturates at about 0.9. The observed decrease in $\alpha_{\rm{shot}}$ in the first breakdown regime ($V_{\rm{th1}}<V_{\rm{in}}<V_{\rm{th2}}$) is interpreted as the result of the interchannel electron tunneling [see Fig.~\ref{fig4_1}(a)]. Tunneling of spin-up electrons decreases ${\cal{T}}_{\uparrow}$ from 1 while that of spin-down electrons increases ${\cal{T}}_{\downarrow}$ from 0~\cite{ChidaPRB2012}. In the second breakdown regime ($V_{\rm{in}}>V_{\rm{th2}}$), on the other hand, saturation of $\alpha_{\rm{shot}}$ suggests that a different mechanism causes the nonlinear behavior. 

Figure~\ref{fig4_1}(c) compares $\alpha_{\rm{shot}}$ with the spin polarization in the constriction $P_{\rm{NMR}} \equiv (n_{\uparrow}-n_{\downarrow})/(n_{\uparrow}+n_{\downarrow})$, where $n_{\uparrow}$ ($n_{\downarrow}$) is spin-up (spin-down) electron density, evaluated from the Knight shift of NMR~\cite{HashisakaPRB2020}. One observes that $P_{\rm{NMR}}$ monotonically decreases with increasing $V_{\rm{in}}$ over the entire range. This result indicates that the saturation of $\alpha_{\rm{shot}}$ reflects a mechanism different from the decrease in the spin polarization, that is, the breakdown of the incompressibility of the local $\nu=1$ state. In the second breakdown regime, spin-down electrons frequently tunnel through the local $\nu=1$ region, leading to a decrease in $P_{\rm{NMR}}$ and the resultant suppression of the exchange energy. Accordingly, the spin gap in the constriction closes and the stochastic electron-tunneling picture breaks down to cause the deviation of current noise from the theoretical shot-noise value [Eq.~(\ref{ShotTheory})]. This scenario was confirmed by solving together three independent equations for the experimental data, i.e., dc conductance, shot noise, and the NMR Knight shift. The solution indicates that the Fano factor $F_C$ of the shot noise monotonically decreases from $F_C \simeq 1$ to 1/3 with increasing $V_{\rm{in}}$, as shown in Fig.~\ref{fig4_1}(d). The value of $F_C \simeq 1/3$ suggests that a classical diffusive conductor~\cite{BeenakkerPRB1992,NagaevPRB1995,KozubPRB1995,OppenPRB1997,SteinbachPRL1996} or a local $\nu=1/3$ fractional QH state~\cite{RoddaroPRL2003,RoddaroPRL2004,HashisakaPRL2015} is formed in the second nonlinear regime. Although the electron dynamics in this regime is still unclear, the experimental results unambiguously signal the two-step breakdown mechanism, that is, the electron tunneling through the local $\nu=1$ state in the first step and the breakdown of the incompressibility of the $\nu=1$ state in the second step. 

Here, we again emphasize that combining the three measurement techniques enables us to identify the two-step QHE breakdown. The above experiment clearly indicates that current-noise measurements provide essential information for understanding complicated nonlinear phenomena in nonequilibrium systems. Shot-noise measurements have also served as efficient probes for QHE breakdown in other experiments performed on GaAs/AlGaAs heterostructures~\cite{ChidaPRB2014,HataJPCM2016} and graphene~\cite{YangPRL2018,LaitinenJLTP2018}, where collective excitations, referred to as magneto-excitons, play an important role in the breakdown mechanism~\cite{YangPRL2018}.

\subsubsection{Coherent tunneling}
\label{subsub:MTJ}
A tunnel junction composed of a thin insulator layer between metals is a representative example of multichannel systems. In contrast to a QPC, where the charge current flows through only a few conduction channels, a large number of channels of small transmission probabilities carry a current through a conventional tunnel junction. This has been confirmed by shot-noise measurements demonstrating the Fano factor $F=1$ of the Poisson processes. On the other hand, ``coherent tunneling'' through magnetic tunnel junctions (MTJs)---a highly transmissive tunneling process conserving all of the energy, momentum, and spin---identified in dc transport measurements requires further shot-noise studies for ensuring the highly transmissive nature of the tunneling process. Here, we introduce a shot-noise measurement performed on MTJs showing coherent tunneling.

\begin{figure}[tbp]
\center 
\includegraphics[width=8.5cm]{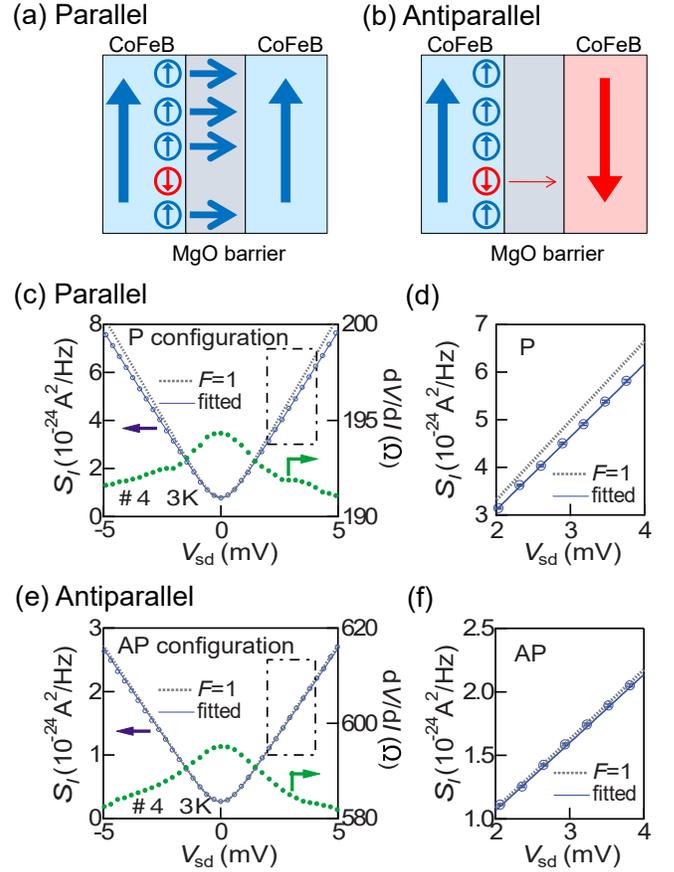} 
\caption{(Color online) Schematic of a CoFeB/MgO/CoFeB MTJ (a) in the parallel (P) configuration and (b) in the antiparallel (AP) configuration. (c) Source-drain bias $V_{\rm{sd}}$ dependence of $dV/dI$ (solid mark, right axis) and $S_I$ (open mark, left axis) in the P configuration. The solid curve fits the current-noise data with $F=0.91 \pm 0.01$, showing deviation from the dashed line assuming $F=1.0$. (d) A part of the graph of Fig.~\ref{MTJ_Arakawa}(c) surrounded
by a dotted-dashed rectangle is enlarged to show that the experimental result clearly deviates from the $F=1.0$ case. (e) and (f) are the counterparts for the AP configuration of Figs.~\ref{MTJ_Arakawa}(c) and (d), respectively. The shot noise fits well with the curve assuming $F=1.0$. Reprinted from Ref.~[\onlinecite{ArakawaAPL2011}], with the permission of AIP Publishing.} \label{MTJ_Arakawa}
\end{figure}

An MTJ is a junction consisting of a tunnel barrier between ferromagnetic metal layers. The tunneling resistance depends on whether the configuration of the magnetization directions is parallel or antiparallel. The resistance in the former case is lower than that in the latter one, as schematically shown in Figs.~\ref{MTJ_Arakawa}(a) and (b). The magnetization-configuration dependence of resistance, referred to as the tunneling magnetoresistance (TMR) effect, is a vital topic in spintronics.

Compared with an MTJ with an amorphous AlO$_x$ barrier~\cite{GuerreroPRL2006,ScolaAPL2007,GuerreroAPL2007,CascalesPRL2012}, an MTJ composed of a crystallized magnesium-oxide (MgO) barrier shows a huge magnetoresistance, exceeding 1,000\%~\cite{YuasaNatMat2004,ParkinNatMat2004,YuasaJPSJ2008}. Theory explains that the presence of coherent-tunneling process only in the parallel configuration is responsible for the huge magnetoresistance~\cite{ButlerPRB2001,MathonPRB2001}. Shot-noise measurements performed on MgO-based MTJ devices provide evidence of coherent tunneling~\cite{SekiguchiAPL2010,ArakawaAPL2011}. Figures~\ref{MTJ_Arakawa}(c) and (e) show the results of shot-noise measurements in the parallel and antiparallel configurations, respectively (MgO layer thickness of 1.05~nm). The solid curves are fits to the experimental data using Eq.~(\ref{ShotTheory}). Figures~\ref{MTJ_Arakawa}(d) and (f) present magnified views of a part (surrounded by a dotted-dashed rectangle) of Figs.~\ref{MTJ_Arakawa}(c) and (e), respectively. Figures~\ref{MTJ_Arakawa}(e) and (f) tell that $F$ is very close to 1 ($F=0.98 \pm 0.01$) in the antiparallel configuration, indicating that the Schottky-type tunneling carries the current; namely, all the tunneling paths have small transmission probabilities (${\cal T}_n \ll 1$). On the other hand, the Fano factor is $F=0.91 \pm 0.01$ in the parallel configuration, as seen in Fig.~\ref{MTJ_Arakawa}(d). The decrease in $F$ suggests the presence of highly transmissive paths due to the coherent tunneling~\cite{ButlerPRB2001,MathonPRB2001}. A first-principles calculation for a realistic MgO barrier quantitatively explains the observed shot-noise reduction~\cite{LiuPRB2012}. 

After the MgO-based MTJ experiments, a similar experiment was performed on an epitaxial-spinel-barrier junction (MgAl$_2$O$_4$) to observe the presence of coherent tunneling~\cite{TanakaAPEX2012}.

\subsubsection{Atomic and single-molecule junctions}
\label{subsec:molecularjunction}
Shot-noise measurements have also been performed to investigate atomic or single-molecule junctions that show conductance quantization~\cite{AgraitPR2003}. Such junctions are often fabricated using mechanically controllable break-junctions (MCBJs), which enables us to form an ultimately small gap between two metal electrodes and hold atoms or molecules in the gap. Various intriguing phenomena appear in such a junction, depending on the transport properties of both the held atoms or molecules and metal electrodes. For example, Cron \textit{et al.} measured charge transport through an aluminum MCBJ holding a few aluminum atoms and observed multiple Andreev reflections at the junction~\cite{CronPRL2001}. The multiple Andreev reflections result in rich features in the current-voltage ($IV$) characteristics, from which Cron \textit{et al.} extracted the entire set of transmission probabilities ${\cal{T}}_n$, which are referred to as mesoscopic PIN (personal-identification-number) codes. The experiment measured the shot noise to evaluate the effective charge ($2e, 4e\cdots$) associated with the multiple Andreev reflections.

\begin{figure}[!t]
\center 
\includegraphics[width=8cm]{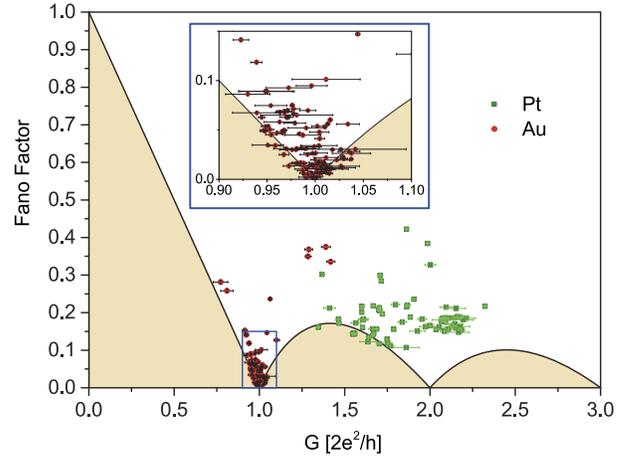} 
\caption{(Color online) Fano factor of 200 different Au or Pt MCBJs. Reprinted from Ref.~[\onlinecite{TewariRSI2017}], with the permission of AIP Publishing.}
\label{AtomicContact_Tewari}
\end{figure}

Shot-noise measurements have also been performed on other atomic or molecular junctions. For example, Fig.~\ref{AtomicContact_Tewari} shows the Fano factors measured for gold or platinum atomic junctions~\cite{TewariRSI2017}. The data obtained from 200 different MCBJs are scattered close to the theoretical shot-noise curve (see also Fig.~\ref{QPC_Fano}), manifesting the appearance of quantized channels in such atomic contacts.

The coupling of an electronic system with other degrees of freedom, such as phonons (or vibration modes) of MCBJs, can be probed by shot-noise measurements~\cite{TalPRL2008,KumarRPL2012,ChenSR2014}. For such measurements, high-stability and high-conductivity molecular junctions, such as the benzene molecular junction~\cite{KiguchiPRL2008}, are fascinating targets. Another direction of the shot-noise study of atomic junctions is to combine it with scanning tunneling microscopy (STM)~\cite{MasseeRSI2018}.

\subsubsection{Quantum dots}
\label{subsub:QD}
Let us consider electron transport through a QD connected to metallic leads. When the capacitance $C$ between the QD and the environment, e.g., leads and gate electrodes, is small, the energy $e^2/2C$ required to add one electron to the QD can be larger than electron temperature $k_\textrm{B}T_\textrm{e}$. In this case, the number of electrons in the QD changes one by one as a function of the applied gate voltage (Coulomb blockade), and finite conductance through the system is observed when the energy level of the QD and the chemical potential of the leads coincide (Coulomb oscillation). Furthermore, when the QD is as small as the de Broglie wavelength of electrons, separation between discrete energy levels exceeds $k_\textrm{B}T_\textrm{e}$. In this case, electron transport occurs through each discrete level.

One may expect that Coulomb repulsion in a QD always suppresses the shot-noise intensity to be sub-Poissonian ($F<1$), as the Pauli exclusion principle does in a QPC. Actually, sub-Poissonian shot noise was observed in the single-electron tunneling regime~\cite{BirkPRL1995}. However, in practice, the shot noise generated in a QD is sometimes super-Poissonian ($F>1$)~\cite{GustavssonPRL2006,OnacPRL2006,ZhangPRL2007,kiesslichPRL2007,FrickePRB2007,ZarchinPRL2007,OkazakiPRB2013,UbbelohdePRB2013,HarabulaPRB2018,SeoPRL2018}, indicating that electrons are ``bunched'' when they transmit through a QD.

One of the mechanisms to enhance the shot noise is a non-Markovian process in a QD~\cite{SukhorukovPRB2001,BelzigPRB2005,ThielmannPRL2005}. For example, let us consider a situation where multiple discrete levels exist in the energy window in a voltage-biased QD. When an electron stays in one of the levels, electrons cannot use the other levels to pass through the QD due to the Coulomb blockade. Suppose that the dwell time of each level differs. When a long-dwell-time level traps an electron, electron transport is suppressed. Otherwise, the current is enhanced from the average in time. In this way, transmitted electrons are bunched in the time domain. Another mechanism is cotunneling, where multiple electrons are involved in a tunneling process.

As a last note, the mechanism of electron transport through a QD is much simpler when only one discrete level contributes to it. This situation is seen, for example, in a small QD fabricated in a carbon nanotube, where the energy separation between discrete levels is large. In this case, the shot noise generated in the QD is well explained by the standard shot-noise formula $S = 2e\vert\langle I \rangle\vert \left(1-\cal{T}\right)$ [see Figs.~\ref{FerrierNatPhysFig2}(d) and~\ref{FerrierNatPhysFig2noise}(a)]~\cite{FerrierNatPhys2016}.

\subsection{Fermion quantum optics}
\label{sec:fermion_optics}
The factor $f_\alpha(\varepsilon)[1-f_\beta(\varepsilon)]$ in Eq.~(\ref{NoiseSingleChannel}) reflects the Pauli exclusion principle of electrons; in the experiments presented above, the fermionic nature of electrons manifests itself in this factor. In contrast, in the research field referred to as ``fermion quantum optics,'' the fermionic nature is observed more directly~\cite{ButtikerScience1999}.

Let us consider a simple example where each of two particles, A and B, randomly takes one of two states, $\vert 1\rangle$ or $\vert 2\rangle$. In this case, possible states are the following four: $\vert 1\rangle_{\rm{A}} \vert 1\rangle_{\rm{B}}, \vert 2\rangle_{\rm{A}}\vert 2\rangle_{\rm{B}}, \vert 1\rangle_{\rm{A}}\vert 2\rangle_{\rm{B}},$ and $\vert 2\rangle_{\rm{A}}\vert 1\rangle_{\rm{B}}$. When the two particles are distinguishable, they take the state $\vert 1\rangle_{\rm{A}} \vert 1\rangle_{\rm{B}}$ ($\vert 2\rangle_{\rm{A}}\vert 2\rangle_{\rm{B}}$) with the probability $P_{\rm{11}}(P_{\rm{22}})=25\%$ independent of their quantum statistical nature. When they are indistinguishable, in contrast, the probability depends on the quantum statistics; the two particles never take one state together in the case of fermions, while they tend to take the same state in the case of bosons. Thus, compared with classical particles, fermions avoid each other (antibunching), while bosons tend to bunch up (bunching). The quantum statistical nature of particles has a vital influence on the shot-noise generation in their scattering processes. 

\begin{figure}[tb]
\center
\includegraphics[width=6cm]{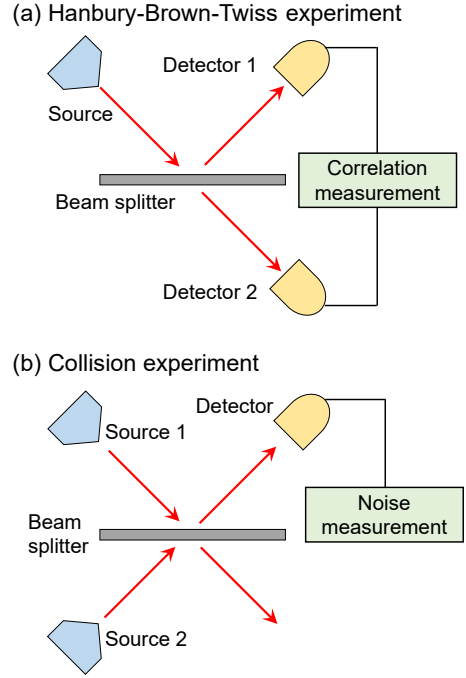}
\caption{(Color online) Schematic of exchange-interference experiments: (a) Hanbury-Brown-Twiss experiment and (b) collision experiment.}
\label{2typeexp}
\end{figure}

Bosonic bunching was first observed in 1954 by Hanbury Brown and Twiss. They estimated the angular diameter of stars by measuring the intensity correlation of light~\cite{HanburyBrown1954,HanburyBrownNature1956}. Purcell interpreted the experimental result as reflecting the bosonic bunching of photons~\cite{PurcellNature1956}. Since the development of the laser, the Hanbury-Brown-Twiss (HBT) setup [Fig.~\ref{2typeexp}(a)] has been widely examined in quantum optics.

The electron-collision experiment in 1998~\cite{LiuNature1998} and the HBT interference experiment in 1999~\cite{HennyScience1999,OliverScience1999} are well-known early fermion-quantum-optics experiments. In the former, electrons randomly ejected from two different sources, 1 and 2, sometimes collide at a beam splitter, as shown in Fig.~\ref{2typeexp}(b). The collisions, which deterministically output one electron each to the two exits, decrease the number of random scattering events at the beam splitter and thus suppress shot-noise generation. Liu {\it{et al}}. observed shot-noise suppression using a beam splitter fabricated in a 2DES in a GaAs/AlGaAs heterostructure~\cite{LiuNature1998}. In the latter, Henny {\it{et al}}. demonstrated the Fermi statistics of electrons in an HBT experiment~\cite{HennyScience1999}. They prepared the HBT setup shown in Fig.~\ref{2typeexp}(a) using a quantum Hall (QH) device and observed negative current-noise cross-correlation reflecting the fermionic nature of electrons. On the other hand, Oliver {\it{et al}}. measured the cross-correlation between the two outputs of a beam splitter in the time domain~\cite{OliverScience1999}. While these HBT experiments were performed using GaAs/AlGaAs semiconductor devices, similar experiments were later conducted using graphene~\cite{TanSciRep2018,EloPRB2019} and free electrons in a vacuum~\cite{KieselNature2002}.

\begin{figure}[tb]
\center
\includegraphics[width=7cm]{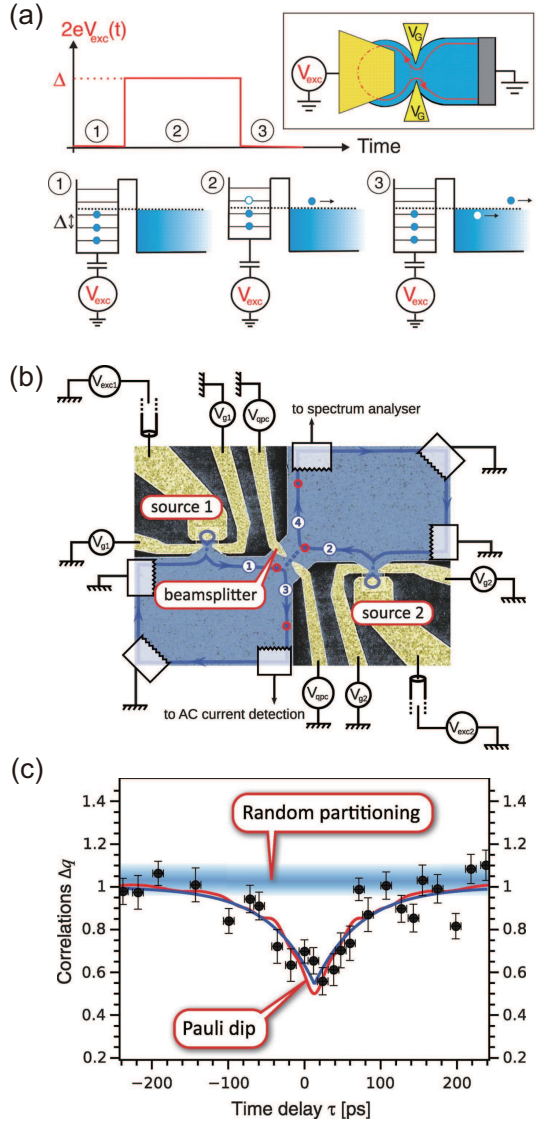}
\caption{(Color online) (a) Schematic of a QD single-electron source. An electron and a hole are ejected one by one from a QD by applying square-wave voltage pulses $V_{\rm{exc}}$ to the gate electrode. Reprinted figure with permission from Ref.~[\onlinecite{FeveScience2007}] Copyright (2007) by American Association for the Advancement of Science. (b) Scanning electron micrograph of an electron-collision device fabricated on an AlGaAs/GaAs 2DES. Electrons ejected from two single-electron sources collide at a beam splitter. (c) Current noise generated at a beam splitter measured as a function of the time delay $\tau$ between the electron ejections. Suppression of the noise at $\tau \simeq 0$ reflects antibunching of electrons due to the Pauli exclusion principle. Panels (b) and (c) are reprinted with permission from Ref.~[\onlinecite{BocquillonScience2013}]. {\copyright} (2013) American Association for the Advancement of Science.}
\label{fig4_x3}
\end{figure}

Another well-known example of a fermion-quantum-optics experiment is Mach-Zehnder interferometry using QH edge channels~\cite{JiNature2003}. This experiment has confirmed the long coherence length of electron waves in a solid-state device and has stimulated various studies on electron-wave interferometry. Recent experiments have demonstrated coherent electron transport over a long distance of 100 $~\mu$m~\cite{DuprezPRX2019}. A significant example of current-noise studies on such interferometers is the one by Neder {\it{et al.,}} who observed exchange interference in a two-particle interferometer~\cite{NederNature2007}. Their study is based on a theoretical proposition of demonstrating electron entanglement in a solid-state device by observing a violation of the Bell inequality~\cite{SamuelssonPRL2004}. 

These experiments may lead to unique developments in fermion quantum optics beyond the mere analogy of quantum optics because electronic systems often produce peculiar quantum many-body states. A recent remarkable example is the demonstration of anyonic statistics of fractionally charged quasiparticles in fractional QH states (for details, see Sect.~\ref{sec:anyonic_statistics})~\cite{BartolomeiScience2020}.

Whereas the above experiments have examined the fermionic nature of electrons by applying a direct current to a mesoscopic device, recent experiments using high-speed electronics have succeeded in observing scattering processes of individual electrons. One of the core technologies in such experiments is a single-electron source, of which several types have been reported~\cite{BauerleRPP2018,FeveScience2007,MaireAPL2008,UbbelohdeNNANO2015,DuboisNature2013,FletcherNatCommun2019}. Since shot noise is generated due to the charge discreteness, the shot-noise measurement plays an essential role in evaluating these single-electron sources. Here, we present two experiments demonstrating single-electron sources, one using a quantum-dot device~\cite{FeveScience2007,BocquillonScience2013} and another using a Lorentzian electron wave packet~\cite{DuboisNature2013}.

\begin{figure}[tb]
\center
\includegraphics[width=7cm]{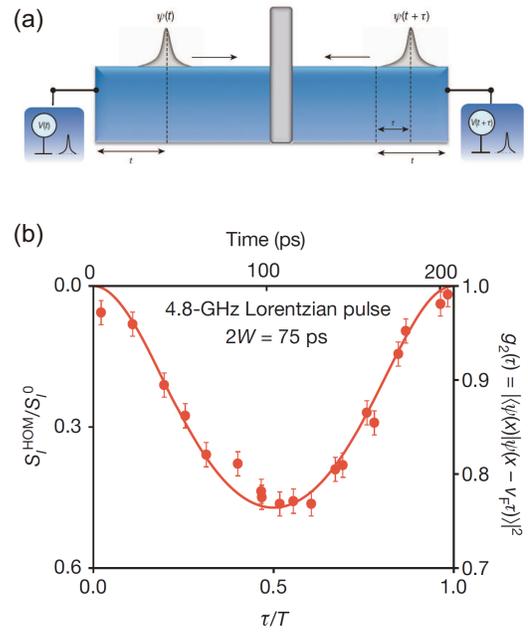}
\caption{(Color online) (a) Schematic of a Leviton-collision experiment. (b) Current noise observed as a function of time delay. The noise suppression at $\tau /T=0$ reflects the HOM interference of Levitons. Reprinted with permission from Ref.~[\onlinecite{DuboisNature2013}]. {\copyright} (2013) Sprinter Nature.}
\label{fig4_x4}
\end{figure}

Figure~\ref{fig4_x3}(a) shows a schematic of a single-electron source using a QD~\cite{FeveScience2007}. A single electron (hole) is ejected from a QD into the lead when a negative (positive) gate-voltage step is applied to the QD to control the number of electrons. Although the ejected electron interacts with electrons in the lead below the Fermi energy to excite many electron-hole pairs after a long propagation time, it propagates coherently within a short time. Figure~\ref{fig4_x3}(b) shows a schematic of an electron-collision experiment using two quantum-dot single-electron sources~\cite{BocquillonScience2013}. When two electrons are incident on the central QPC simultaneously, they collide with each other causing the exchange interference. Figure~\ref{fig4_x3}(c) presents the measured current-noise cross-correlation between the two outputs from the QPC as a function of the time difference $\tau$ between the electron ejections. One observes a suppression of the cross-correlation at $\tau \simeq 0$~ps, which indicates the Pauli exclusion principle of electrons. This experiment can be regarded as a fermionic version of the Hong-Ou-Mandel (HOM) coincidence measurement~\cite{HongPRL1987}.

The fermionic HOM interference effect has also been observed in a Leviton-collision experiment. A Lorentzian pulse excites a collective excitation of electrons without holes. Levitov {\it{et al}}. proposed that a minimal charge excitation, referred to as a Leviton, transferring an elementary charge is possible by controlling the Lorentzian pulse size~\cite{LevitovJMP1996,IvanovPRB1997,KeelingPRL2006}. One can expect to observe the exchange interference of two electrons when two Levitons collide with each other at a QPC [Fig.~\ref{fig4_x4}(a)]. Figure~\ref{fig4_x4}(b) demonstrates the current-noise cross-correlation measured as a function of time delay $\tau$ normalized by the Leviton-ejection period $T$. The measured cross-correlation agrees well with the theoretical curve (solid line), in which the noise suppression at $\tau/T=0$ reflects the Pauli exclusion principle.

\section{Current noise in quantum liquids}
\label{sec:quantumliquid}
\subsection{Quantum liquids and their non-equilibrium}

The behavior of a single particle, an electron, for example, can be explained by solving the Schr\"{o}dinger equation. On the other hand, it is usually impossible to solve the equation rigorously when many particles correlate. Exotic behaviors of such many-particle systems, which cannot be expected from the single-particle picture, have attracted great attention from researchers in condensed-matter physics. We call such a many-particle system, where many indistinguishable particles correlate, showing liquid-like behaviors,  ``quantum liquid''~\cite{NozieresTQL1999}.

Quantum liquids have long been an important topic in condensed-matter physics, and now we can understand the equilibrium properties of several quantum liquids to a considerable extent. However, we do not have any generalized method for predicting their non-equilibrium properties; constructing a canonical way for describing non-equilibrium behavior is one of the most significant challenges in modern physics.

Non-equilibrium phenomena are everywhere: light-matter interaction, transistors in electronic devices, chemical reactions, and life. Despite their familiarity with us, such phenomena are inherently challenging to analyze due to their complexity. Quantum liquids provide a quantum-mechanical prototype of such intriguing non-equilibrium issues and serve as good touchstones for understanding non-equilibrium phenomena.

This section discusses shot-noise measurements on three types of quantum liquids, namely those formed by, respectively, the Kondo effect, fractional quantum Hall effect, and superconductivity.

\subsection{Non-equilibrium fluctuations in the Kondo effect}
\label{Subsec:KondoNoise}
\subsubsection{Kondo effect and local Fermi liquid}
The Kondo effect is a typical quantum many-body phenomenon. The state created by this effect (Kondo state) is a type of quantum liquid called ``local Fermi liquid''. In this subsection, we briefly introduce the Kondo effect and discuss the shot noise in a quantum dot (QD) where the Kondo effect emerges.

We start from the Kondo effect in bulk materials~\cite{KondoPro1988}. Usually, the electrical resistivity of nonmagnetic metals decreases with decreasing temperature because the electron-phonon scattering is suppressed at low temperature. However, we often observe that the resistivity of nonmagnetic metals with a small number of magnetic impurities (say, 0.1--0.001\%) starts to increase at low temperatures, showing a resistivity minimum at a particular temperature. Since the 1930s, this phenomenon had been a long-standing mystery called the ``resistivity minimum phenomenon'' (for historical background, see Ref.~[\onlinecite{KondoJSPS2005}]). In 1964, Kondo theoretically solved this problem by considering a spin-dependent electron scattering at a single magnetic impurity atom~\cite{KondoPTP1964}. 

\begin{figure*}[t]
\center 
\includegraphics[width=12.5cm]{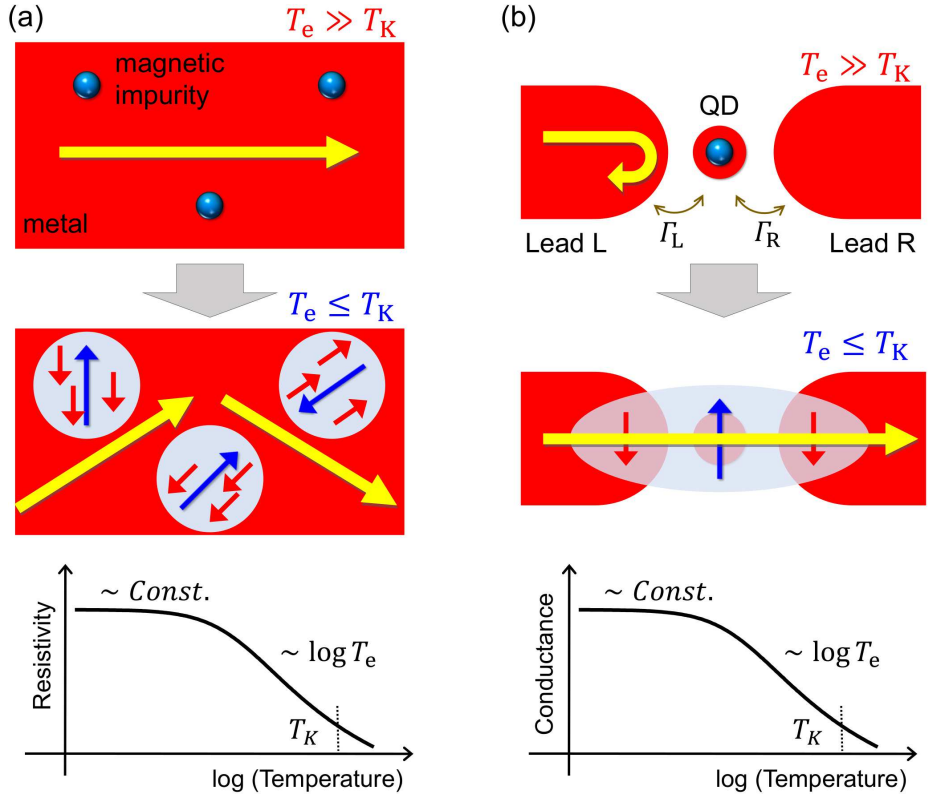} 
\caption{(Color online) (a) Conceptual view of magnetic impurity atoms randomly distributed in a nonmagnetic metal. (top) At high temperature, $T_\textrm{e} \gg T_\textrm{K}$, the localized spins are paramagnetic. (middle) At low temperatures, $T_\textrm{e} \leq T_\textrm{K}$, the spin in the impurity atom and the conduction-electron spin begin to form a bound state, and the system becomes non-magnetic. The motion of the conduction electrons becomes inhibited. (bottom) As a result, the resistivity logarithmically increases. At even lower temperature, the resistivity becomes constant (unitary limit). (b) Similar phenomenon occurs in a QD. Consider a situation where the QD has a single level and contains only one electron. (top) At high temperature, $T_\textrm{e} \gg T_\textrm{K}$, no other electrons are allowed to enter due to $U$ (Coulomb blockade). (middle) At lower temperature, $T_\textrm{e} \leq T_\textrm{K}$, the Kondo state forms, and electrons can pass the QD. (bottom) The conductance of the QD increases logarithmically in decreasing temperature and shows a constant value in the low-temperature limit (unitary limit).} \label{KondoSchematicFig}
\end{figure*}

Figure~\ref{KondoSchematicFig}(a) shows a random distribution of magnetic impurity atoms with localized spins in a nonmagnetic metal. We assume that each impurity atom has a single discrete level. The level forms a resonant state with a finite width $\Gamma$ due to hybridization with the surrounding conduction electrons. Electrons move in and out of the level on a time scale characterized by $\Gamma$. Now, when the Coulomb energy $U$ between the electrons occupying the level is sufficiently large, $U\gg \Gamma$, only one electron can enter it at a time.

At high temperature, the electrons rapidly enter and exit the level one by one on a time scale characterized by $\Gamma$ ($\gg k_\textrm{B}T_\textrm{e}$), leading to the fluctuation of the direction of the spin in the level. Thus, the spins in the magnetic impurity atoms are paramagnetic. However, a different situation arises at low temperature: the spin direction of the electrons entering and exiting the level becomes correlated, which can be described by the second-order perturbation in the Anderson impurity model~\cite{YosidaTM1996}. The correlation becomes increasingly significant as temperature decreases, and finally, the spin in the impurity atom and the conduction-electron spin begin to form a spin-singlet bound state (Kondo state), making the system non-magnetic. The resistivity logarithmically increases due to the scattering by the Kondo state that develops at each impurity atom [see the middle panel of Fig.~\ref{KondoSchematicFig}(a)]. The Kondo temperature $T_\textrm{K}$ is the temperature at which the Kondo state starts to form. At sufficiently low temperature ($T_\textrm{e} \ll T_\textrm{K}$), the resistivity approaches a constant value, meaning that the scattering by the Kondo state is the dominant factor to determine the resistivity. This situation is called the unitary limit, where the Kondo state is a perfect spin-singlet formed around the discrete level of each impurity atom.

The Kondo effect is a phenomenon where the magnetism and resistivity of a nonmagnetic metal gradually change with decreasing temperature due to magnetic impurities. The essence of the Kondo effect lies in that the levels that initially have a resonance state of width $\Gamma$ newly behave according to an energy scale $k_\textrm{B}T_\textrm{K}$ ($k_\textrm{B}T_\textrm{K} \ll \Gamma, U$) due to the presence of the many-body effect $U$ [see Eq.~(\ref{TKexpression}) for the expression of $k_\textrm{B}T_\textrm{K}$]. Kondo calls the emergence of this nontrivial energy scale $k_\textrm{B}T_\textrm{K}$ ``Fermi surface effect'' because the abrupt change in the occupation number at the Fermi surface peculiar to the Fermi-Dirac distribution function is responsible for the logarithmic behavior~\cite{KondoPro1988}. Note that the Kondo effect is not a phase transition but a crossover across $T_\textrm{K}$.

It is also possible to understand the Kondo effect in terms of the ``Fermi liquid'', a kind of quantum liquids: Landau proposed a phenomenological Fermi liquid theory in 1956 and gave its microscopic proof based on many-body quantum theory in 1958~\cite{LandauJETP1956}. Roughly speaking, as long as the low energy physics concerns, the Fermi liquid theory enables us to treat an interacting fermion system as if it is a ``free'' fermion system by renormalizing the interaction. More accurately, due to the renormalization, we have to consider quasi-particles rather than free fermions because there exists residual interaction between quasi-particles. 

Landau's phenomenology describes many-body quantum states in the interacting system using an energy functional. It assumes that the low-energy eigenvalues of the system are the functional of the quasi-particle distribution function's deviation from the ground state. The excitation energy spectrum expressed in the functional form enables us to predict several observables, such as the effective mass of a quasi-particle and the magnetic susceptibility of the system. For example, in liquid~$^3$He, a representative Fermi liquid, we can experimentally determine these parameters and quantitatively predict many-body effects in other physical quantities~\cite{LeggettRPP2016}. Such a method to describe many-body states by a few parameters was also successful in Kondo physics in the 1970s~\cite{NozieresJLTP1974,YamadaPTP1975,YosidaPTP1975,Yamada2PTP1975,ShibaPTP1975,YoshimoriPTEP1976}. In this case, we use the term ``local Fermi liquid'' because we are dealing with a state formed around a localized level. 

\subsubsection{Kondo effect in QDs}
The Kondo effect also occurs in QDs. While the underlying physics is the same between the Kondo effect in bulk metals and that in QDs, it is instructive to discuss the QD case here based on a microscopic model. Now, consider a QD with only a single level and only one electron occupying it, as shown in Fig.~\ref{KondoSchematicFig}(b). This situation is described by the impurity Anderson model $\mathcal{H}_A=\mathcal{H}_0+\mathcal{H}_T+\mathcal{H}_I$:
\begin{equation}
\begin{split}
\mathcal{H}_0 &= \sum_{k\alpha\sigma}\varepsilon_{k\alpha}c^{\dagger}_{k\alpha\sigma}
c_{k\alpha\sigma}+ \sum_{\sigma}\epsilon_d d^{\dagger}_{\sigma}d_{\sigma}, \\ 
\mathcal{H}_T &= \sum_{k\alpha\sigma} (v_\alpha d^{\dagger}_{\sigma}c_{k\alpha\sigma}
+v_\alpha^* c^{\dagger}_{k\alpha\sigma}d_{\sigma}),\\
\mathcal{H}_I  &= U d^{\dagger}_{\uparrow} d_{\uparrow}d^{\dagger}_{\downarrow} d_{\downarrow},
\end{split}
\label{eq:AndersonModel}
\end{equation}
where $c^{\dagger}_{k\alpha\sigma}$ is an operator that creates an electron with wavenumber $k$ and spin $\sigma = \uparrow, \downarrow$ in the left and right leads $\alpha =\textrm{L}$, and $\textrm{R}$, respectively. $d^{\dagger}_{\sigma}$ is an operator that creates an electron with spin $\sigma$ in the level $\epsilon_d$ of the QD. The electrons move between the lead $\alpha$ and the QD with a tunneling matrix element $v_\alpha$, and by this tunneling the level has a line width $\Gamma = \Gamma_\textrm{L} + \Gamma_\textrm{R}$, where $\Gamma_\alpha = 2\pi \rho_c |v_\alpha|^2$ and $\rho_c$ is the density of states of the conduction electrons of the leads. In addition, the electron in the QD has a Coulomb repulsion $U$. The chemical potential of the left and right leads is set to $\mu_\textrm{L/R}=\pm eV/2$, and a voltage $V\geq 0$ is applied between the leads.

If the energy $U$ is sufficiently large such that $U \gg \Gamma$, no other electrons can enter the QD, resulting in a Coulomb blockade, and conduction is inhibited. However, at low temperature $T \lesssim T_\textrm{K}$, a different situation emerges due to the Kondo effect. Even if an electron occupies the QD, another electron with the opposite spin can enter it from either lead, allowing two electrons to coexist, treated by the second-order perturbation in Eq.~(\ref{eq:AndersonModel}). Although this state is energetically unstable, the two electrons can still coexist as long as Heisenberg's uncertainty relation about time and energy allows. As temperature decreases, these virtual processes become more frequent, which leads to forming a new resonant state that bridges the left and right leads through the QD, despite its being in a Coulomb blockade state [see the middle panel of Fig.~\ref{KondoSchematicFig}(b)]. This state is nothing more than the spin-singlet bound state (the Kondo state), which we discussed already. With the formation of the Kondo state, the conductance of the QD increases logarithmically with decreasing temperature [see the bottom panel of Fig.~\ref{KondoSchematicFig}(b)]. The Kondo effect in a QD was first realized experimentally in 1998~\cite{Goldhaber-GordonNature1998,CronenwettScience1998,SchmidPB1998}.  The conductance reaches $2e^2/h$ at sufficiently low temperature $T \ll T_\textrm{K}$, signaling the unitary limit~\cite{vanderWielScience2000}.

In the Kondo effect in bulk metals, the resistivity increases with decreasing temperature, as shown in Fig.~\ref{KondoSchematicFig}(a). The Kondo effect occurs due to the formation of Kondo states around magnetic impurities, which inhibits the transport of the conduction electrons. In contrast, in the Kondo effect in QDs, the transport through the QDs, which are the magnetic impurities themselves, is relevant. In this case, the conductance increases with the formation of the Kondo state, as shown in Fig.~\ref{KondoSchematicFig} (b).

\begin{figure}[!t]
\center 
\includegraphics[width=6.5cm]{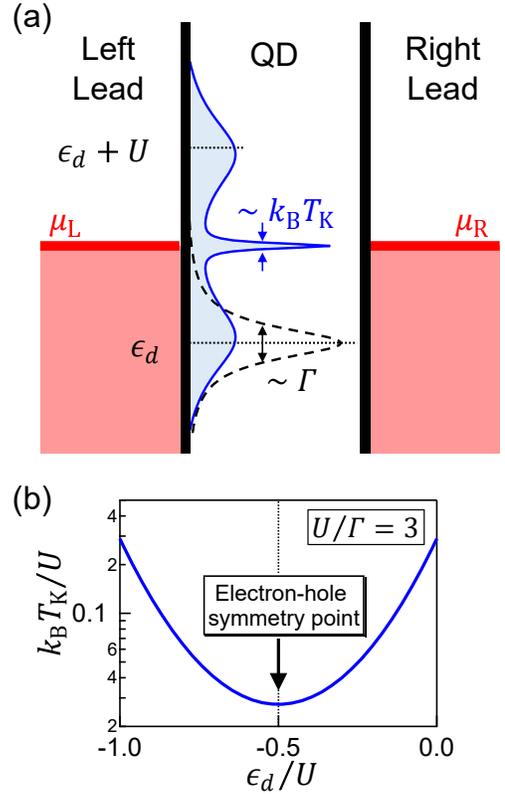} 
\caption{(Color online) (a) Schematic energy diagram of a QD in the Kondo regime. The discrete energy level $\epsilon_d$, which is localized inside the double potential barrier, has a finite resonance width $\Gamma$ due to the tunneling of the conduction electrons (dashed curve). If $\epsilon_d$ is lower than the chemical potential of the conduction band of the leads ($\mu_\textrm{L}$ and $\mu_\textrm{R}$), the resonance peak becomes sharper and shifts closer to the Fermi level due to the repulsion $U$ between the electrons occupying this level (solid curve). The figure shows a situation in which the electron-hole symmetry holds, $\epsilon_d/U = -0.5$. (b) Kondo temperature $k_\textrm{B}T_\textrm{K}/U$ for $U/\Gamma =3$ is shown as a function of $\epsilon_d/U$.} 
\label{KondoEnergyFig}
\end{figure}

Figure~\ref{KondoEnergyFig}(a) shows the energy diagram of a QD. A discrete energy level $\epsilon_d$, localized inside the double potential barrier, has a finite resonance width $\Gamma$ (dashed curve) due to the tunneling of the conduction electrons from the left and right leads. If the energy level is lower than the chemical potential of the leads ($\mu_\textrm{L}, \mu_\textrm{R}$), the Kondo effect occurs because of the repulsion $U$ between the electrons occupying this level. The resonance peak becomes sharper such that $k_\textrm{B}T_\textrm{K}\ll \Gamma$ and shifts very close to the Fermi level (solid curve). The resonance appears as a peak at zero bias in differential conductance. Figure~\ref{KondoEnergyFig}(a) shows a situation in which electron-hole symmetry holds, $\epsilon_d/U = -0.5$, while it is possible to control $\epsilon_d$ by controlling the gate voltage. In this case, $T_\textrm{K}$ varies as~\cite{HaldanePRL1978,Goldhaber-GordonPRL1998,vanderWielScience2000}
\begin{equation}
k_\textrm{B}T_\textrm{K}=\frac{\sqrt{\Gamma U}}{2} \exp \left[\frac{\pi \epsilon_d (\epsilon_d +U)}{\Gamma U} \right].
\label{TKexpression}
\end{equation}
Figure~\ref{KondoEnergyFig}(b) shows the Kondo temperature $k_\textrm{B}T_\textrm{K}/U$ when $U/\Gamma =3$ as a function of $\epsilon_d/U$.

\subsubsection{Non-equilibrium transport}
The realization of the Kondo effect in QDs has a significant meaning. Since the 1960s, the Kondo effect has been one of the central topics in the study of strongly correlated electron systems (e.g., heavy fermion systems and high-temperature superconductors). Experimentally, many studies have used macroscopic samples to measure the properties of the ensemble average of many spins. By using QDs, however, we can now control all the parameters related to the Kondo effect in a single site, such as the Kondo temperature, the number of electrons in the QD, the spin states, the orbital states, and the bias voltage to drive the QD to the nonequilibrium. For example, the sensitivity of the Kondo effect to the even/odd parity of the number of electrons in the QD was demonstrated by controlling the discrete level position~\cite{Goldhaber-GordonNature1998,CronenwettScience1998,SchmidPB1998} and thus $T_\textrm{K}$ [see Fig.~\ref{KondoEnergyFig}(b)]~\cite{Goldhaber-GordonPRL1998,vanderWielScience2000}. In addition, researchers have observed Zeeman splitting of the Kondo resonance by applying a magnetic field~\cite{Goldhaber-GordonNature1998,CronenwettScience1998,SchmidPB1998}. Thus, the Kondo effect in QDs is ideal for accurately verifying the theories on the strong electron correlation and quantum liquids.

The new opportunity to study non-equilibrium states is particularly remarkable. A quantitative understanding of the excited states of the local Fermi liquid becomes possible by precisely investigating the universal behavior of the Kondo effect, which appears in transport phenomena. For this purpose, it is necessary to consider the impact of the bias voltage on the quasi-particle lifetime due to electron correlation. Such theories include phenomenological Fermi liquid theory, microscopic Fermi liquid theory, and renormalized perturbation theory. The renormalized perturbation theory was introduced for the impurity-Anderson model by Hewson~\cite{HewsonPRL1993,HewsonJPCM2001} and extended to low-bias steady states~\cite{OguriPRB2001,OguriJPSJ2005}. Three important parameters in the impurity-Anderson model [Eq.~(\ref{eq:AndersonModel})] are  $\epsilon_d$, $\Gamma$, and $U$. Due to the Kondo effect, these quantities are renormalized to $\tilde{\epsilon_d}$, $\tilde{\Gamma}$, and $\tilde{U}$, respectively. For example, $\tilde{U}$ corresponds to a residual interaction between quasi-particles. The renormalized level width $\tilde{\Gamma}$ corresponds to the Kondo temperature $k_\textrm{B} T_\textrm{K} = \pi \tilde{\Gamma}/4$ in the Kondo regime~\cite{OguriJPSJ2005}. Following the spirit of the Fermi liquid theory, the dynamics of the Kondo effect at low energy is expected to be described by these parameters alone.

In the following, we consider the electron-hole symmetry case, namely $\epsilon_d/U = -0.5$ and $v_\textrm{L}=v_\textrm{R}$, in the Hamiltonian given by Eq.~(\ref{eq:AndersonModel}). In this case, the differential conductance of the QD can be calculated exactly up to the square of the bias voltage $V$, temperature $T_\textrm{e}$, and magnetic field $B$ so that
\begin{equation}
\begin{split}
\frac{d\langle I \rangle}{dV} &=\frac{2e^2}{h}\left[ 1 - c_V \left( \frac{eV}{\tilde{\Gamma}} \right)^2 
\right.  \\
&\left. - c_T \left( \frac{\pi k_\textrm{B} T_\textrm{e}}{\tilde{\Gamma}} \right)^2  - c_B \left( \frac{g\mu_\textrm{B} B}{\tilde{\Gamma}} \right)^2\right],
\end{split}
\label{kondononeq}
\end{equation}
where $g$ represents the $g$-factor and $\mu_\textrm{B}$ represents the Bohr magneton. Here,
\begin{equation}
\begin{split}
c_V &= \frac{1+5(R-1)^2}{4}, \\
c_T &= \frac{1+2(R-1)^2}{3}, \\
c_B &= \frac{R^2}{4}.
\end{split}
\end{equation}
The expression of $c_V$ was derived by Oguri~\cite{OguriPRB2001,OguriJPSJ2005}. That of $c_T$ originates from Refs.~[\onlinecite{YamadaPTP1975}] and~[\onlinecite{YoshimoriPTEP1976}]. For $c_B$, refer to Refs.~[\onlinecite{FerrierNatPhys2016}] and~[\onlinecite{MoraPRB2015}]. Equation~(\ref{kondononeq}) depicts how the Kondo resonance, which appears in the differential conductance near zero bias, behaves at finite bias, finite temperature, and finite field. $R$ is a quantity called the Wilson ratio and is represented by the magnetic susceptibility $\chi_s$ and the electronic specific heat coefficient $\gamma$ as follows:
\begin{equation}
R \equiv \frac{4\pi k_\textrm{B}^2}{3g^2\mu_\textrm{B}^2}\frac{\chi_s}{\gamma}   =1+\frac{\tilde{U}}{\pi \tilde{\Gamma}}.
\end{equation}
$R$ is a measure of the strength of the electron-electron interaction at the fixed point of the Fermi liquid. It increases monotonically from $R=1$ in the case of $U/\Gamma=0$, where there is no interaction, to $R=2$ in the Kondo limit $U/\Gamma \rightarrow \infty$. 

The conductance at $V=0$ and $B=0$ is the unitary-limit value $2e^2/h$ at $T_\textrm{e}=0$, as shown in Eq.~(\ref{kondononeq}). At very low bias, the Kondo resonance looks to electrons just a resonance state centered at the Fermi level with the width of $\sim k_\textrm{B}T_\textrm{K}$ [see Fig.~\ref{KondoEnergyFig}(a)]. As the resonance symmetrically couples to both leads ($v_\textrm{L}=v_\textrm{R}$), the transmission probability becomes 100\% at the level position.  The Kondo state becomes ``invisible'' to the conduction electrons. This fact is also concordant with the spirit of the Fermi liquid theory. By renormalizing the interaction, we can treat an interacting fermion system as if it were a free-particle system. 

On the other hand, at finite bias, finite temperature, and finite magnetic field, the conductance decreases from the unitary limit, since the current starts to be reflected by the Kondo state. Thus, this backscattered current contains information on the excited states corresponding to the backflow of the Fermi liquid~\cite{YamadaPTP1986}. A lot of theoretical work has been done on the universal behavior of the differential conductance  based on Eq.~(\ref{kondononeq})~\cite{RinconPRB2009,RouraBasPRB2010,SelaPRB2009,MoraPRB2009_2,MoraPRB2015,FilipponePRB2017,OguriPRB2018,OguriPRL2018,TerataniPRL2020}. Several experiments have also been reported~\cite{GrobisPRL2009,ScottPRB2009,DelattreNatPhys2009,YamauchiPRL2011,KretininPRB2011,KretininPRB2012,FerrierNatPhys2016,HataNatComm2021}. Such studies are an essential step towards understanding the Fermi liquid in the non-equilibrium regime.

\subsubsection{Backscattering and shot noise}
The realization of the Kondo effect in QDs in 1998~\cite{Goldhaber-GordonNature1998,CronenwettScience1998,SchmidPB1998} has also triggered a great deal of interest in shot noise. In quantum many-body systems, effective charge states are formed as a result of electron-electron correlations. As we discuss in this review, in the fractional quantum Hall effect, the fractional charge $e/3$ was observed~\cite{SaminadayarPRL1997,de-PicciottoNature1997} in 1997, and later $e/5$ and $e/4$ were also observed~\cite{ReznikovNature1999,DolevNature2008}. In a normal-metal-superconductor junction, the formation of the Cooper pair charge $2e$ was detected through shot noise~\cite{JehlNature2000,KozhevnikovJLTP2000,KozhevnikovPRL2000} in 2000. It is an essential and exciting fact that shot noise provides us direct and quantitative information about the non-equilibrium quantum state impossible to obtain with other experimental methods.

The expressions for the shot noise at sufficiently low temperature and low bias compared to the Kondo temperature were obtained by several groups~\cite{SelaPRL2006,GogolinPRL2006,VitushinskyPRL2008,MoraPRL2008,SelaPRB2009,FujiiJPSJ2010,SakanoPRB2011,SakanoPRB2011_2}. Here, before going to the shot noise, we discuss the current at $T_\textrm{e}=0$ and $B=0$ to give a physical picture of the backscattering. We assume the electron-hole symmetry case ($\epsilon_d/U = -0.5$ and $v_\textrm{L}=v_\textrm{R}$) again.  In this case, the current is given up to the cubic order of the bias voltage $V$ as follows,
\begin{equation}
\langle I\rangle = \frac{2e^2}{h}V-e P_{\textrm{b}0}-e P_{\textrm{b}1}-(2e) P_{\textrm{b}2}.
\label{noneqcurrent_scattering}
\end{equation}
$P_{\textrm{b}0}$, $P_{\textrm{b}1}$, and $P_{\textrm{b}2}$, all of which are proportional to $V^3$, represent the probabilities per unit of time of the different backscattering processes~\cite{SelaPRL2006,SakanoPRB2011_2}. While Eq.~(\ref{noneqcurrent_scattering}) is consistent with the conductance given by Eq.~(\ref{kondononeq}) at $T_\textrm{e}=0$ and $B=0$, it is easier to understand the detailed backscattering processes.

\begin{figure}[tbhp]
\center 
\includegraphics[width=8.5cm]{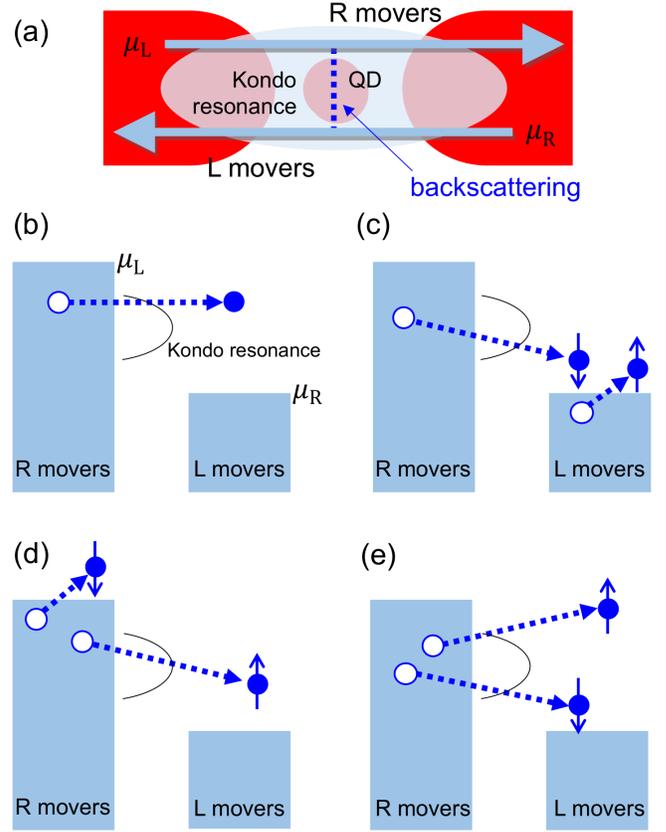} 
\caption{(Color online) (a) Electron transport through the Kondo resonance is schematically shown. Near the unitary limit, electrons incident from the left (right) are almost totally transmitted to the right (left), which is represented by the motion of the R (L) movers~\cite{SelaPRL2006}. The vertical dashed line to connect the R and L movers indicates the backscattering at finite bias. (b) Elastic scattering between the R and L movers by the Kondo resonance centered at the Fermi level is schematically depicted with the resonance peak superposed. One R mover tunnels into the L movers, creating one particle-hole pair. (c) In addition to one particle-hole pair creation, the second particle-hole pair appears in the L movers. (d) The second particle-hole pair appears in the R movers. (e) Both particle-hole pairs are created separately in the R and L movers, corresponding to the two-particle backscattering process.} 
\label{KondoScatteringSchematicFig}
\end{figure}

After Ref.~[\onlinecite{SelaPRL2006}], we describe the backscattering processes near the unitary limit by using the term of ``right movers (R movers)'' and ``left movers (L movers)'', as shown in Fig.~\ref{KondoScatteringSchematicFig}(a). The R (L) movers correspond to electron propagation from the left (right) lead to the right (left) lead.  The chemical potential of the R (L) movers is $\mu_\textrm{L}$ ($\mu_\textrm{R}$) [Readers may remind that the current operator is expressed in terms of the R movers ($a_{\textrm{L}, k}^\dagger a_{\textrm{L}, k'}$) and the L movers ($b_{\textrm{L}, k}^\dagger  b_{\textrm{L}, k'}$) in Eq.~(\ref{Eq_current_operator}). Also see Fig.~\ref{fig_setup}]. There is no scattering between the two movers at zero bias, as the transmission probability is 100\%. On the other hand, the conductance decreases at finite bias, which can be interpreted that some R (L) movers are backscattered into L (R) movers, as indicated by the vertical dashed line in Fig.~\ref{KondoScatteringSchematicFig}(a). Figures~\ref{KondoScatteringSchematicFig}(b), (c), (d), and (e) illustrate several backscattering processes between the two movers.

$P_{\textrm{b}0}$ in Eq.~(\ref{noneqcurrent_scattering}) expresses the probability of the elastic scattering by the quasi-particle level in the QD. As $V$ becomes finite, the Kondo resonance starts to reflect electrons because the electron energy shifts from the resonance peak. Figure~\ref{KondoScatteringSchematicFig}(b) shows that one R mover stochastically tunnels into the L movers, creating one (quasi-)particle-hole pair. This process decreases the current and generates the Poissonian shot noise corresponding to the backscattered current. As the resonance peak has a parabolic energy dependence, the scattering occurs with the probability that is proportional to $V^3$, which defines the $V$-dependence of $P_{\textrm{b}0}$. This process occurs even without the residual interaction ($\tilde{U}=0$).

In contrast, the probabilities $P_{\textrm{b}1}$ and $P_{\textrm{b}2}$ reflect the residual interaction. As shown in Figs.~\ref{KondoScatteringSchematicFig}(c), (d), and (e), two particle-hole pairs are excited by the second-order process of $\tilde{U}$. A hole and a particle in one pair always appear separately in each mover. For the other pair, there are two cases. 
\begin{enumerate}
 \item As shown in Figs.~\ref{KondoScatteringSchematicFig}(c) and (d), the second pair appears in one of the two movers with probability $P_{\textrm{b}1}$, and therefore it does not contribute to current. This process is the backscattering of a single charge of $e$ in the end. 
 \item As shown in Fig.~\ref{KondoScatteringSchematicFig}(e), the hole and the particle in the second pair appear separately in the R and L movers, respectively, as the first pair does. This process produces the backscattering of unit charge $2e$. It is peculiar to Kondo physics called ``two-particle backscattering'', which occurs with probability $P_{\textrm{b}2}$.
\end{enumerate}
The phase-space restrictions for particle-hole-pair creation require $P_{\textrm{b}1}$ and $P_{\textrm{b}2}$ to be proportional to $V^3$, as well as $P_{\textrm{b}0}$.
All these backscattering processes contribute to Eq.~(\ref{noneqcurrent_scattering}).

Accordingly, the shot noise $S$ is obtained. Since $P_{\textrm{b}0}$ and $P_{\textrm{b}1}$ are the backscattering of the unit charge $e$, and $P_{\textrm{b}2}$ is the process of $2e$, the following holds,
\begin{equation}
S = 2 \left[ e^2 P_{\textrm{b}0}+e^2 P_{\textrm{b}1}+(2e)^2 P_{\textrm{b}2}   \right].
\label{Kondo_shot_scattering}
\end{equation}
Because there is not only $e^2$ but also $(2e)^2$, the noise is enhanced compared to the case of a simple backscattering of $e$.

It is useful to think of a quantity defined by the ratio between the shot noise and the backscattered current 
\begin{equation}
    I_\textrm{b} = \frac{2e^2}{h}V-I
\label{eq:backscattered}
\end{equation}
rather than the ordinary Fano factor defined by Eq.~(\ref{Fano_classic}). Let's call this ratio the effective charge $e^*$. Eq.~(\ref{noneqcurrent_scattering}) and Eq.~(\ref{Kondo_shot_scattering}) yield
\begin{equation}
e^* \equiv \frac{S}{2\vert\langle I_\textrm{b}\rangle\vert}=\frac{e^2 P_{\textrm{b}0}+e^2 P_{\textrm{b}1}+(2e)^2 P_{\textrm{b}2}}{e P_{\textrm{b}0}+e P_{\textrm{b}1}+(2e) P_{\textrm{b}2}}.
\label{Kondo_shot_estar}
\end{equation}
In the limit of strong electron correlation $U/\Gamma\rightarrow \infty$, we can show that the probability that the unit charge $e$ is scattered ($\propto P_{\textrm{b}0}+P_{\textrm{b}1}$) and the probability that two particles $2e$ are scattered ($\propto P_{\textrm{b}2}$) are the same~\cite{SelaPRL2006,SakanoPRB2011_2}, resulting in
\begin{equation}
e^* \rightarrow \frac{5}{3}e.
\label{eq:5_3}
\end{equation}
This result can also be expressed using the Wilson ratio $R$~\cite{SelaPRB2009,FujiiJPSJ2010,SakanoPRB2011,SakanoPRB2011_2}.
\begin{equation}
\frac{e^*}{e}=\frac{1+9(R-1)^2}{1+5(R-1)^2} \rightarrow \frac{5}{3} \quad (R\rightarrow 2).
\label{eq_estar_wilson}
\end{equation}
Although we call $e^*$ effective charge in line with the literature, we should not confuse it with an exotic charge like that in the fractional quantum Hall effect. The present $e^*$ is the consequence of several scattering processes.

The above discussion illustrates that we can obtain the Wilson ratio $R$ from the current and the shot noise in the non-equilibrium state. In particular, the shot noise directly provides information on the two-particle backscattering due to the residual interaction, i.e., information corresponding to the ``internal structure'' of the Kondo state. As already mentioned, in the unitary limit, the transmission of the Kondo QD is 100\%: the conduction electrons cannot ``see'' the Kondo effect in the equilibrium. However, in the non-equilibrium state, the entity of the many-body interaction that produces the Kondo state reappears.

\subsubsection{Orbital degeneracy}
The discussion so far treats the most common Kondo effect, called the SU(2) Kondo effect, to which only the spin degree of freedom contributes. However, when there exist other degrees of freedom, such as the orbital one, a more exotic SU($n$) Kondo effect may occur~\cite{CoqblinPRB1969}. In the mesoscopic research field, such QDs have been realized and actively studied. The SU(4) Kondo effect in carbon nanotubes (CNTs) is a representative example~\cite{ChoiPRL2005,Jarillo-HerreroNature2005,DelattreNatPhys2009,LairdRMP2015}. The electrons in a CNT have two orbitals, one clockwise and the other counterclockwise with respect to the axis of the tube. They are doubly degenerate when the effects to lift the orbital degeneracy, such as an external magnetic field, are absent. Each of these two orbitals is also degenerate with respect to the spin, leading to the SU(4) Kondo effect due to the four-fold degenerate electron levels in total.

Shot noise has also been theoretically studied for the general SU($n$) Kondo effect. In the case of electron-hole symmetry, no magnetic field, and sufficiently low temperature, the effective charge is theoretically shown as follows~\cite{GogolinPRL2006,SelaPRB2009,FujiiJPSJ2010}
\begin{equation}
\frac{e^*}{e}=\frac{1+9(n-1)(R-1)^2}{1+5(n-1)(R-1)^2}.
\label{eq_estar_SUn}
\end{equation}
For example, $e^*/e=3/2$ is predicted for the SU(4) Kondo effect ($R\rightarrow 4/3$ for $U/\Gamma \rightarrow \infty$). In Sect.~\ref{subsub:su2-su4}, we discuss the experimental results to validate this formula in addressing the crossover between the SU(2) and SU(4) Kondo effects.

\subsubsection{Shot-noise experiments}
\label{subsub:kondonoise}
As mentioned earlier, many theoretical studies have been conducted on the shot noise in the Kondo state. In contrast, there have been only a limited number of experimental studies. In 2008, Zarchin \textit{et al.} reported the first shot-noise measurement in the Kondo state using a lateral QD fabricated in a two-dimensional electron system (2DES) in a GaAs/AlGaAs heterostructure~\cite{ZarchinPRB2008}. They argued that, although the unitary limit was not reached, $\frac{5}{3}e$ was measured as the effective charge, as predicted by the theory. Delattre \textit{et al.} observed the current noise due to the SU(4) Kondo effect in the $1/4$-filling region in a CNT QD~\cite{DelattreNatPhys2009}. They found that the noise is larger than the shot noise expected from a single-particle picture and explained this enhancement quantitatively by the slave boson method as being due to the Kondo effect. They measured in a bias region higher than $k_\textrm{B}T_\textrm{K}$ and did not address the low energy excited states inherent to the Fermi liquid~\cite{EggerNatPhys2009}.

One of the authors performed measurements using a lateral QD fabricated in a 2DES~\cite{YamauchiPRL2011} in 2011 and found that the noise increases with the development of the Kondo effect as lowering the temperature. This finding is qualitatively consistent with the increase in two-particle backscattering as described above. However, the effective charge obtained from the shot noise exceeded $5/3$, which is not compatible with the theory. Regarding this discrepancy, it was pointed that finite transport through other levels in the QD contributes to the shot noise~\cite{YamauchiPRL2011}. Because $U$, $\Gamma$, and the spacing between the discrete levels were on the same order in that experiment, transport via the other adjacent levels, which are irrelevant to the Kondo state but are strongly coupled with the leads, might be possible. Even if the transmission probabilities of those levels are small, such multi-level transport enhances the shot noise, as we have seen in Sect.~\ref{subsub:QD}.

\paragraph{Realization of the unitary limit Kondo effect} To overcome the above problem, one of the authors performed shot-noise measurements in the Kondo state using a CNT QD~\cite{FerrierNatPhys2016,FerrierPRL2017,FerrierJLTP2019,HataSpringer2019}. CNT QDs have the advantage of being much smaller in volume than QDs in a GaAs/AlGaAs 2DES described above, so compared to $k_\textrm{B}T_\textrm{K}$, the discrete level spacing is large enough to neglect the effects of adjacent discrete levels. Fortunately, in our experiments, we could observe the unitary limit in perfect accordance with theory since our QD was symmetric: the two barriers forming it are almost the same, i.e., the symmetric condition $v_\textrm{L}=v_\textrm{R}$ in Eq.~(\ref{eq:AndersonModel}) is satisfied. This situation allowed us to unambiguously test the theory for both the SU(2) and SU(4) Kondo effects. In the following, until the end of Sect.~\ref{Subsec:KondoNoise}, we discuss the results obtained with this device~\cite{FerrierNatPhys2016,FerrierPRL2017}.

\begin{figure}[tbhp]
\center 
\includegraphics[width=8.5cm]{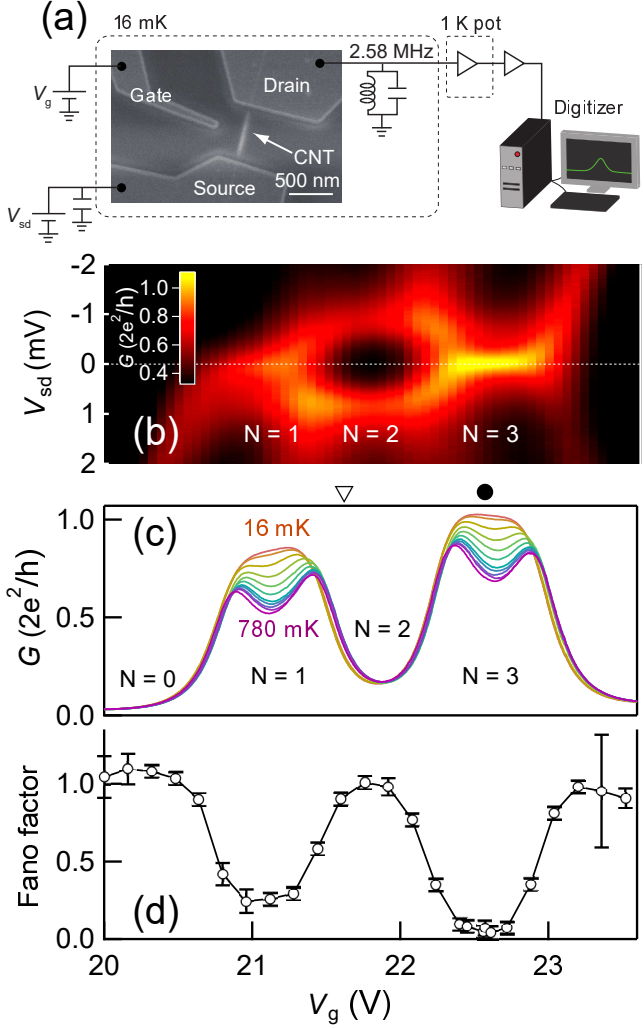} 
\caption{(Color online) (a) SEM image of a CNT connected to two metallic leads on a silicon wafer along with the noise measurement setup. (b) Intensity plot of the conductance as a function of the bias voltage $V (=V_\textrm{sd})$ and $V_\textrm{g}$. Kondo ridges appear as bright horizontal lines at $V=0$ for $N=1$ and $N=3$ electrons. (c) Conductance of the Kondo region at zero bias as a function of $V_\textrm{g}$ for several different temperatures between 16 and 780 mK. Two Kondo ridges at $N=1$ and $N=3$ are clearly visible. The differential conductance and the current noise obtained at the gate voltage positions indicated by $\bigtriangledown$ and $\bullet$ are shown in Figs.~\ref{FerrierNatPhysFig2noise}(a) and (b), respectively. (d) Fano factor extracted from the linear part of the current noise in the regime $eV\ll k_\textrm{B}T_\textrm{K}$ ($I\leq 5$ nA) using the definition $S = 2eF \vert I\vert$. Figures are reproduced from Ref.~[\onlinecite{FerrierNatPhys2016}]. {\copyright} (2015) Nature Publishing Group.} 
\label{FerrierNatPhysFig2}
\end{figure}

Figure~\ref{FerrierNatPhysFig2}(a) shows a scanning electron microscope (SEM) image of the CNT QD and a schematic diagram of the measurement setup. The device consists of a single CNT sandwiched between the two leads. The Kondo effect was obtained by controlling the gate voltage $V_\textrm{g}$ applied to the gate electrode close to the QD. Figure~\ref{FerrierNatPhysFig2}(b) shows an intensity plot of the differential conductance as a function of the bias voltage $V (=V_\textrm{sd})$ and $V_\textrm{g}$. This figure, the so-called stability diagram of QD, shows a Coulomb diamond in every fourfold degenerate shell with combined spin and orbit, characteristic of a CNT QD. $N (=0, 1, 2, 3)$ represents the number of electrons in the outer-most shell.

In Fig.~\ref{FerrierNatPhysFig2}(c), the horizontal and vertical axes of the graph represent $V_\textrm{g}$ and the equilibrium conductance of the QD, respectively. At 780 mK, four peaks appear in the conductance as $V_\textrm{g}$ is increased. This behavior indicates that $N$ changes one by one such that $N=0 \rightarrow 1\rightarrow 2\rightarrow 3$. Now, in the $N=1$ and $N=3$ regions, lowering the temperature from 780 to 16 mK leads to the conductance increase, signaling the Kondo effect. On the other hand, in the $N=2$ region, which is in between, the conductance remains small and almost temperature-independent, meaning that the QD is in an ordinary Coulomb blockade. The Kondo effect is usually expected when $N$ is odd, which is consistent with the observation. It is also important to note that for $N=3$, the conductance at 16 mK is almost the quantized conductance $2e^2/h$. This value indicates the unitary limit. The detailed analysis of the temperature dependence of the conductance reveals that $U=6\pm 0.5$~meV, $\Gamma = 1.8 \pm 0.2$~meV, and $T_\textrm{K} = 1.6 \pm 0.05$~K at the electron-hole symmetry point for the case of $N=3$ [indicated by $\bullet$ in Fig.~\ref{FerrierNatPhysFig2}(c)]~\cite{FerrierNatPhys2016}.

The unitary limit is expected to occur in the $N=1$ region as well. However, the conductance does not rise above $0.85 \times 2e^2/h$ even at the lowest temperature. In this case, the coupling strength between the left lead and QD is different from that between the right lead and QD, namely asymmetric lead-QD coupling ($v_\textrm{L} \neq v_\textrm{R}$). The variation of $V_\textrm{g}$ to change $N$ may affect the spatial distribution of the wave function, and consequently, the coupling strength between the QD and the leads would change. The unitary limit has occurred for $N=3$, which, with hindsight, implies that the two couplings are almost the same.

\begin{figure}[tbhp]
\center 
\includegraphics[width=8cm]{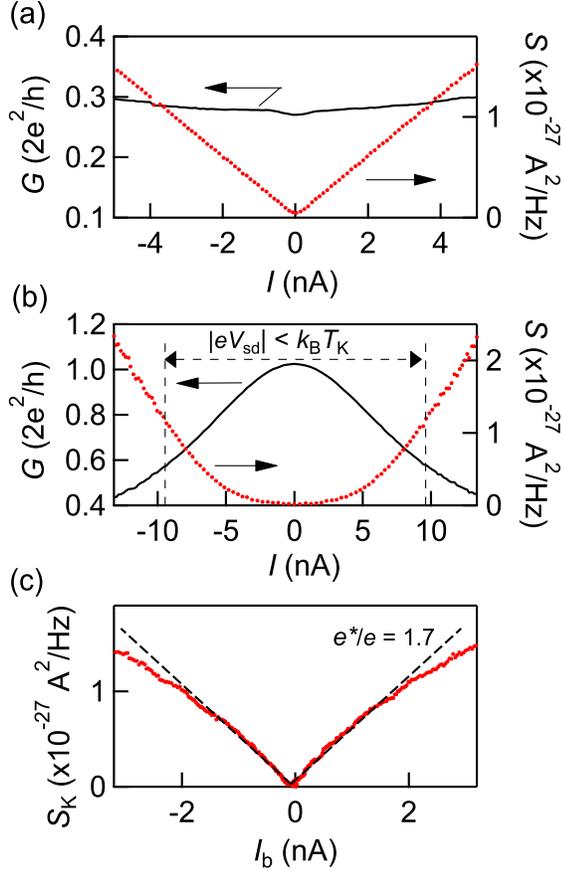} 
\caption{(Color online) (a) Conductance (black line, left axis) and noise (red dots, right axis) as a function of $\langle I \rangle$ for the Coulomb blockade region ($N=2$). $V_\textrm{g}$ is set to the location indicated by $\bigtriangledown$ in Fig.~\ref{FerrierNatPhysFig2}(c). The noise is linear to $\vert \langle I \rangle\vert$, with a slope around $2e$. (b) Conductance (black line, left axis) and noise (red dots, right axis) as a function of $\langle I \rangle$ on the Kondo ridge ($N=3$). $V_\textrm{g}$ is set to the location indicated by $\bullet$ in Fig.~\ref{FerrierNatPhysFig2}(c). The slope of $S$ around $I=0$ is almost zero owing to the perfect transmission ($F=0.06$). (c) The Kondo-effect-induced shot noise $S_\textrm{K}$ as a function of the backscattered current $I_\textrm{b}$ is plotted for the data shown in (b). Around $I_\textrm{b}=0$, $S_\textrm{K} \propto \vert I_\textrm{b} \vert$ holds. Figures are reproduced from Ref.~[\onlinecite{FerrierNatPhys2016}]. {\copyright} (2015) Nature Publishing Group.} 
\label{FerrierNatPhysFig2noise}
\end{figure}

\paragraph{Shot noise in Coulomb blockade regime} We discuss the results of the shot-noise measurements. First, we examine the simple case where the Kondo effect does not occur. The gate voltage is set to the location indicated by $\bigtriangledown$ in Fig.~\ref{FerrierNatPhysFig2}(c), where conventional Coulomb blockade ($N=2$) occurs. In Fig.~\ref{FerrierNatPhysFig2noise}(a), the horizontal axis shows the current $\langle I\rangle$ flowing through the QD, the left vertical axis shows the differential conductance $G$, and the right vertical axis shows the current noise $S$. Regarding $S$, we represent the value after subtracting the contribution of the thermal noise as usual (see Sect.~\ref{subsub:generalshot}). $G$ is almost independent of $\langle I\rangle$, reflecting that the QD is in Coulomb blockade. Clearly, the current noise is proportional to the absolute value of the current $\langle I\rangle$, that is, $S\propto \vert \langle I\rangle\vert$. 

The Fano factor $F=S/2e\vert \langle I\rangle\vert$ obtained from the shot-noise measurements is shown in Fig.~\ref{FerrierNatPhysFig2}(d) as a function of $V_\textrm{g}$. In the Coulomb blockade regime ($N=0$ and $2$), $F\sim 1$: the shot noise is in the Poisson limit, as expected for a common tunneling barrier. The overall behavior of the Fano factor in Fig.~\ref{FerrierNatPhysFig2}(d) is just an upside-down version of the behavior of the conductance at the lowest temperature, 16~mK, shown in Fig.~\ref{FerrierNatPhysFig2}(c). This is consistent with the theoretically expected behavior of $F=1-{\cal T}$.

In Sect.~\ref{subsub:QD}, we introduced several experimental results regarding the Fano factor in QDs. In particular, we raised the possibility of electron bunching leading to $F>1$. The present results show that for a QD in the resonant tunneling effect regime, the noise behaves perfectly following the theory described in Sect.~\ref{subsec:noise_in_quantum_transport}, even if the QD has a fixed number of electrons because of the charging effect. This agreement also reflects coherent electron transport through the QD.

\paragraph{Shot noise in the Kondo state} Figure~\ref{FerrierNatPhysFig2noise}(b) shows the conductance and shot noise obtained at the gate voltage marked with $\bullet$ shown in Fig.~\ref{FerrierNatPhysFig2}(c). This gate voltage corresponds to the vicinity of the electron-hole symmetry point of the Kondo state ($N=3$). When the current is around zero ($\vert I \vert <2$ nA), the conductance is $2e^2/h$, reflecting that the Kondo state is perfect. In this case, the current noise $S$ is almost zero, and thus the Fano factor is close to zero [see Fig.~\ref{FerrierNatPhysFig2} (d)]. The absence of shot noise means that electrons can pass through the QD as if there were no scatterer at all. Consequently, this situation is similar to a QPC with a quantized conductance of $2e^2/h$, where the shot noise disappears. When the current is close to zero, the Kondo state is a ``non-viscous'' liquid that does not reflect any electrons.

As shown in Fig.~\ref{FerrierNatPhysFig2noise}(b), the conductance decreases rapidly as the current $\vert I \vert$ is increased. We consider the high bias region within the range satisfying $k_\textrm{B} T_\textrm{K}/2<\vert eV_{sd} \vert < k_\textrm{B} T_\textrm{K}$. In Fig.~\ref{FerrierNatPhysFig2noise}(b), the conductance at $ \vert I\vert=10$~nA is almost half ($0.5 \times 2e^2/h$) of that in the equilibrium. As electrons are constantly injected, the Kondo state gradually breaks down, making it harder for electrons to flow. Conversely, the backscattering starts. It is also evident that the noise increases as this backscattered current increases.

For more quantitative analysis, we utilize the backscattered current $I_\textrm{b} = \frac{2e^2}{h}V-I$ as we did in Eq.~(\ref{eq:backscattered}). Additionally, we define the increase of the shot noise as 
\begin{equation}
S_\textrm{K} = S -2eF \vert \langle I\rangle\vert.
\label{eq_def_SK}
\end{equation}
The second term $-2eF \vert \langle I\rangle\vert$ is intended to subtract the shot-noise contribution that does not originate from the Kondo effect, but instead is caused by slight lead-dot asymmetry. In this way, we extract the shot noise purely due to the Kondo effect discussed in Eq.~(\ref{Kondo_shot_scattering}). 

\begin{figure}[!b]
\center 
\includegraphics[width=7.5cm]{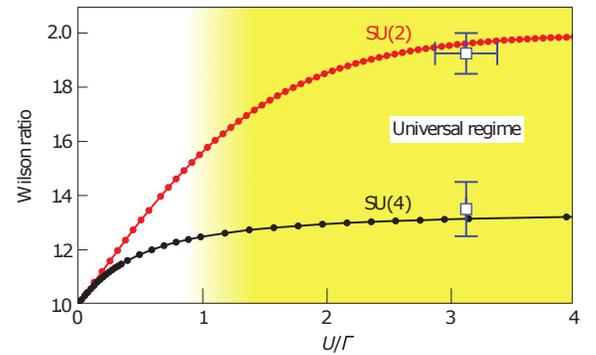}
\caption{(Color online) Theoretical Wilson ratio for the SU(2) and SU(4) Kondo effects is shown by solid curves as a function of $U/\Gamma$~\cite{SakanoPRB2011}. The experimental results plotted with the mark $\square$ are almost on the theoretical curves. This consistency indicates that our experimental result agrees with the Fermi liquid theory. The yellow part of the graph represents the region of universality: all the properties depend on a single parameter $T_\textrm{K}$~\cite{OguriJPSJ2005}. Reprinted from Ref.~[\onlinecite{FerrierNatPhys2016}]. {\copyright} (2015) Nature Publishing Group.} 
\label{Fig_FerrierNatPhys2016Fig3}
\end{figure}

In Fig.~\ref{FerrierNatPhysFig2noise}(c), the noise $S_\textrm{K}$ is plotted as a function of the backscattered current $\langle I_\textrm{b} \rangle$. We obtain $S_\textrm{K}/2e \vert \langle I_\textrm{b} \rangle\vert =1.7\pm 0.1$, thus $e^*/e=1.7\pm 0.1$. This value is consistent with $5/3$ [Eq.~(\ref{eq:5_3})] and validates the two-particle backscattering process in the Kondo regime. In addition, we derive $R=1.95 \pm 0.1$ from Eq.~(\ref{eq_estar_wilson}), which is consistent with the value expected from the experiment ($U=6 \pm 0.5$~meV and $\Gamma = 1.8 \pm 0.2$~meV). Figure~\ref{Fig_FerrierNatPhys2016Fig3} presents the behavior of the Wilson ratio expected from the theory as a function of $U/\Gamma$. The experimental result corresponds to the mark $\square$ [SU(2)], which agrees with the theoretical prediction. Since $R=2$ is the limit of strong correlations in the Kondo effect, the result here clearly verifies that the present QD is genuinely close to a quantum liquid in the strong correlation limit.

Note that in the $N=1$ region, the unitary limit is not reached due to asymmetric lead-dot coupling. We obtained $e^*/e=1.2\pm 0.08$ from the shot noise in this region. The lead-dot coupling asymmetry $\delta$ is defined by $G(V=0)= (1-\delta )2e^2/h$. The result of the conductance measurement is $G(V=0)=0.85(2e^2/h)$, yielding $\delta=0.15$. The experimental result $1.2\pm 0.08$ is consistent with the theoretical prediction $e^*/e=5/3-(8/3)\delta =1.26$~\cite{MoraPRB2009_2}.

\paragraph{Intuitive picture of the enhanced noise} Why is the effective charge larger than 1 in the limit of strong electron correlation? It is essentially impossible to describe the quantum many-body effect intuitively with a classical picture. However, we believe that such a description can sometimes provide helpful physical intuition, and we try it here. The experimental achievement is an observation of the two-particle backscattering process, which we explained in Fig.~\ref{KondoScatteringSchematicFig}. Consider this phenomenon very intuitively from the standpoint of the electrons passing through the QD. In the non-equilibrium state, electrons constantly injected from one lead escape into the other, feeling strong interaction as they pass through the quantum liquid. As seen from the decrease in the conductance, they collide with the quantum liquid to cause backscattering into the original lead. Thus, to electrons, the quantum liquid is no longer felt as a ``non-viscous'' state, but as a ``viscous'' entity. This viscosity due to the many-body interaction results in the ``two-electron'' bunching~\cite{ZarchinPRB2008}, like a ``water splash'', and increases the current noise [see Fig.~\ref{KondoScatteringSchematicFig}(e)]. This phenomenon is unique to the strongly correlated quantum liquid in the non-equilibrium: the Kondo effect creates the quantum liquid, but this fact is somehow hidden in the equilibrium, as the Fermi liquid theory tells us. Only in the non-equilibrium state does the true nature to emerge the quantum liquid manifest itself.

\subsubsection{SU(2)-SU(4) crossover}
\label{subsub:su2-su4}
The significance of studying the Kondo effect in QDs lies not only in exploring its non-equilibrium properties but also in the potential to achieve more exotic Kondo effects using other degrees of freedom. As mentioned earlier, in CNT QDs, the SU(4) Kondo effect has been realized using orbital degrees of freedom~\cite{FerrierNatPhys2016,ChoiPRL2005,Jarillo-HerreroNature2005,DelattreNatPhys2009,LairdRMP2015}. One of the authors was able to experimentally verify $e^*/e=3/2$ predicted by Eq.~(\ref{eq_estar_SUn}) with an accuracy of $\pm 0.1$ by realizing the SU(4) Kondo effect close to the unitary limit~\cite{FerrierNatPhys2016}. The obtained Wilson ratio is consistent with the theory as plotted by the $\square$ mark in Fig.~\ref{Fig_FerrierNatPhys2016Fig3}.

Consider what happens if a sufficiently strong magnetic field is applied to lift only the spin degeneracy in the SU(4) Kondo state realized at zero magnetic field. Naively, it would lead to a novel SU(2) Kondo state, which would involve only orbital degeneracy. In actuality, the situation is not so simple because the orbital state of the CNT is also affected by the magnetic field. Nevertheless, the crossover from SU(4) to SU(2) occurs by applying the magnetic field to the CNT at a specific angle~\cite{TerataniJPSJ2016,FerrierPRL2017,TerataniPRB2020}.

\begin{figure}[!t]
\center 
\includegraphics[width=8.5cm]{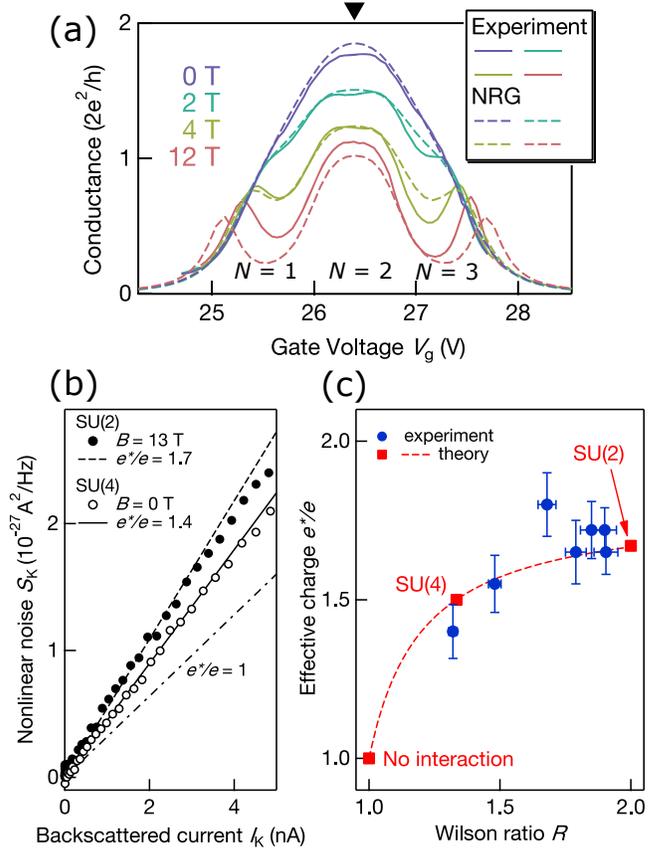} 
\caption{(Color online) (a) Comparison of the zero-bias conductance between the experiment (solid curves) and the numerical renormalization group (NRG) calculations (dashed curves) for several magnetic fields. $\blacktriangledown$ shows the electron-hole symmetry point at $N=2$. (b) Kondo-effect-induced noise $S_\textrm{K}$ as a function of the backscattered current $I_K$ [$=I_\textrm{b}$ defined by Eq.~(\ref{eq:backscattered})] at $B = 0$~T [SU(4) state] and $B = 13$~T [SU(2) state]. (c) The filled circles show the effective charge $e^*/e$ as a function of $R$, which quantifies the strength of quantum fluctuations. The three square symbols represent the theoretical prediction for SU(4), SU(2), and noninteracting particles. The dashed curve is the extended theoretical prediction based on Eq.~(\ref{eq_estar_SUn}).  Figures are reproduced from Ref.~[\onlinecite{FerrierPRL2017}]. {\copyright} (2017) American Physical Society.} 
\label{Fig_FerrierPRL2017}
\end{figure}

Figure~\ref{Fig_FerrierPRL2017}(a) shows the conductance when the gate voltage is varied to change $N$ in the QD. Around 0 T, the conductance at $N=2$ reaches $1.85(2e^2/h)$, which is close to $2(2e^2/h)$, the conductance of the unitary limit of the SU(4) Kondo effect. The dashed curves in Fig.~\ref{Fig_FerrierPRL2017}(a) show the results of the numerical renormalization group (NRG) calculations. From these theoretical calculations and conductance measurements, we confirm that the SU(4) Kondo state appears in $N=1, 2$, and $3$.

By applying a magnetic field at a specific angle to the CNT QD in this situation, we found that the SU(4) Kondo effect can gradually transform into a SU(2) Kondo effect~\cite{FerrierPRL2017}. Experimental results and theoretical simulations of the conductance in several magnetic fields are shown in Fig.~\ref{Fig_FerrierPRL2017}(a). The results reveal that the Kondo effect at $N=1$ and $N=3$ disappears but the Kondo effect at $N=2$ remains. The conductance at a high magnetic field of 12~T is $2e^2/h$ expected in the unitary limit of the SU(2) Kondo effect. Thus, in 12~T, a perfect SU(2) Kondo state is realized. This SU(2) Kondo state is a special one with two electrons arising from the hybridization of orbital and spin degrees of freedom~\cite{TerataniJPSJ2016,TerataniPRB2020}.

The shot noise is measured at the gate voltage where the electron-hole symmetry holds [indicated by $\blacktriangledown$ in Fig.~\ref{Fig_FerrierPRL2017}(a)], as shown in Fig.~\ref{Fig_FerrierPRL2017}(b). At zero magnetic field, the SU(4) Kondo state exists. The noise purely associated with the Kondo effect, $S_\textrm{K}$ defined by Eq.~(\ref{eq_def_SK}), is plotted as a function of the backscattered current $I_\textrm{b}$ [denoted as $I_K$ in Fig.~\ref{Fig_FerrierPRL2017}(b)]. There is a clear linear relationship from which $e^*/e=1.4 \pm 0.1$ is obtained: this value is consistent with $3/2$ expected for the SU(4) Kondo effect. The relationship between $S_\textrm{K}$ and $I_\textrm{b}$ in a strong magnetic field (13~T) is also shown in the same figure. This yields $e^*/e=1.7 \pm 0.1$ as expected for the SU(2) Kondo effect, which guarantees that the Kondo effect in high magnetic fields is indeed the SU(2) Kondo effect. Importantly, as mentioned already, this SU(2) Kondo effect is no longer the conventional one. Nevertheless, the result is consistent with the theoretically expected value of $5/3$ exemplifies the universality of Kondo physics.

This result means that we can control the symmetry of the Kondo effect from SU(4) to SU(2) by varying the magnetic field. In Fig.~\ref{Fig_FerrierPRL2017}(c), the effective charge $e^*/e$ obtained from the shot noise is plotted as a function of $R$ derived from the NRG simulation of the conductance shown in Fig.~\ref{Fig_FerrierPRL2017}(b). In a free electron system, $R=1$ and $e^*/e=1$. As $R$ increases, many-body correlations develop, two-particle backscattering from the QD occurs, and $e^*/e$ grows. The limit of the effective charge is $e^*/e=5/3$ at $R=2$. By changing the magnetic field, we can see how the Kondo effect evolves continuously from SU(4) to SU(2).

According to the theory of the SU($n$) Kondo effect~\cite{SakanoPRB2011}, in the Kondo limit $U \rightarrow \infty$, the Wilson ratio can be expressed as $R=1+1/(n-1)$ and the effective charge as $e^*/e=(n+8)/(n+4)$. Therefore, from Eq.~(\ref{eq_estar_SUn}), $e^*/e =[1+9(R-1)]/[1+5(R-1)]$ is expected. This is plotted in Fig.~\ref{Fig_FerrierPRL2017}(c) by the dashed curve. It explains the experimental results well, which has experimentally established a convincing link between the nonlinear noise and Wilson ratio.

\subsubsection{Short summary}
Throughout this Sect.~\ref{Subsec:KondoNoise}, we have discussed the shot-noise study in a QD in the Kondo regime. The Kondo effect creates a strongly correlated quantum liquid, but this is unseen in the equilibrium state as the transmission probability is unity. However, by injecting electrons to the QD to drive the system into the non-equilibrium state, the nontrivial behavior of the shot noise is revealed, and the true nature of the quantum liquid with residual interaction emerges via two-particle backscattering. The shot-noise measurement was also successfully demonstrated to explore the symmetry crossover of the Kondo effect.

\subsection{Noise in quantum Hall systems}
\label{subsec:qhe}
The detection of fractional quasiparticles in fractional quantum Hall (FQH) systems is one of the most epoch-making experimental accomplishments in mesoscopic physics~\cite{LaughlinPRL1983,SaminadayarPRL1997,de-PicciottoNature1997}. Current-noise measurements proved the effective charge $e/3$ of tunneling quasiparticles through the Landau-level filling factor $\nu=1/3$ and $2/3$ FQH systems. After the discovery of $e/3$ quasiparticles, current-noise measurements detected various fractional quasiparticles in other FQH systems~\cite{ReznikovNature1999,ChungPRL2003,DolevNature2008}. Furthermore, they have also revealed various phenomena peculiar to quantum Hall (QH) systems, such as anyonic quantum statistics of $e/3$ quasiparticles~\cite{BartolomeiScience2020}, Josephson relations for fractional charges~\cite{KapferScience2019,BisogninNatCom2019}, heat transport along QH edge channels~\cite{BidNature2010,DolevPRL2011,GrossPRL2012,InoueNatCommun2014,SaboNatPhys2017,BanerjeeNature2017,BanerjeeNature2018,CohenNatCommun2019,JezouinScience2013,SivreNatPhys2017,SivreNatCommun2019}, and the Tomonaga-Luttinger (TL) liquid nature of QH edge channels~\cite{InouePRL2014}. These observations unambiguously indicate the excellence of current-noise measurements for investigating not only QH systems but also topological quantum many-body systems in the near future. This section summarizes the current-noise measurements performed on QH systems after the early fractional-charge-detection experiment in 1997~\cite{SaminadayarPRL1997,de-PicciottoNature1997}. In the following, we overview the basics of current-noise measurements on a QH device and then review recent topics.

\subsubsection{Current-noise measurements based on chiral edge transport}
\label{sec:current_noise_chiral_edge}
When a 2DES is subjected to a strong perpendicular magnetic field, the Hall conductance of the system takes an integer or a rational fractional value in a unit of $e^2/h$. The former is the integer quantum Hall (IQH) effect, caused by the Landau-level formation and the Anderson localization~\cite{vonKlitzingPRL1980}. The latter is the FQH effect, resulting from the energy gap opening due to the many-body Coulomb interaction. Both of these effects reflect the incompressibility of the bulk 2DES and the Landauer-B\"uttiker's edge transport picture~\cite{DattaETMS}.

\begin{figure}[bt]
\center 
\includegraphics[width=8.5cm]{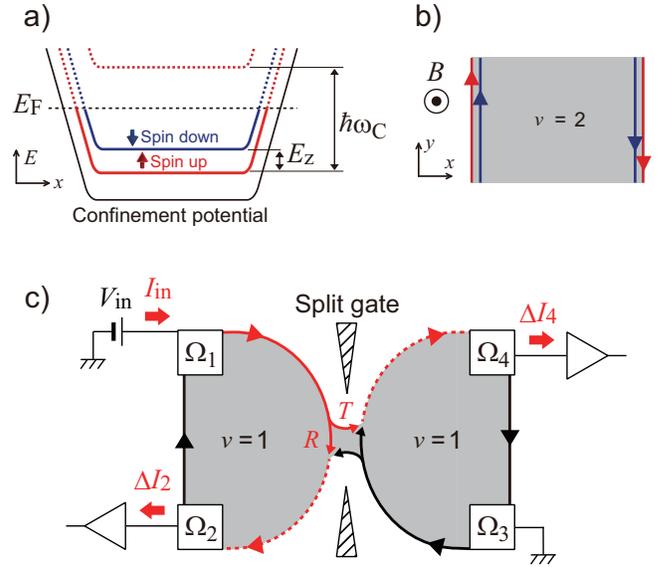}
\caption{(Color online) (a) Schematic of Landau levels in the $\nu=2$ IQH system. These levels are lifted by a confinement potential and cross the Fermi energy at the sample edge. (b) Chiral edge channels in the $\nu=2$ IQH system. The red and blue arrows indicate the spin-up and spin-down channels, respectively. (c) Schematic of a shot-noise measurement in the $\nu=1$ IQH systems. Electronic current fed from $\Omega_1$ is partitioned and generates shot noise at a narrow constriction formed by applying a split-gate voltage. The current noise in $\Omega_2$ and $\Omega_4$ is measured to evaluate the shot noise.}
\label{fig5_x1}
\end{figure}

QH edge channels are unidirectional one-dimensional (1D) electronic states arising at the edge of QH regions. Figure ~\ref{fig5_x1}(a) shows a schematic of the $\nu=2$ IQH state in a 2DES confined by electrostatic potential. The lowest Landau levels of spin-up and spin-down electrons are filled in the bulk region, while the confinement potential lifts them at the sample edges to cross the Fermi energy, forming conductive edge channels. The red and blue arrows in Fig.~\ref{fig5_x1}(b) are schematics of the spin-up and spin-down edge channels, respectively. Because electrons coherently flow along an edge channel without backscattering, the channel is sometimes regarded as an electronic analog of an optical laser path; one can construct Fabry-P\'erot~\cite{ChamonPRB1997,CaminoPRL2005} or Mach-Zehnder~\cite{JiNature2003} interferometers using the edge channels. Thus, an IQH edge channel is a promising platform for fermion-quantum-optics experiments (see Sect.~\ref{sec:fermion_optics})~\cite{GrenierModPhsLett,BocquillonAnnPhys,RousselPhysStatSolidiB,GlattliPhysStatSolidiB}.

A QPC fabricated in a QH system works as a beam splitter for an edge channel. Let us consider a QPC formed in the $\nu=1$ IQH system, as schematically shown in Fig.~\ref{fig5_x1}(c). A source-drain bias $V_{\rm{in}}$ lifts the Fermi energy of the edge channel stemming from the ohmic contact $\Omega_{\rm{1}}$ to $E_{\rm{F}}=eV_{\rm{in}}$. On the other hand, $\Omega_{\rm{3}}$ is connected to the ground so that the Fermi energy of the corresponding channel is $E_{\rm{F}}=0$. These two channels approach each other at the QPC to cause stochastic electron tunneling between them, generating shot noise. The outputs flow along the transmitted and reflected channels to reach the contacts $\Omega_{\rm{2}}$ and $\Omega_{\rm{4}}$, respectively, to raise current noise in these contacts. Importantly, in this setup, the current noise reflects only the tunneling process occurring between the channels from $\Omega_{\rm{1}}$ and $\Omega_{\rm{3}}$. The channels from $\Omega_{\rm{2}}$ and $\Omega_{\rm{4}}$ do not influence the measurement since they are well separated from the other channels by wide incompressible bulk regions.

\begin{figure*}[t]
\center 
\includegraphics[width=17cm]{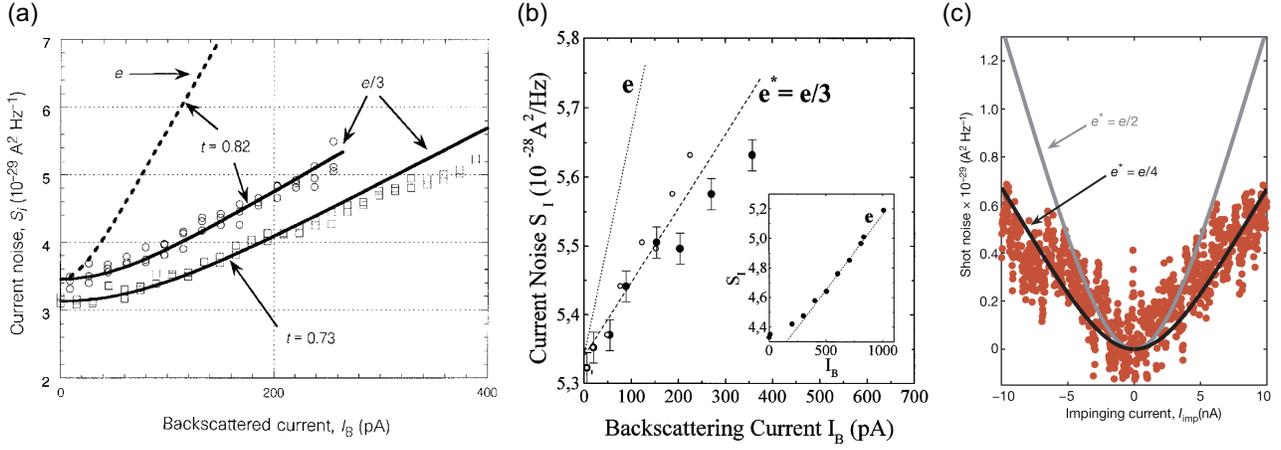}
\caption{(Color online) (a) Shot-noise data for the $e/3$ quasiparticle tunneling in the $\nu=1/3$ FQH state. (b) Shot-noise data in the $\nu=2/3$ state. (c) Shot-noise data of the $e/4$ quasiparticle tunneling through the $\nu=5/2$ state. Panels (a) and (c) are reprinted with permission from Ref.~[\onlinecite{de-PicciottoNature1997}] ({\copyright} 1997 Springer Nature) and Ref.~[\onlinecite{DolevNature2008}] ({\copyright} 2008 Springer Nature), respectively. Panel (b) is reprinted with permission from Ref. ~[\onlinecite{SaminadayarPRL1997}] ({\copyright} 1997 American Physical Society).}
\label{fig5_x2}
\end{figure*}

Within the single-particle picture, current-noise auto-correlation PSD in $\Omega_2$ is described as [see Eq.~(\ref{auto_Sbb})]
\begin{equation}
\begin{split}
S_{22}=2e\langle I_{\rm{in}}\rangle \times {\cal{T}}(1-{\cal{T}}),
\label{qhe_s22}
\end{split}
\end{equation}
at zero temperature. Similarly, the cross-correlation PSD between $\Omega_2$ and $\Omega_4$ is given by [see Eq.~(\ref{crosscorr_Sbc})]
\begin{equation}
\begin{split}
S_{24}=-2e\langle I_{\rm{in}}\rangle\times {\cal{T}}(1-{\cal{T}}).
\label{qhe_s24}
\end{split}
\end{equation}
Here, ${\cal{T}}$ is the transmission probability through the QPC, and $\langle I_{\rm{in}}\rangle = GV_{\rm{in}}$ is the impinging current fed from $\Omega_{\rm{1}}$. The factor ${\cal{T}}(1-{\cal{T}})$ reflects the partitioning process at the QPC. These equations can be modified using the tunneling current $\langle I_{\rm{T}}\rangle  = {\cal{T}} \times \langle I_{\rm{in}}\rangle $ through the QPC as 
\begin{equation}
\begin{split}
S_{22}=2e\langle I_{\rm{T}}\rangle\times \frac{{\cal{T}}(1-{\cal{T}})}{{\cal{T}}}=2e\langle I_{\rm{T}}\rangle(1-{\cal{T}}),
\label{qhe_s22_2}
\end{split}
\end{equation}
\begin{equation}
\begin{split}
S_{24}=-2e\langle I_{\rm{T}}\rangle\times \frac{{\cal{T}}(1-{\cal{T}})}{{\cal{T}}}=-2e\langle I_{\rm{T}}\rangle(1-{\cal{T}}).
\label{qhe_s24_2}
\end{split}
\end{equation}
Equations (\ref{qhe_s22_2}) and (\ref{qhe_s24_2}) are derived from Eqs.~(\ref{shot_47}) and (\ref{FanoTheory}) in Sect.~\ref{sec:current_noise} with the Fano factor $F=1-{\cal{T}}$. The negative sign of Eq.~(\ref{qhe_s24_2}) reflects the negative correlation between the two outputs due to the binomial partitioning at the QPC [see Fig.~\ref{fig3_5}(c)].

Thus, one can perform shot-noise measurements by focusing on scattering processes between selected channels. This is true not only in the $\nu=1$ IQH system but also in other QH systems at different filling factors. When a QH system has more than two channels at an edge, e.g., as shown in Fig.~\ref{fig5_x1}(b) for the $\nu=2$ case, the complexity of the system increases due to the increase in degrees of freedom and inter-channel Coulomb interaction and tunneling. However, even in such multiple-channel cases, one can measure the shot-noise generation between selected edge channels.

\subsubsection{Fractional charge of tunneling quasiparticles}
\label{sec:fractional_charge_tunneling}
The presence of fractionally charged quasiparticles was one of the most important predictions in early theories of the FQH effect~\cite{LaughlinPRL1983}. Whereas the charging energy of an antidot measured in an FQH system suggested the presence of fractional quasiparticles~\cite{GoldmanScience1995}, evidence was obtained by shot-noise measurements in the $\nu=1/3$ and $\nu=2/3$ systems in 1997~\cite{de-PicciottoNature1997,SaminadayarPRL1997}. When quasiparticles tunnel through the FQH systems stochastically, namely without correlation, zero-temperature shot noise is described as 
\begin{equation}
\begin{split}
S=2e^*\langle I_{\rm{B}}\rangle.
\label{qhe_s1/3}
\end{split}
\end{equation}
Here, $\langle I_{\rm{B}}\rangle$ is the backscattered current, and $e^*$ is the effective charge of tunneling quasiparticles. Equation (\ref{qhe_s1/3}) is modified at finite temperatures, $T_e$, due to the crossover between the thermal noise and the shot noise as [see Eq.~(\ref{shot_calib_finite})]~\cite{MartinPRB1992,MartinBook}
\begin{equation}
\begin{split}
S=2e^*\langle I_{\rm{B}}\rangle\times \left[\coth\left(\frac{e^*V}{2k_{\rm{B}}T_{\rm{e}}}\right)-\frac{2k_{\rm{B}}T_{\rm{e}}}{e^*V}\right].
\label{qhe_s1/3_2}
\end{split}
\end{equation}

Figures~\ref{fig5_x2}(a) and~\ref{fig5_x2}(b) compare the experimental shot-noise data at $\nu=1/3$ and $\nu=2/3$, respectively, with the theoretical one simulated using Eq.~(\ref{qhe_s1/3_2})~\cite{de-PicciottoNature1997,SaminadayarPRL1997}. When the backscattered current $\langle I_{\rm{B}}\rangle$ increases with the source-drain bias, the shot noise also increases, agreeing with the theoretical curves for $e^* = e/3$. These observations are the hallmark of Laughlin's $e/3$ quasiparticles in FQH systems. After these observations, shot-noise measurements demonstrated various fractional charges, e.g., $e/5$ quasiparticles in the $\nu=2/5$ state~\cite{ReznikovNature1999,ChungPRL2003} and the $e/7$ quasiparticles in the $\nu=3/7$ state~\cite{ChungPRL2003}, clearly exhibiting the significance of shot-noise measurements for observing exotic charge carriers.

In addition to the odd-denominator fractional charges, the $e/4$ charge of quasiparticles in the $\nu=5/2$ even-denominator FQH system was observed in 2008~\cite{DolevNature2008}. As discussed in the next subsection, theories predict the non-Abelian nature of the $\nu=5/2$ state as a candidate for fault-tolerant quantum computing. The $e/4$ charge is a necessary condition for the non-Abelian nature; hence its experimental confirmation is essential. Figure~\ref{fig5_x2}(c) shows the shot-noise data obtained from a $\nu=5/2$ system formed in a split-gate device. The data agree well with the calculation assuming $e^*=e/4$ charges, signaling the presence of $e/4$ quasiparticles in the $\nu=5/2$ state. Note that the $e/4$ charge was also confirmed by analysis of the bias dependence of the direct tunneling current~\cite{RaduScience2008}, and measurement of the charging energy of microscopic $\nu=5/2$ regions~\cite{VenkatachalamNature2011}. 

\begin{figure}[!b]
\center 
\includegraphics[width=8cm]{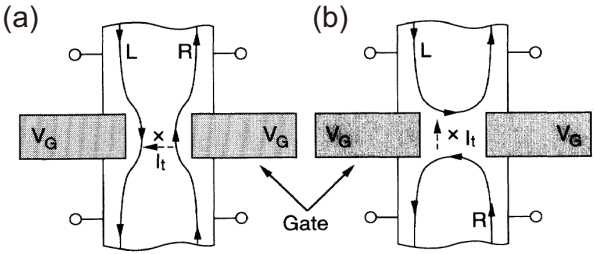}
\caption{(a) Schematic of fractional-charge tunneling in the weak-backscattering limit. (b) Electron tunneling in the strong-backscattering limit. Reprinted figures with permission from Ref.~[\onlinecite{ChamonPRB1995}]. {\copyright} (1995) American Physical Society.}
\label{fig5_x3}
\end{figure}

In the above experiment [Fig.~\ref{fig5_x2}(a)], the measurements were performed in the weak-backscattering limit (transmission probability through the constriction: $\cal{T}\simeq ~$1), where a backscattering event corresponds to a quasiparticle tunneling process through the FQH region [see Fig.~\ref{fig5_x3}(a)~\cite{ChamonPRB1995}]. In this limit, the tunneling process is so infrequent that there is no correlation between the quasiparticles, allowing one to compare the experimental data with simulations using Eq.~(\ref{qhe_s1/3_2}). A similar stochastic tunneling process occurs in the strong-backscattering limit ($\cal{T}\simeq ~$0) [Fig.~\ref{fig5_x3}(b)~\cite{ChamonPRB1995}], where a forward scattering event is considered as a tunneling process through the depleted region. Because tunneling quasiparticles are electrons, in this case, we observe $e^*=e$ tunneling charge~\cite{GriffithsPRL2000}.

In intermediate backscattering regimes, tunneling events occur so frequently that the correlation between each tunneling event becomes relevant. In this case, one cannot use Eq.~(\ref{qhe_s1/3_2}) for evaluating the effective charge of tunneling quasiparticles~\cite{FendleyPRL1995,FendleyPRB1996,TrauzettelPRL2004}. Conversely, however, Eq.~(\ref{qhe_s1/3_2}) allows us to evaluate the correlation of tunneling quasiparticles using the ``effective charge'' $e^*$  as an index. Indeed, $e^*$ continuously varies with the backscattering strength when shot-noise data are analyzed using Eq.~(\ref{qhe_s1/3_2})~\cite{GriffithsPRL2000,ChungPRL2003,HeiblumBook,FeldmanPRB2017}.

\begin{figure}[tb]
\center 
\includegraphics[width=7cm]{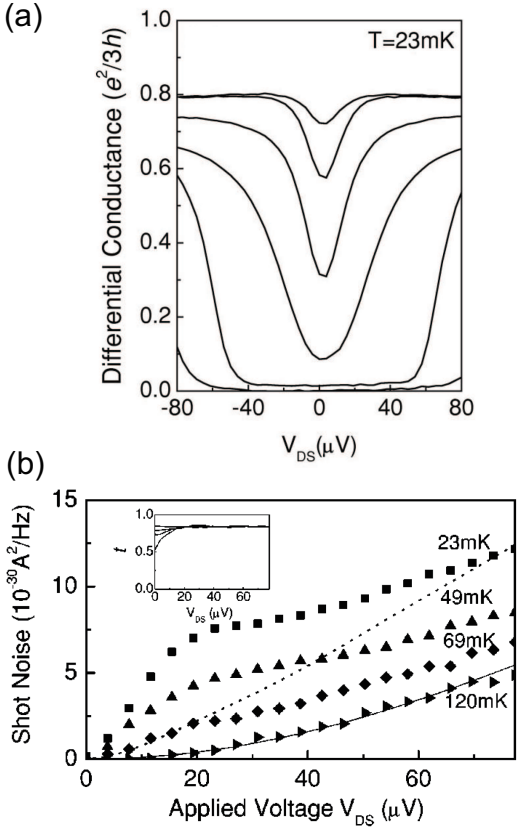}
\caption{(a) Differential conductance through a narrow constriction formed in the $\nu=1/3$ state. Each trace corresponds to the result measured at a different gate voltage that varies the backscattering strength. (b) Source-drain bias dependence of the shot noise at several temperatures. Reprinted figures with permission from Ref.~[\onlinecite{ChungPRB2003}]. {\copyright} (2003) American Physical Society.}
\label{fig5_x4}
\end{figure}

A quasiparticle tunneling between FQH edge channels can be modeled as tunneling between chiral Tomonaga-Luttinger (TL) liquids~\cite{KanePRL1994,ChamonPRB1995,ChamonPRB1996,FendleyPRL1995,FendleyPRB1996,MartinBook}. Within this model, both the dc transport and shot-noise properties can be calculated analytically over the entire range of the backscattering strength~\cite{FendleyPRL1995,FendleyPRB1996}. The TL-liquid nature of FQH edge channels manifests itself in the power-law behaviors observed in transport properties~\cite{ChangRevModPhys2003}. Figure~\ref{fig5_x4}(a) shows the bias $V_{\rm{DS}}$ dependence of differential conductance $g$ through a constriction formed in the $\nu = 1/3$ state~\cite{ChungPRB2003}. The several black curves, showing the power-law behaviors of $g$, were measured at different split-gate voltages applied to form the constriction and modulate the backscattering strength~\cite{ChungPRB2003}. At a fixed gate voltage, $g$ varies with $V_{\rm{DS}}$ from the strong-backscattering regime ($g \simeq 0$) to the weak-backscattering regime ($\simeq 0.8 \times e^2/3h$).

The nonlinear bias dependence of the backscattering strength gives rise to the irregular behaviors of the shot noise. Figure~\ref{fig5_x4}(b) shows the shot-noise data obtained from the same $\nu = 1/3$ FQH device as that in Fig.~\ref{fig5_x4}(a)~\cite{ChungPRB2003}. The shot noise below 100 mK shows an irregular increase with $V_{\rm{DS}}$: a steep increase at $0 < V_{\rm{DS}} < 20~\mu$V and a slowing down of the increase at $V_{\rm{DS}} > 20~\mu$V. This behavior qualitatively agrees with a simulation using the TL-liquid model~\cite{TrauzettelPRL2004}, indicating the relevance of the model~\cite{GlattliPhysicaE2000,ChungPRB2003}. The irregular behavior becomes more significant at lower temperatures [see Fig.~\ref{fig5_x4}(b)], corresponding to the prediction by the TL-liquid theory, in which the impact of electron correlation becomes more pronounced at low temperatures. 

While the experimental data qualitatively agree with the predictions by the TL-liquid theory, as discussed above, they disagree with them quantitatively. The disagreement may reflect the difference between an actual device and an ideal point-like scatterer assumed in the TL-liquid theory. For example, additional Coulomb interaction between the channels and unintentional localized states may be responsible for the disagreement~\cite{RosenowPRL2002}.

\begin{figure}[tb]
\center 
\includegraphics[width=6cm]{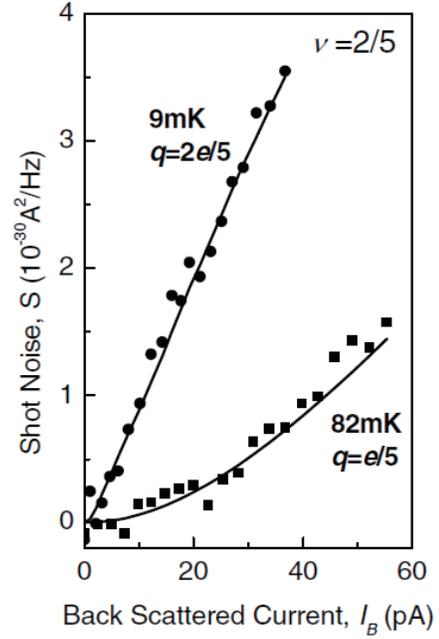}
\caption{Shot noise measured in the $\nu=2/5$ FQH system. Increase in the shot-noise intensity at low temperature suggests the bunching of fractional quasiparticles. Reprinted figure with permission from Ref.~[\onlinecite{ChungPRL2003}]. {\copyright} (2003) American Physical Society.}
\label{fig5_x5}
\end{figure}

Interestingly, some minor points in the results of several inter-channel tunneling experiments even qualitatively differ from the predictions by the TL-liquid model~\cite{FendleyPRL1995,FendleyPRB1996}. For example, in the weak-backscattering regime, conductance through a constriction decreases with increasing a bias in experiments, while the TL-liquid theory predicts the monotonic increase. The disagreement may result from the switching of edge configurations between the ones shown in Fig.~\ref{fig5_x3}(a) and Fig.~\ref{fig5_x3}(b)~\cite{RoddaroPRL2005,DolevNature2008,HeiblumBook}, while the TL-liquid model only considers a change in the coupling strength in the latter configuration. More interestingly, an increase in effective tunneling charges, unexpected in the TL-liquid theory, is observed at low temperatures. Figure~\ref{fig5_x5} shows a representative result observed in the $\nu=2/5$ state, where the effective charge, $e/5$ at 82 mK, is doubled to $2e/5$ at 9 mK~\cite{ChungPRL2003}. Similar increases in the effective charges also occur in the other FQH states, e.g., $\nu=3/7$~\cite{ChungPRL2003} and $\nu=5/2$~\cite{DolevPRB2010} states. Bunching of tunneling quasiparticles has been discussed as a possible cause of these observations. 

We have discussed the tunneling experiments by implicitly assuming that the whole 2DES is in an FQH state. However, because the shot noise reflects the effective charge of tunneling quasiparticles, only the barrier region needs to be in the FQH state for the fractional-charge detection, not the whole sample. Actually, $e/4$ charge [see Fig.~\ref{fig5_x2}(c)] was measured in a local $\nu=5/2$ region formed in a bulk $\nu=3$ IQH system~\cite{DolevNature2008}. Such local FQH systems are often observed in split-gate devices~\cite{RoddaroPRL2003,RoddaroPRL2004,RoddaroPRL2005,MillerNatPhys2007}.

\begin{figure*}[bt]
\center 
\includegraphics[width=17cm]{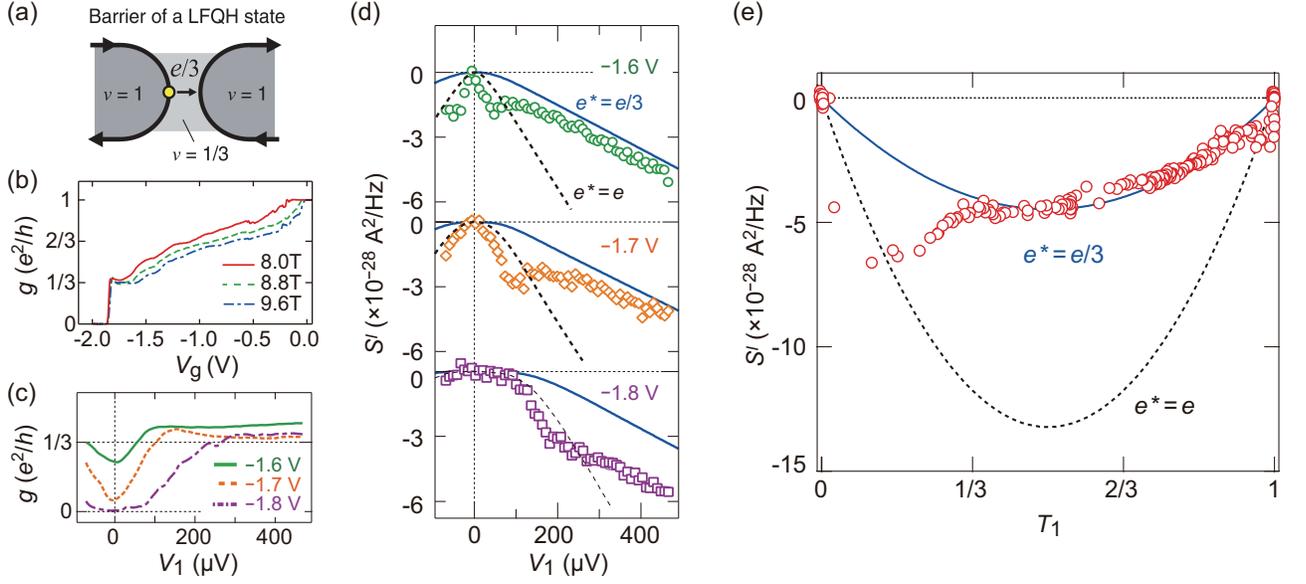} 
\caption{(Color online) (a) Schematic of $e/3$ quasiparticle tunneling through a local $\nu=1/3$ state formed in a bulk $\nu=1$ system. (b) Split-gate voltage $V_{\rm{g}}$ dependence of $g$ through a local $\nu=1/3$ state measured at $V_{1}=450~\mu V$ at several magnetic fields. (c) Source-drain bias $V_1$ dependence of $g$ and (d) $S^I$ obtained at several $V_{\rm{g}}$. (e) Transmission probability $T_1$ dependence of $S^I$ measured at $V=450~\mu V$. Reprinted figures with permission from Ref.~[\onlinecite{HashisakaPRL2015}]. {\copyright} (2015) American Physical Society.}
\label{fig5_x6}
\end{figure*}

Here, we introduce a striking example of such quasiparticle-tunneling experiments through a local FQH state~\cite{HashisakaPRL2015}. Figure~\ref{fig5_x6}(a) shows a schematic of a tunneling experiment through a local $\nu=1/3$ state formed in a bulk $\nu=1$ system. A split-gate voltage applied to constrict the $\nu=1$ system decreases electron density in the constriction, sometimes forming a local $\nu=1/3$ state. When a source-drain voltage is applied across such a constriction, the electronic current flowing between the separated $\nu=1$ regions may be carried by quasiparticle tunneling through the incompressible $\nu=1/3$ region. In this case, we can expect to observe the $e/3$ fractional charge via shot-noise measurements.

The experiment was performed on a QPC fabricated in a 2DES in an AlGaAs/GaAs heterostructure~\cite{HashisakaPRL2015}. Figure~\ref{fig5_x6}(b) presents the split-gate voltage $V_{\rm{g}}$ dependence of the differential conductance $g$ at the source-drain bias $V_{1}=450~\mu V$ at several magnetic fields. When $V_{\rm{g}}$ decreases from zero, $g$ decreases from $e^2/h$, showing a plateau at $e^2/3h$ around $-1.7~\rm{V}$, to zero below $V_{\rm{g}} = -1.9~\rm{V}$. The $e^2/3h$ plateaus signal the formation of a local $\nu=1/3$ state in the constriction. The plateau structure becomes more pronounced at higher magnetic fields, suggesting the increased FQH energy gap of the $\nu=1/3$ state.

The $V_{1}$ dependence of $g$ measured near $V_{\rm{g}}=-1.7~\rm{V}$ at 8 T ($e^2/3h$ plateau region), shown in Fig.~\ref{fig5_x6}(c), exhibits conductance suppression near $V_{1} = 0$. Despite the electronic current flowing between the $\nu=1$ regions, the measured zero-bias anomaly reminds us of the TL-liquid nature of the FQH edge channels [see Fig.~\ref{fig5_x4}(a)]. The formation of strip-like FQH edge states at the smooth edge of the $\nu=1$ regions is responsible for the observation~\cite{RoddaroPRL2003,RoddaroPRL2004,ParadisoPRL2012}. In the low $g$ region at low bias, the transmitted current is carried by electron tunneling through the depleted region between the strip-like FQH edge states [see Fig.~\ref{fig5_x3}(b)]. On the other hand, at high bias, the conductance increases to saturate at $g\simeq e^2/3h$, suggesting the $\nu=1/3$ state develops over the constriction region, as shown in Fig.~\ref{fig5_x6}(a).

While the $e/3$ charge tunneling through the $\nu=1/3$ system has been observed by analyzing the differential conductance~\cite{RoddaroPRL2003,RoddaroPRL2004,RoddaroPRL2005}, shot-noise measurements provide further evidence of it~\cite{HashisakaPRL2015}. Figure~\ref{fig5_x6}(d) shows the shot-noise data measured simultaneously with $g$ presented in Fig.~\ref{fig5_x6}(c). In this measurement, current-noise cross-correlation $S^I$ between the transmitted and reflected channels were evaluated. The negative sign of $S^I$ originates from the partitioning of the tunneling quasiparticles, as expressed in Eq.~(\ref{qhe_s24}). Whereas the experimental data are close to the theoretical shot-noise curve with $e^*=e$ at low bias, they approach the curve with $e^*=e/3$ at high bias. The latter observation corresponds to the fractional-quasiparticle tunneling picture illustrated in Fig.~\ref{fig5_x6}(a).

Figure~\ref{fig5_x6}(e) shows the transmission probability $T_{1}=G\times h/e^2$ dependence of $S^I$ measured at $V_{1}=450~\mu V$, where $G$ is the conductance through the constriction. The shot-noise data near $T_{1}=0$ and $T_{1}=1$ are close to the theoretical curve with $e^*=e$ (dashed black curve), indicating the electron tunneling through the depleted region and the incompressible $\nu = 1$ region, respectively. Meanwhile, the data in the intermediate $T_{1}$ region agrees well with the $e^*=e/3$ curve (solid blue curve). The latter result indicates that the charge-transfer process over the broad intermediate $T_{1}$ regime is the stochastic $e/3$ tunneling between the $\nu = 1$ edge channels. The stochasticity, namely the absence of correlation between quasiparticles, can be interpreted as the result of the 1D free-electron-system nature of the $\nu = 1$ edge channels. The above experimental results demonstrate that the effective tunneling charge corresponds to the charge of elementary excitations in the barrier region and also that correlation between tunneling quasiparticles reflects the interaction in the edge channels.

While the above data are obtained at highly non-equilibrium, finite current noise on plateaus is also observed at low bias in a similar setup~\cite{BidPRL2009,RosenblattNatComm2017}. Upstream charge-neutral modes in QH systems are considered to be responsible for the current noise at low bias. The relation between the observations in the high- and low-bias regimes is unclear and requires more studies.

\subsubsection{Anyonic statistics of fractional quasiparticles}
\label{sec:anyonic_statistics}

A marked feature of FQH quasiparticles is not only their fractional charge but also their anyonic quantum statistics~\cite{WilczekPRL1982}. Unlike elementary excitations in three-dimensional systems, quasiparticles in 2D systems can be neither bosons nor fermions but anyons. When the wave function of the system obtains the phase $\theta \neq \pi$ or $2\pi$ by an exchange operation of two quasiparticles, they are referred to as ``Abelian anyons''. On the other hand, when the operation is described not by the phase evolution but by an arbitrary unitary transformation, they are called ``non-Abelian anyons''. Theories predict that the non-Abelian statistics provide the basis of fault-tolerant quantum computing, stimulating intensive studies on the quantum statistical nature of FQH quasiparticles~\cite{MooreNuclPhysB1991,ReadPhysicaB,KitaevAnnalsPhys2003,NayakRevModPhys2008}. While quasiparticles in some FQH states, such as the well-known $\nu=1/3$ and $\nu=2/5$, are Abelian anyons, other FQH states, e.g., $\nu=5/2$ and $\nu=12/5$, may support non-Abelian anyons. Although the wave function of the $\nu=5/2$ state is still under debate, some candidates are considered to support quasiparticles having $e/4$ charge and non-Abelian statistics. The shot-noise measurement performed in the $\nu=5/2$ state provides evidence of the $e/4$ charge [see Fig.~\ref{fig5_x2}(c)], which is the necessary condition for the $\nu=5/2$ state being non-Abelian~\cite{DolevNature2008}. Thus, in FQH systems, various Abelian and non-Abelian anyons can appear in a single device by varying the filling factor using external parameters such as the magnetic field; hence the FQH state is the promising testbed for investigating anyons. 

The anyonic statistics, both Abelian and non-Abelian, of FQH quasiparticles, were not confirmed in experiments until more than three decades after the first theoretical prediction~\cite{WilczekPRL1982}. However, very recently, the Abelian statistics of the $\nu=1/3$ quasiparticles were observed by shot-noise measurement~\cite{BartolomeiScience2020} and Fabry-P\'erot interferometry~\cite{NakamuraNatPhys2020}. Here, we introduce the shot-noise experiment, where $e/3$ quasiparticles collide to show their Abelian anyonic statistics.

\begin{figure*}[bt]
\center 
\includegraphics[width=16cm]{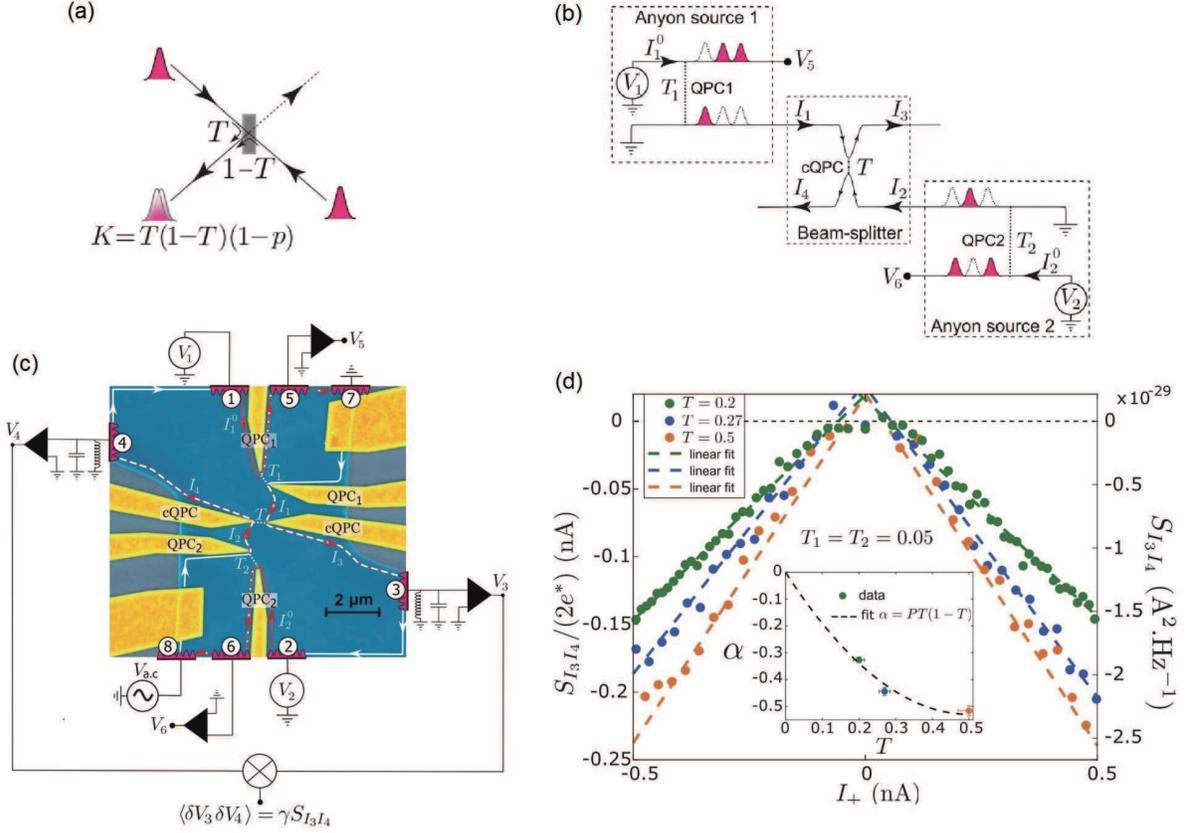}
\caption{(Color online) (a) Schematic of an anyon-collision experiment. (b) Experimental setup for the collision experiment using three QPCs. Some of the anyons randomly emitted from QPC1 and QPC2 collide at cQPC and cause cross-correlation between the output currents $I_3$ and $I_4$. (c) False-color scanning electron micrograph of the three-QPC device and the measurement setup. Electronic currents injected from ohmic contacts 7 and 8 flow along with the white arrows and are partitioned at QPC1 and QPC2, respectively, to generate currents $I_1$ and $I_2$ accompanied by anyon excitations (dashed red/white allows). The anyons randomly impinge on cQPC from both the top and bottom sides and sometimes collide to experience the exchange interference. The output signals are measured through contacts 3 and 4 to evaluate current-noise cross-correlation. (d) Input current $I_+$ dependence of the cross-correlation $S_{I_3I_4}/(2e^*)$ measured at three different cQPC transmission probabilities ${\cal{T}}$. The dashed lines are linear fits to the $S_{I_3I_4}/2e^*$ data. (Inset) Slope $\alpha$ extracted from the fit. The dashed lines is a fit to $\alpha = PT(1-T)$ with $P = -2.1$. Reprinted with permission from Ref.~[\onlinecite{BartolomeiScience2020}]. {\copyright} (2020) American Association for the Advancement of Science.}
\label{fig5_x7}
\end{figure*}

Figure~\ref{fig5_x7}(a) shows a schematic of a collision experiment, where two quasiparticles impinge on a beam splitter of transmission probability ${\cal{T}}$ at the same time. Here, we define the probability $K$ of both quasiparticles scattered to the left side of the beam splitter. We describe $K={\cal{T}}(1-{\cal{T}})$ in a  classical model considering the quasiparticles to be distinguishable, while $K={\cal{T}}(1-{\cal{T}})(1-p)$ when they are indistinguishable. As discussed in Sect.~\ref{sec:fermion_optics}, one finds $p=1$ and $K=0$ due to the Pauli exclusion principle in the fermion case. In contrast, in the boson case, $p<0$ and $K$ is larger than that in the classical model. If the quasiparticles are anyons, one can expect that $p$ takes an intermediate value between those of fermions and bosons. For example, $e/3$ quasiparticles in the $\nu=1/3$ state are predicted to take $p<0$ since the exchange phase $\theta = \pi/3$ is close to $\theta=0$ of bosons.

Although we considered a single collision event above, in practice, it is difficult to perform such an experiment due to the lack of an on-demand single-anyon source. This difficulty contrasts with the case of electron-collision experiments, which have been achieved using single-electron sources (see Sect.~\ref{sec:fermion_optics})~\cite{BocquillonScience2013, DuboisNature2013}. For observing anyonic statistics of FQH quasiparticles, Rosenow $et~al.$ proposed a different approach using two anyon sources that randomly emit anyons in the time domain~\cite{RosenowPRL2016}. Figure~\ref{fig5_x7}(b) shows a schematic of the approach. In this setup, QPC$_1$ and QPC$_2$ are set in the weak backscattering regime. At finite bias $V_1$ and $V_2$, these QPCs randomly emit $e/3$ quasiparticles due to the tunneling, serving as the anyon sources. Here, ${\cal{T}}_1$ and ${\cal{T}}_2$ are the QPC transmission probabilities. The tunneling currents $I_1$ and $I_2$, carrying $e/3$ quasiparticles, impinge on the center QPC (cQPC) that works as an anyon beam splitter (transmission probability ${\cal{T}}$). The exchange interference between two quasiparticles enhances the shot noise accompanying the output currents $I_3$ and $I_4$. When ${\cal{T}}_1$ = ${\cal{T}}_2$ and $\langle I_1\rangle=\langle I_2\rangle$, current-noise cross-correlation $S_{I_{3}I_{4}}$ between $I_3$ and $I_4$ is described as 
\begin{equation}
\begin{split}
S_{I_3I_4}=2e^*P{\cal{T}}(1-{\cal{T}})I_+,
\label{anyon}
\end{split}
\end{equation}
\begin{equation}
\begin{split}
P=\frac{-2}{m-2},
\end{split}
\end{equation}
where $I_+ = \langle I_1\rangle+\langle I_2\rangle$ is the sum of the currents impinging on cQPC, and $m$ is a factor characterizing the exchange phase $\theta=\pi/m$. In the case of $e/3$ quasiparticles, one can expect to observe $P=-2$ since $m=3$.

Figure~\ref{fig5_x7}(c) displays a false-colored scanning electron micrograph of the sample and the experimental setup examined by Bartolomei $et~al$~\cite{BartolomeiScience2020}. The edge currents $I_1$ and $I_2$ carrying $e/3$ quasiparticles are mixed at cQPC to generate finite $S_{I_{3}I_{4}}$. Figure~\ref{fig5_x7}(d) shows $S_{I_{3}I_{4}}$ measured at ${\cal{T}}_1$ = ${\cal{T}}_2 = 0.05$ (weak-backscattering regime) for three different cQPC transmission probabilities (${\cal{T}}$). The $S_{I_{3}I_{4}}$ data show a negative correlation at finite $I_+$, and the $I_+$ dependence well fits linear functions with $P=-2.1\pm0.1$ in Eq.~(\ref{anyon}). This $P$-value is close to the theoretical prediction $P=-2$, being a hallmark of the anyonic nature of $e/3$ quasiparticles. Note that Bartolomei $et~al$. also evaluated $S_{I_{3}I_{4}}$ in the case of $\langle I_1\rangle \neq \langle I_2\rangle$ and confirmed good agreements between the experimental results and the theoretical predictions~\cite{RosenowPRL2016}.

The above experimental result is the first evidence of anyons. This achievement is a significant milestone toward the realization of future topological quantum computation using FQH anyons.

\subsubsection{Quantum many-body effects in edge channels}

Electron correlation in a QH edge channel sometimes causes peculiar transport phenomena. For example, one observes TL-liquid-like behaviors~\cite{WenPRB1990,ChangRevModPhys2003} in quasiparticle tunneling processes in FQH systems, as discussed in Sect.~\ref{sec:fractional_charge_tunneling}. Here, we introduce other intriguing behaviors originating from electron correlation in QH edge channels. 

While IQH edge channels are often described as 1D free-electron systems, their charge excitations often show the TL-liquid nature due to the long-range intra- and inter-edge Coulomb interaction~\cite{HashisakaRevPhys2018}. For example, in the $\nu=2$ state, transport eigenmodes in copropagating edge channels, charge and charge-neutral (spin) modes with different velocities cause spatial separation of charge and spin excitations. This well-known phenomenon is referred to as the spin-charge separation in the QH TL liquid~\cite{InouePRL2014,FreulonNatCommun2015,HashisakaNatPhys2017}. Current-noise measurements allow us to identify these transport eigenmodes that mix the current noise in the two channels~\cite{InouePRL2014}. 

A larger variety of correlation phenomena are observed in FQH systems. One representative example is the charge-neutral transport in the $\nu=2/3$ FQH state. Below we discuss the $\nu=2/3$ edge channels that contain the essence of quantum many-body physics at the edge of topological quantum liquids.

The $\nu=2/3$ state is particle-hole symmetric with the $\nu=1/3$ state; it is the hole $\nu_{\rm{h}}=1/3$ state in the lowest Landau level. Based on this picture, in 1990, MacDonald proposed a model of the $\nu=2/3$ edge state, in which a hole $\nu_{\rm{h}}=1/3$ edge state and an electron $\nu_{\rm{e}}=1$ edge state counter-propagate~\cite{MacDonaldPRL1990,JohnsonPRL1991}. The formation of such counter-propagating channels is referred to as ``edge reconstruction'' in hole-conjugate FQH states. MacDonald's $\nu=2/3$ edge-state model is theoretically reasonable. However, this model contradicted the experimental results at that time: whereas the model predicts the two-terminal conductance $G=4/3\times e^2/h$ of the $\nu=2/3$ system, experiments reported $G=2/3\times e^2/h$.

Kane $et~al$. studied transport eigenmodes in the reconstructed $\nu=2/3$ edge state using the renormalization-group theory~\cite{KanePRL1994,KanePRB1995}. They found that because of the mode mixing between the counter-propagating channels due to the random inter-channel tunneling and Coulomb interaction, the upstream charge-neutral mode appears, in addition to the charge mode that gives a quantized Hall conductance of $G=2/3\times e^2/h$. This conductance value agrees with the experimental observations, indicating the validity of both the edge-reconstruction and mode-mixing pictures. Moreover, the charge-neutral transport was also observed later, as introduced below. Note that similar peculiar edge transport can occur not only in the $\nu=2/3$ state but also in various FQH systems and in non-QH 2D topological quantum liquids as well. From this perspective, the $\nu=2/3$ edge state has been an significant testbed for studying edge transport in topological systems.

\begin{figure}[tb]
\center 
\includegraphics[width=8.5cm]{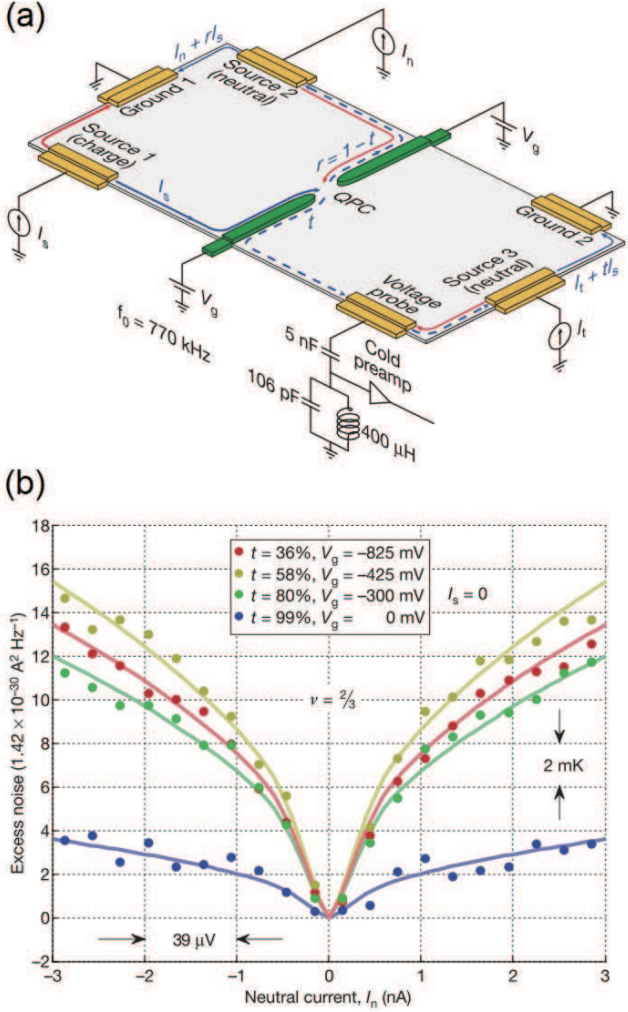}
\caption{(Color online) (a) Schematic of the experimental setup for observing the upstream charge-neutral transport in the $\nu=2/3$ state. Charge and charge-neutral transport are indicated by blue and red arrows, respectively. Charge-neutral mode excited at ``Source 2'' by applying a current $I_n$ impinges on the QPC and generates excess current noise. (b) Excess current noise as a function of $I_n$ measured at several QPC transmission probabilities $t$. Reprinted with permission from Ref.~[\onlinecite{BidNature2010}]. {\copyright} (2010) Springer Nature.}
\label{fig5_x8}
\end{figure}

Current-noise measurements have played an essential role in observing the charge-neutral transport in the $\nu=2/3$ state. Figure~\ref{fig5_x8}(a) is a schematic of the measurement setup in the first experiment~\cite{BidNature2010}. Charge excitations in the counter-propagating $\nu=1$ and $\nu=1/3$ edge channels are mixed via the inter-channel Coulomb interaction and random tunneling to form the charge mode (blue arrows) and the charge-neutral mode (red arrows), propagating clockwise and anticlockwise, respectively. When a current $\langle I_n\rangle$ is applied to an ohmic contact ``Source 2'', the charge mode is grounded at the contact ``Ground 1''. In contrast, the charge-neutral mode impinges on a QPC and generates excess current noise at the ``Voltage probe''.

Figure~\ref{fig5_x8}(b) displays the measured excess current noise as a function of $\langle I_n\rangle$. The monotonic increase in the excess noise with $|\langle I_n\rangle|$ signals the presence of the upstream charge-neutral transport. The excess noise depends on the transmission probability $t$ of the QPC, indicating that the excess noise is generated at the QPC that scatters the charge-neutral excitations. This observation proves both edge reconstruction~\cite{MacDonaldPRL1990,JohnsonPRL1991} and the formation of transport eigenmodes~\cite{KanePRL1994,KanePRB1995} in the $\nu=2/3$ state.

It is noteworthy that the charge-neutral transport in the $\nu=2/3$ state was also confirmed using another method based on QD thermometry~\cite{VenkatachalamNatPhys2012,GurmanNatCommun2012}. Additionally, the charge-neutral transport has been observed in other FQH systems, such as the $\nu=4/3$, $\nu=1/3$, and $\nu=3/5$ states in GaAs/AlGaAs heterostructures~\cite{AltimirasPRL2012,InoueNatCommun2014,SaboNatPhys2017}. While the early theories predicted the presence of the charge-neutral mode only in hole-conjugate FQH systems, the charge-neutral transport observed in the $\nu=4/3$ and $\nu=1/3$ states suggests the edge-reconstruction physics in non-hole-conjugate FQH systems confined by a smooth edge potential.

\begin{figure}[tb]
\center 
\includegraphics[width=8.5cm]{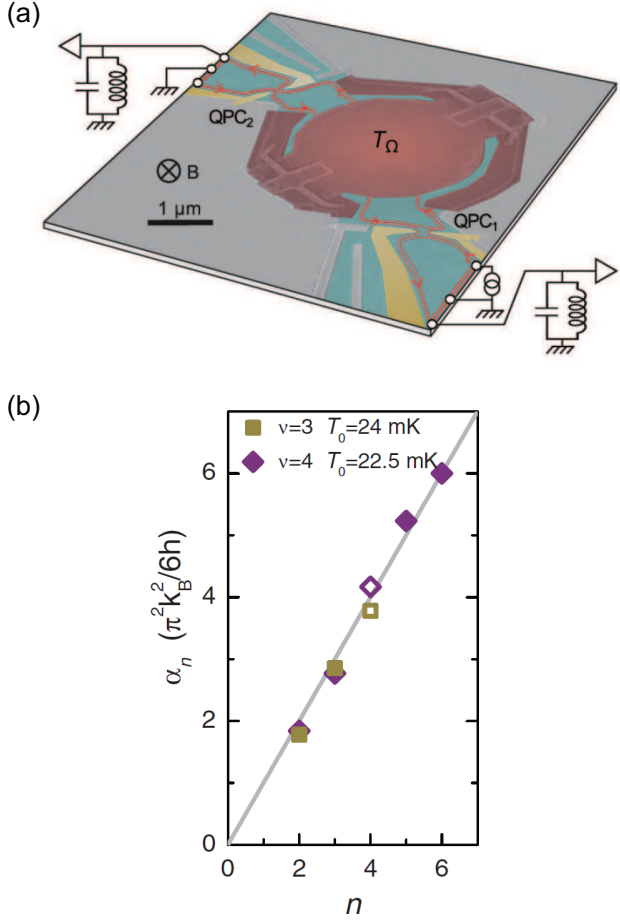}
\caption{(Color online) (a) Experimental setup for measuring quantized heat transport along IQH edge channels. A micrometer-scale ohmic contact (dark red region) divides the 2DES into two regions (light blue regions). Heat transport along edge channels (red arrows) is measured by the two current-noise-measurement setups using LC resonance circuits. (b) Measured heat-current factor $\alpha_n$ versus the number of edge channels $n$. The gray line is the prediction for the quantized heat transport. Reprinted with permission from Ref.~[\onlinecite{JezouinScience2013}]. {\copyright} (2013)  American Association for the Advancement of Science.}
\label{fig5_x9}
\end{figure}

While the above experiment confirmed the presence of upstream heat transport in the $\nu=2/3$ state, recently, the heat flow along QH edge channels has been investigated quantitatively. Such an experiment was first performed to observe the quantized heat transport along IQH edge channels~\cite{JezouinScience2013}. Figure~\ref{fig5_x9}(a) shows the experimental setup, where a micrometer-scale ohmic contact $\Omega$ (dark red region) separates the two IQH systems (blue regions), which are individually equilibrated through different ohmic contacts (temperature $T_0$). A current injected to the 2DES on the right-hand side of $\Omega$ propagates through QPC$_1$ to reach $\Omega$. The dc excitation is equilibrated with electrons fed from the left 2DES through QPC$_2$, leading to the increase in electron temperature $T_{\rm{\Omega}}$. $T_{\rm{\Omega}}$ is evaluated by current-noise thermometry performed on both left and right 2DESs.

The heat flow along the edge channels is quantitatively evaluated from the relation between the Joule heat and increase in $T_{\rm{\Omega}}$. Wiedemann-Franz law predicts that the heat flow $J^e_Q(T_{\Omega},T_0)$ along a single edge channel is described as 
\begin{equation}
\begin{split}
J^e_Q(T_{\Omega},T_0)=\frac{\pi^2k_{\rm{B}}^2}{6h}(T_{\Omega}^2-T_0^2),
\label{heatflow}
\end{split}
\end{equation}
where $T_{\Omega}$ and $T_0$ are temperatures of the reservoirs. Figure~\ref{fig5_x9}(b) shows the heat-current factor $\alpha_n = nJ^e_Q/(T_{\Omega}^2-T_0^2)$ evaluated at several filling factors. One observes that $\alpha_n$ data fall near the prediction by Wiedemann-Franz law (gray line): namely, each edge channel carries the heat current $J^e_Q$. This result proved the quantized heat flow across IQH edge channels. Furthermore, other experiments have revealed the impact of inter-channel Coulomb interaction on heat transport along co-propagating IQH edge channels~\cite{SivreNatPhys2017,SivreNatCommun2019}.

The heat flow along FQH edge channels has also been investigated in several FQH states, e.g., the $\nu=1/3, 2/3, 3/5$, and $4/7$ states~\cite{BanerjeeNature2017}. The experiments confirmed that both IQH and FQH channels carry quantized heat flow. Furthermore, a similar heat-transport measurement was performed on the $\nu=5/2$ state, which may support the presence of non-Abelian quasiparticles~\cite{BanerjeeNature2018}. The experimental result shows a half-integer quantized heat conductance, suggesting the Majorana edge mode of the particle-hole Pfaffian state. This observation contradicts numerical calculation predicting the Pfaffian or the anti-Pfaffian state and therefore requires further theoretical and experimental studies. Although still inconclusive, the experimental result implies a non-Abelian topological order of the $\nu=5/2$ state.

\subsubsection{Short summary}
Current-noise measurements have revealed various phenomena in QH systems, such as fractional charge, anyonic statistics, and quantum-many-body effects in edge channels. There remain many other interesting issues, such as Andreev reflection of fractional quasiparticles at QH interfaces, requiring shot-noise measurements~\cite{HashisakaNatComm2021}. Current-noise measurements will allow us to gain more in-depth insight into QH systems and other 1D electron systems, including helical edge states in 2D topological materials.

\subsection{Noise in superconductor-based junctions}
Superconductivity is one of the most representative examples of the quantum many-body effect. While a supercurrent does not generate shot noise in a bulk superconductor, it can be backscattered at a junction between a superconductor and a normal conductor to generate a shot noise. 

Blanter and B\"uttiker introduced several theoretical treatments for shot-noise generation at superconductor junctions~\cite{BlanterPR2000}. While there were only a few experiments at that time, we now have many examples of such experimental shot-noise studies. Here, we introduce some experiments that studied Andreev reflection, Cooper-pair splitting, the Kondo-Andreev effect, and superconductor-QH junctions. As we will see below, these junctions provide valuable platforms for exploring quantum many-body phenomena.

\subsubsection{Andreev reflection}
Electron injection from a normal metal to a superconductor results in hole reflection to form a Cooper pair in the superconductor. This intriguing phenomenon, directly manifesting the electron pairing in a superconductor, is referred to as Andreev reflection. The Andreev process leads to the charge-$2e$ shot-noise generation at a normal metal-superconductor (NS) junction~\cite{KhlusJETP1987,deJongPRB1994}. 

\begin{figure}[tbp]
\center 
\includegraphics[width=7cm]{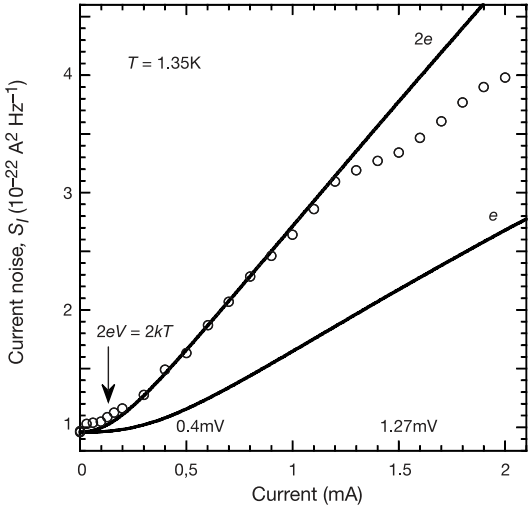} 
\caption{Shot noise at an NS junction as a function of bias current $\langle I \rangle$ at 1.35 K. Reprinted with permission from Ref.~[\onlinecite{JehlNature2000}]. {\copyright} (2000) Springer Nature.} 
\label{Fig:JehlNature2000}
\end{figure}

The charge-$2e$ shot noise in the Andreev process was first observed in 2000~\cite{JehlNature2000,KozhevnikovJLTP2000,KozhevnikovPRL2000}. Figure~\ref{Fig:JehlNature2000} shows the data obtained by Jehl \textit{et al.} for a Nb/Cu junction~\cite{JehlNature2000}. For this measurement, a SQUID was used to detect the current noise in the junction of small resistance ($\sim 0.8~\Omega$)~\cite{JehlRSI1999} [see Sect.~\ref{sec:transimpedance}]. In the low-bias regime, the experimental data (circles) agree well with the following theoretical curve [see Eq.~(\ref{ShotTheory})]:
\begin{equation}
S=\frac{2}{3}\left[\frac{4k_B T}{dV/dI} +e^* \langle I \rangle \coth\left(\frac{e^*V}{2k_\textrm{B}T_\textrm{e}}\right)\right],
\end{equation}
where $e^*=2e$ reflects the Cooper-pair charge and the denominator ``3'' of the factor $2/3$ represents the Fano factor $F=1/3$ of the diffusive mesoscopic device~\cite{NagaevPLA1992,NagaevPRB2001}. The deviation of the experimental data from the $2e$ curve at high bias ($\gtrsim 1.2$~mA) may reflect the occurrence of charge-$e$ quasiparticle tunneling.

Similar charge-$2e$ shot noise was also observed in a superconductor-semiconductor junction~\cite{LeflochPRL2003} and a superconductor-QD-superconductor device~\cite{DasNatComm2012,HataPRL2018}. The former experiment, where electrons are emitted into the superconducting gap with a Poisson distribution, is a superconducting analog of Schottky's experiment on electron emission into a vacuum~\cite{SchottkyAP1918}. The measured shot noise agrees with the theoretical curve of $S=4e\vert\langle I\rangle\vert$ at low bias, indicating the $2e$ effective charge.

Furthermore, multiple Andreev reflection (MAR) occurs at a junction where two superconductors sandwich a normal conductor. When such a junction is voltage-biased ($\vert eV \vert < 2\Delta$, where $2\Delta$ is the superconducting gap), Andreev processes occur many times at the two NS interfaces. In this case, the shot noise generated at the whole junction is sometimes enhanced such that $S/2e\vert\langle I \rangle\vert =1+\Delta/eV$, indicating that multiple charge quanta, $2e, 3e$, etc., carry the transmitted current~\cite{DielemanPRL1997,HossPRB2000,CronPRL2001,RonenPNAS2016}. On the other hand, another recent experiment demonstrated shot-noise suppression with a factor less than one at high bias $\vert eV\vert \sim 2\Delta$, where the quasiparticle tunneling can arise~\cite{RonenPNAS2016}. 

\subsubsection{Cooper-pair splitting}
When two normal conductors contact a superconductor with a spacing shorter than the superconducting coherence length, a correlation appears between the charge-scattering processes at the contacts. When an electron is incident from one of the normal conductors to the superconductor, a hole is ejected from the other contact to form a Cooper pair. This process is referred to as the crossed Andreev reflection (CAR), or nonlocal Andreev reflection. In a similar setup, an inverse process of CAR occurs. A Cooper pair in the superconductor splits into two electrons, and each of them is ejected to the two normal conductors one by one. The Cooper-pair splitting was observed, for example, in a ferromagnet-superconductor junction~\cite{BeckmannPRL2004}.

Because two electrons generated by a single Cooper-pair splitting are entangled quantum mechanically, this process has been intensively studied for realizing a solid-state entangled-pair generator. In experiments, a Cooper-pair-splitting device is often constructed by attaching two QDs to a superconductor~\cite{HofstetterNature2009}. Thanks to the Coulomb blockade, each QD traps only a single electron at a time so that the device enables us to observe a single Cooper-pair splitting process. Time-correlation measurement for the currents flowing through the QDs provides striking evidence for the Cooper-pair splitting. Das \textit{et al.} performed a current-noise cross-correlation measurement between InAs QDs attached to aluminum~\cite{DasNatComm2012}. The observed positive cross-correlation indicates that almost all the current is ejected from the superconductor by the Cooper-pair splitting processes.

\subsubsection{Kondo-Andreev effect}
The coexistence of and competition between superconductivity and the Kondo effect has attracted attention since the 1960s~\cite{SodaPTP1967}. Superconductivity in $s$-wave superconductors emerges via a macroscopic wave function of spin-singlet Cooper pairs. The Kondo effect, in contrast, occurs due to the spin-singlet (Kondo-singlet) formation between a localized spin and conduction electrons. Because these two quantum many-body effects originate from different electronic states, one may think they cannot coexist. Alternatively, because electrons screen magnetic impurities in the Kondo state, another may think that the Kondo effect enhances superconductivity. In reality, the coexistence of the two effects can be observed in bulk materials, such as heavy-fermion systems. 

\begin{figure}[tbhp]
\center 
\includegraphics[width=7cm]{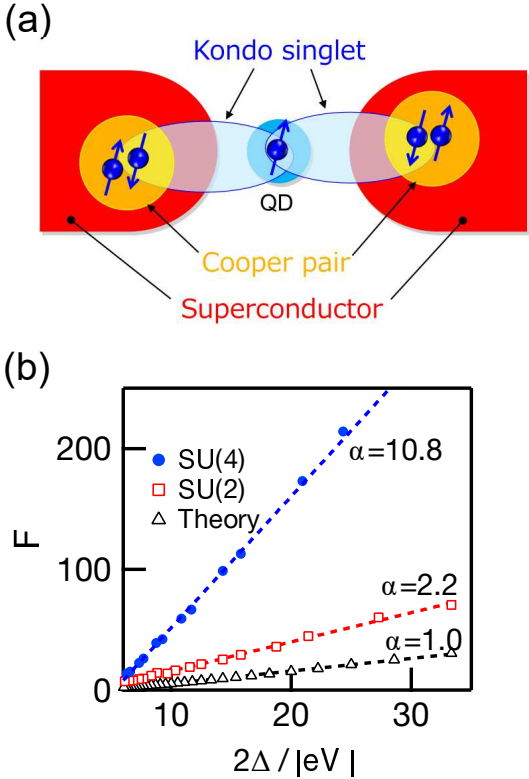} 
\caption{(Color online) (a) Schematic of a S-QD-S device, where the Kondo effect and superconductivity coexist. (b) $2\Delta/|eV|$ dependence of Fano factor $F =S/2e|\langle I\rangle |$ at the electron-hole-symmetry point [circles and squares: experimental data at the SU(4) and SU(2) Kondo states, respectively. Triangles: numerical calculation result for a QPC]. The data points fit well with the simulations using Eq.~(\ref{eq:hataPRL2018}). $\alpha$ is the noise-enhancement factor in the Kondo regime. Panel (b) is reproduced from Ref.~[\onlinecite{HataPRL2018}]. {\copyright} (2018) American Physical Society.} 
\label{Fig:HataAndreevKondo}
\end{figure}

A QD-superconductor device provides a vital platform for observing the Kondo-Andreev effect because a QD behaves as a controllable magnetic impurity. Figure~\ref{Fig:HataAndreevKondo}(a) shows a schematic of such a device, a superconductor-QD-superconductor (S-QD-S) junction, where we can examine the relationship between the two quantum many-body effects in a controlled way~\cite{BuitelaarPRL2002,KimPRL2013,HataPRL2018}.

Recently, one of the authors of this review reported a shot-noise measurement performed on the QD device shown in Fig.~\ref{Fig:HataAndreevKondo}(a)~\cite{HataPRL2018}. The QD was adjusted to the SU(2) or SU(4) Kondo state, while a weak magnetic field was applied to break superconductivity and prepare a normal state in the leads. When the magnetic field is turned off, the lead wires enter the superconducting state, and the Kondo-Andreev state can appear.

Figure~\ref{Fig:HataAndreevKondo}(b) presents the shot-noise results at the electron-hole-symmetry point. The Fano factor $F$ measured at low bias voltage, $V$ below the superconducting gap $2\Delta/e$, is plotted as a function of $2\Delta/|eV|$. The experimental data demonstrate strong shot-noise enhancement ($F\propto V^{-1}$) in both the SU(2) and SU(4) Kondo states. These observations qualitatively agree with the theory of $n$th-order multiple-Andreev-reflection (MAR) processes ($n\sim 2\Delta/eV$), where a single transport process is considered to carry the $e^*=ne$ effective charge~\cite{CuevasPRL1999}.

On the other hand, if we take a more quantitative look at the experimental data, we find that the observed shot-noise enhancement is much larger than the theoretical prediction. Using the factor $\alpha$, the enhanced Fano factor is defined as
\begin{equation}
    F\equiv \frac{2\Delta}{|eV|}\alpha.
\label{eq:hataPRL2018}
\end{equation}
For a simple junction without the Kondo correlation, $\alpha$ equals one, whereas the experiment yielded $\alpha =2.2$ for SU(2) and $\alpha =10.8$ for SU(4) as shown in Fig.~\ref{Fig:HataAndreevKondo}(b). This means that MAR enhances the noise in the Kondo state by several times. As seen in Sect.~\ref{Subsec:KondoNoise}, the shot noise in the Kondo state connected to the normal lead is in quantitative agreement with theory for both the SU(2)- and SU(4)- Kondo states. Although there is a theory of the noise in the Kondo-Andreev state~\cite{AvishaiPRB2003}, further theoretical development is necessary to explain the experimental data. Elucidating the non-equilibrium behavior in systems where two different singlet states interplay remains an important issue for the future.

\subsubsection{Junction with quantum Hall states}
Before closing this section, we discuss a junction between a topological edge state (including a chiral edge state in a QH system) and a superconductor. While such systems have been studied theoretically since the 1990s~\cite{MaEPL1993,FisherPRB1994}, recent theories predicting the emergence of non-Abelian anyons at a superconductor-edge state interface~\cite{FuPRL2008,MongPRX2014} have stimulated many theoretical and experimental studies.

Despite the fundamental importance, quantum Hall-superconductor (QH-S) junctions had not been fabricated in experiments due to technical difficulties until recently. The difficulty was that, first, superconductivity usually disappears at high magnetic fields where the QH effect occurs. However, superconductors with a high critical magnetic field solved this problem. A more severe problem was the Schottky barrier that degrades the proximity effect at a QH-S interface. Recently, this problem was also solved by using novel 2DESs such as graphene~\cite{KomatsuPRB2012} and ZnO-based 2DES~\cite{KozukaJPSJ2018}.

A shot-noise measurement, which enables us to evaluate the effective charge of an elementary excitation, serves as a powerful probe for a superconducting correlation in QH edge channels. Recently, Sahu \textit{et al.} reported a shot-noise measurement in a bilayer graphene-superconductor junction~\cite{SahuPRB2019}. They observed the shot-noise enhancement by a Fano factor of about two due to the Andreev reflection. The enhancement is more significant than that in a metal-superconductor interface formed in the same device at a zero magnetic field (the Fano factor is about 1.5 in this case). These observations may be an indication of a superconducting correlation in QH edge channels.

\section{Fluctuation Theorem and Current Noise}
\label{sec:fluctuationtheorem}
\subsection{Fluctuation Theorem}
We have so far discussed current noise in mesoscopic systems based on the Landauer-B\"{u}ttiker picture; we would now like to discuss from a different perspective, namely the Fluctuation Theorem (FT). In the 1950s, the linear response theory was formulated in the field of statistical mechanics~\cite{KuboJPSJ1957}. Whereas this theory provides a powerful methodology for studying the response of a system to an external field at near equilibrium, it cannot be applied to highly non-equilibrium systems. There have been many attempts over the years to investigate non-equilibrium systems beyond the linear response theory. One of the significant achievements is the FT reported in 1993~\cite{EvansPRL1993}.

\begin{figure}[!t]
\center 
\includegraphics[width=7cm]{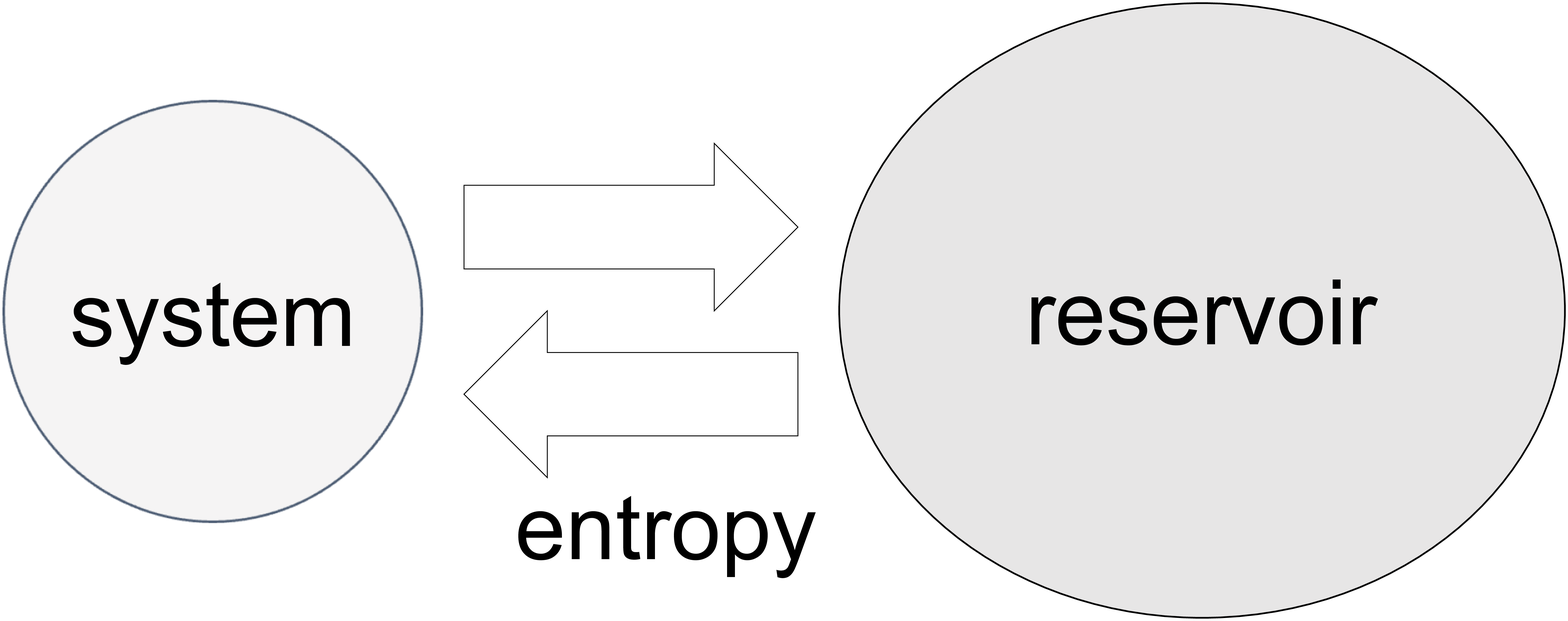} 
\caption{FT considers a small system connected to a reservoir.} 
\label{Fig_FTconcept}
\end{figure}

Let us consider a small system connected to a reservoir as shown in Fig.~\ref{Fig_FTconcept}~\cite{EvansAP2002}. While the entropy of the entire system does not decrease due to the second law of thermodynamics, the entropy in the small system fluctuates since it exchanges energy, particles, heat, work, entropy, etc., with the reservoir. We are interested in the rate of the entropy generation $\sigma$ in the small system and its average over a finite time $t$, $\overline{\sigma}_t\equiv \frac{1}{t}\int_0^t ds \sigma(s)$. The FT claims that the probability $P(\overline{\sigma}_t)$, i.e., the probability such that the entropy generation rate is $\overline{\sigma}_t$, strictly satisfies the following equation
\begin{equation}
\frac{P(\overline{\sigma}_t=A)}{P(\overline{\sigma}_t=-A)}=\exp\left(\frac{At}{k_\textrm{B}}\right).
\label{Eq_FT}
\end{equation} 
This equation is derived from the microreversibility and conservation laws. According to this equation, for a large $t$, the probability of the entropy increase (numerator of the left-hand side) is overwhelmingly larger than that of the entropy decrease (denominator of the left-hand side) because $A>0$. In this sense, this equation corresponds to the second law of thermodynamics. The FT has been studied as a new guiding principle in statistical mechanics because Eq.~(\ref{Eq_FT}) can reproduce the fundamental equations of linear response theory, such as the fluctuation-dissipation relations~\cite{GallavottiPRL1996} and the Onsager-Casimir reciprocity~\cite{SaitoPRB2008,UtsumiPRB2009}.

Experimentally, the FT has been verified by observing the motion of particles in fluids and RNA molecules, for example~\cite{WangPRL2002,BustamantePhysToday2005}. It has also been confirmed in transport measurements, e.g., performed on bulk resistors~\cite{GarnierPRE2005} and in electron-counting experiments using quantum dots (QDs)~\cite{UtsumiPRB2010,KungPRX2012}.

\subsection{FT in quantum transport}
While the experiments mentioned above have verified the FT in classical systems, the authors reported the first experiment examining the FT in a quantum system~\cite{NakamuraPRL2010,NakamuraPRB2011}, based on a theoretical proposal~\cite{UtsumiPRB2009}. Below, we briefly introduce this experiment.

We consider a general situation where a voltage induces a current $\langle I \rangle$ in a sample. The current-voltage characteristics can be written as a polynomial of $V$ as follows:
\begin{equation}
\langle I \rangle= G_1V + \frac{1}{2!} G_2 V^2 + \frac{1}{3!} G_3 V^3 + \cdots,
\label{Eq_IinV}
\end{equation}
where the first term represents Ohm's law, and $G_1$ is the conductance. When the current $\langle I \rangle$ is described only by the first term, which is often the case at low bias, the conventional linear response theory holds. On the other hand, at high bias, the system generally goes into a non-equilibrium state showing a non-linear response (suppose a largely stretched spring deviates from Hooke's law, as an example). In that case, higher-order terms in Eq.~(\ref{Eq_IinV}), characterized by the response coefficients $G_2, G_3$, and so on, become significant. In mesoscopic systems, $G_1$ is directly related to the transmission probability, as shown in Eq.~(\ref{LandauerConductance}). On the other hand, the higher-order response coefficients reflect electron correlation under non-equilibrium conditions~\cite{SanchezPRL2004,SpivakPRL2004,WeiPRL2005,LeturcqPRL2006}. Unlike $G_1$, which has Onsager-Casimir reciprocity, these quantities are not symmetric with respect to the magnetic field reversal and cause the nonreciprocal transport.

Similarly, the current noise $S$ can also be expressed by a polynomial of the applied voltage $V$ such that
\begin{equation}
S = S_0 + S_1V + \frac{1}{2!} S_2 V^2 + \cdots,
\label{Eq_SinV}
\end{equation}
where $S_0$ is the thermal noise. 

The fluctuation-dissipation relation $S_0 = 4 k_\textrm{B}T_\textrm{e}G_1$ holds between the first terms of Eq.~(\ref{Eq_IinV}) and Eq.~(\ref{Eq_SinV}), as we saw in Eq.~(\ref{thermalnoise})~\cite{JohnsonPR1928,NyquistPR1928}. This suggests analogous relations between the higher-order terms. Actually, the aforementioned FT predicts~\cite{UtsumiPRB2009} 
\begin{equation}
S_1=2k_\textrm{B} T_\textrm{e}G_2,
\label{Eq_FT_S1G2}
\end{equation}
for example. This relation between the second terms can be understood as follows. Electron transport occurs in a ``conductor-device-conductor'' system, which can be viewed as an exchange of electrons between two reservoirs via the device. A finite voltage $V$ between the two reservoirs causes a difference between their chemical potentials with $eV$. Let us consider the probability $P(N)$ of $N$ electrons flowing from the left to the right lead. Due to the time-reversal symmetry, the conservation of the particle number, and the energy conservation, the following equation holds~\cite{UtsumiPRB2009,NakamuraPRB2011}:
\begin{equation}
\frac{P(N)}{P(-N)}=\exp \left( \frac{eV}{k_\textrm{B}T_\textrm{e}}N \right).
\label{Eq_FT_electron}
\end{equation}
The original FT, expressed in Eq.~(\ref{Eq_FT}), relates the probabilities of the entropy generation and reduction processes. In the electron transport, the entropy generation is related to Joule heating. When $N$ electrons move across a potential difference of $eV$, the Joule heat of $NeV$ is produced after equilibration in the reservoir at temperature $T_\textrm{e}$ so that the entropy generation is $NeV/T_\textrm{e}$. Thus, we can evaluate the probability of the transfer of $N$ electrons and obtain Eq.~(\ref{Eq_FT_electron}). The equation gives a strong constraint on electrical conduction, from which Eqs.~(\ref{thermalnoise}) and (\ref{Eq_FT_S1G2}) are derived, as described in Ref.~[\onlinecite{NakamuraPRB2011}].

\begin{figure*}[bthp]
\begin{center}
\includegraphics[width=12.5cm]{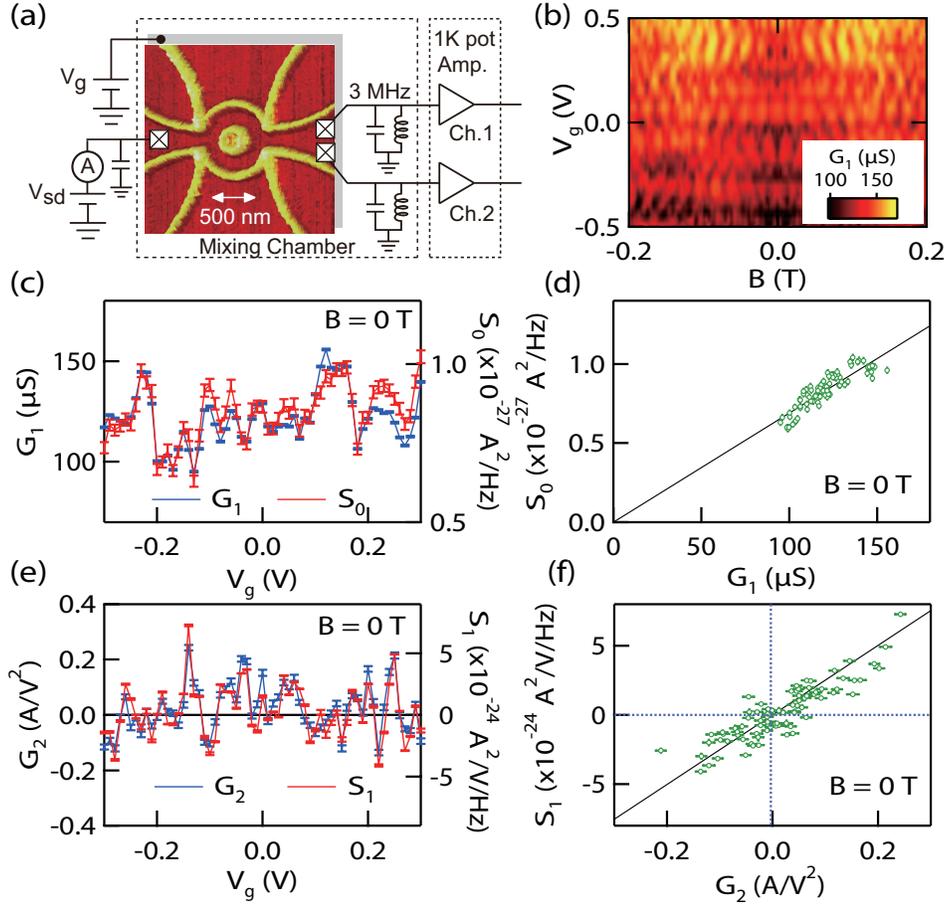} 
\end{center}
\caption{(Color online) (a) Atomic-force micrograph of the AB ring with the DC and noise measurement setup in a dilution refrigerator. The in-plane gates defined by the oxide lines are grounded in this experiment. The carrier density in the AB ring can be controlled by the gate voltage $V_\textrm{g}$ applied to the back gate of the substrate. (b) Conductance of the AB ring as a function of $V_\textrm{g}$ and $B$. (c) $G_1$ (left axis) and $S_0$ (right axis) as a function of $V_\textrm{g}$ at $B=0$~T. (d) $S_0$ are plotted as a function of $G_1$. The solid line indicates the fluctuation-dissipation relation of $S_0 = 4 k_\textrm{B}T_\textrm{e}G_1$ with $T_\textrm{e}=125$~mK. (e) $G_2$ (left axis) and $S_1$ (right axis) as a function of $V_\textrm{g}$. (f) $S_1$ are plotted as a function of $G_2$. The solid line is the linear fit ($S_1 = 3.64 \times 4k_\textrm{B}T_\textrm{e}G_2$ with $T_\textrm{e}=125$~mK). Figures are reprinted from Ref.~[\onlinecite{NakamuraPRL2010}].{\copyright} (2010) American Physical Society.} 
\label{Fig_NakamuraPRL2010}
\end{figure*}

Electron counting experiments to verify Eq.~(\ref{Eq_FT_electron}) have been performed~\cite{UtsumiPRB2010,KungPRX2012}. They used QDs coupled to a nearby QPC as a charge detector. This charge-detection technique was discussed in Sect.~\ref{sec:FCS}. In the counting experiments, the electron transport is not coherent but in the incoherent tunneling regime. 

Relation Eq.~(\ref{Eq_FT_S1G2}) is for nonlinear non-equilibrium regime and is a new ``nonequilibrium fluctuation relation'' that goes beyond the known fluctuation-dissipation relation. In our experiment, we aimed to demonstrate this equation in the quantum coherent regime. The device is an Aharonov-Bohm (AB) ring (460 nm in diameter) fabricated in a GaAs/AlGaAs 2DES, as shown in Fig.~\ref{Fig_NakamuraPRL2010}(a). The electron interference in the ring was controlled by applying magnetic field $B$ or gate voltage $V_\textrm{g}$. Figure~\ref{Fig_NakamuraPRL2010}(b) shows the conductance in the $B$-$V_\textrm{g}$ plane, where the periodic oscillations of the conductance as a function of the magnetic field signal the AB oscillation, manifesting the coherent electron transport.

First, we discuss the results obtained at equilibrium ($V=0$). Figure~\ref{Fig_NakamuraPRL2010}(c) shows the conductance $G_1$ and the thermal noise $S_0$ when $V_\textrm{g}$ is varied. As the fluctuation-dissipation relation tells us, the behaviors of $G_1$ and $S_0$ agree with each other. In Fig.~\ref{Fig_NakamuraPRL2010}(d), $S_0$ is plotted as a function of $G_1$. From this linear fit, we estimate electron temperature $T_\textrm{e}$ = 125~mK.

Then, we estimated $G_2$ and $S_1$ from the current-voltage characteristics and the voltage dependence of the current noise, respectively. The $V_\textrm{g}$ dependence of these two quantities is shown in Fig.~\ref{Fig_NakamuraPRL2010}(e). We see that the behaviors of $G_2$ and $S_1$ are coincident with each other. Figure~\ref{Fig_NakamuraPRL2010}(f) tells that the experimental data yields $S_1 = 3.64 \times 4k_\textrm{B}T_\textrm{e}G_2$. Although the numerical factor does not agree with that of the theoretical prediction, $S_1=2k_\textrm{B} T_\textrm{e}G_2$, Fig.~\ref{Fig_NakamuraPRL2010}(f) manifests the presence of nontrivial proportionality between them ($S_1 \propto G_2$).

Some readers may consider that the current noise should be expressed by Eq.~(\ref{ShotTheory}) even in the non-equilibrium state. Equation (\ref{ShotTheory}) is an expression derived for free fermion systems and does not include the influence of electron-electron correlation, except for the Pauli exclusion principle. In Sect.~\ref{subsub:kondonoise}, we have already seen that Eq.~(\ref{ShotTheory}) is not sufficient to calculate the noise in the Kondo effect. The correlation is phenomenologically taken into account as an ``effective charge'', while, microscopically, there are multiple scattering mechanisms. More generally, when a system is driven out of equilibrium, many-body effects often play significant roles~\cite{SanchezPRL2004,SpivakPRL2004,WeiPRL2005,LeturcqPRL2006}. Such correlation effects are not included in the framework of the Landauer-B\"{u}ttiker formula based on the single-particle picture. 

In the present experiment, we also observed a distinct nonreciprocity arising from the quantum many-body effect in a non-equilibrium state at a finite magnetic field~\cite{LeturcqPRL2006,NakamuraPRL2010}. We demonstrated that the relation derived from FT is relevant even in this case~\cite{NakamuraPRL2010,NakamuraPRB2011}. 

\section{Conclusion and Future Perspectives}
\label{sec:closing}
In this review, we have presented the advances in mesoscopic shot-noise experiments over the past two decades. We hope that this review helps in convincing readers of the advantages of shot-noise measurements and, for experimentalists, to perform the measurements themselves. 

While we have made every effort to cover a wide range of topics, unfortunately, several important ones could not be included in this review. Below, we would introduce some of them.

The first one is higher-order cumulants. In this review, we have mainly focused on current noise given by $\langle \Delta I^2 \rangle$, which is the second-order cumulant, i.e., the variance of the number of transmitted electrons. However, as briefly mentioned in Sect.~\ref{sec:FCS}, higher-order cumulants such as $\langle \Delta I^3\rangle$ and $\langle \Delta I^4\rangle\cdots$, which can be evaluated in full-counting statistics, also provide fruitful information on transport phenomena~\cite{LevitovJMP1996}. Experimentally, third-order and higher-order ones have been measured for several devices, e.g., tunnel junctions~\cite{ReuletPRL2003}, quantum dots~\cite{GustavssonPRL2006,FujisawaScience2006}, avalanche diodes~\cite{GabelliPRB2009}, and short diffusive conductors~\cite{PinsollePRL2018}. However, measuring higher-order cumulants is still challenging for most mesoscopic systems. We believe that advances in experimental techniques will allow us to measure full-counting statistics in various samples and that the results will provide deeper insight into their transport properties.

The second is shot noise at high frequencies. While we have mainly focused on shot noise in the low-frequency limit, shot noise becomes frequency-dependent and sometimes provides essential information on electron dynamics at high frequencies. For example, the Josephson relation of fractional quasiparticles in the fractional quantum Hall state is a representative result obtained by measuring high-frequency shot noise~\cite{KapferScience2019,BisogninNatCom2019}.
High-frequency shot noise in the Kondo state~\cite{DelagrangePRB2018,FerrierJLTP2019} is a challenging issue for future experiments. Such a measurement will allow us to evaluate the dynamics of the Kondo singlet below and above the Kondo temperature.

Thirdly, expanding the scope of shot-noise measurements is a promising direction. Most of the shot-noise studies in mesoscopic physics have been performed on semiconductor devices fabricated in a 2DES. However, as described in this review, the noise measurement is applicable to other systems such as magnetic tunnel junctions, atomic/molecular junctions, carbon nanotubes, and spintronics devices. In this context, the shot-noise measurement performed on copper-oxide high-temperature superconductors~\cite{ZhouNature2019} is a notable example. In that study, an increase in the effective charge at a temperature above the superconducting transition temperature was observed, implying a precursor phenomenon of Cooper-pair formation. 

Finally, shot-noise measurements can be combined with other experimental techniques, e.g., scanning microscopy~\cite{MasseeRSI2018,WengScience2018}. Such experiments will provide unique information on charge, heat, and spin transport phenomena at the surface of a solid-state device.

In the 1990s, a shot-noise measurement itself was challenging. Today, however, it is a standard experimental technique in mesoscopic physics. There are still many fascinating theoretical proposals left unexplored, which are attracting much attention from experimentalists. We hope that this review will encourage many researchers to join this rich research field.

\begin{acknowledgments}
For the shot-noise theory described in Sect.~\ref{subsec:noise_in_quantum_transport}, we thank Takeo Kato for his enlightening lecture note~\cite{KatoBussei2014}. We acknowledge instructive comments on the writing from Heorhii Bohuslavskii and fruitful discussions in the meeting of the Cooperative Research Project of RIEC, Tohoku University. This work was partially supported by JSPS KAKENHI (Grants No. JP19H00656, JP19H05826, JP16H06009, 19H05603) and JST PRESTO (Grant No. JP17940407).
\end{acknowledgments}

\bibliography{noise_review.bib}
\end{document}